\documentclass[11pt]{article}
\pdfoutput=1
\usepackage{float,jheppub,slashed}
\usepackage{braket,commath}
\usepackage{subcaption,siunitx}
\usepackage{dcolumn}
\usepackage{multirow}
\usepackage{stackengine}
\usepackage{booktabs}
\usepackage{subcaption}
\usepackage{lscape}
\usepackage{mathrsfs}

\newcommand{\dd}{{\mathrm d}}

\title{Entanglement and Bell Nonlocality in $\tau^+ \tau^-$ at the LHC using Machine Learning for Neutrino Reconstruction}

\author[4]{Yulei Zhang,}
\author[1,2,3]{Bai-Hong Zhou,}
\author[1,2,4]{Qi-Bin Liu,}
\author[5]{Tong Arthur Wu,}
\author[1,2,3]{Shu Li,}
\author[5]{Tao Han,}
\author[4]{Shih-Chieh Hsu,}
\author[5]{Matthew Low}

\affiliation[1]{\it Tsung-Dao Lee Institute, Shanghai Jiao Tong University, Shanghai 201210, China}
\affiliation[2]{\it Institute of Nuclear and Particle Physics, School of Physics and Astronomy, Shanghai Jiao Tong University, Shanghai 200240, China}
\affiliation[3]{\it State Key Laboratory of Dark Matter Physics, Key Laboratory for Particle Astrophysics and Cosmology (KLPPAC-MoE), Shanghai Key Laboratory for Particle Physics and Cosmology (SKLPPC), Shanghai 201210, China
}
\affiliation[4]{\it University of Washington, Seattle, WA 98195, United States of America}
\affiliation[5]{PITT PACC, Department of Physics and Astronomy,\\ University of Pittsburgh, 3941 O’Hara St., Pittsburgh, PA 15260, USA}

\emailAdd{yulei.zhang@cern.ch, schsu@uw.edu, mal431@pitt.edu}

\abstract{
Experiments at the CERN Large Hadron Collider (LHC) have accumulated an unprecedented amount of data corresponding to a large variety of quantum states. Although searching for new particles beyond the Standard Model of particle physics remains a high priority for the LHC program, precision measurements of the physical processes predicted in the Standard Model continue to lead us to a deeper understanding of nature at high energies. 
We carry out detailed simulations for the process $pp \to \tau^+\tau^- X$ to perform quantum tomography and to measure the quantum entanglement and the Bell nonlocality of the $\tau^+\tau^-$ two qubit state, including both statistical and systematic uncertainties.  By using advanced machine learning techniques for neutrino momentum reconstruction, we achieve precise measurements of the full spin density matrix, a critical advantage over previous studies limited by reconstruction challenges for missing momenta. Our analysis reveals a clear observation of Bell nonlocality with high statistical significance, surpassing 5$\sigma$, establishing $\tau^+ \tau^-$ as an ideal system for quantum information studies in high-energy collisions. Given its experimental feasibility and the high expected sensitivity for Bell nonlocality, we propose that $\tau^+ \tau^-$ should be regarded as the new benchmark system for quantum information studies at the LHC, complementing and extending the insights gained from the $t\bar{t}$ system.}

\makeatletter
\def\@fpheader{\vspace{1.5em}}
\makeatother

\begin{document}

\maketitle
\flushbottom

\section{Introduction}
\label{sec:introduction}

Experiments at the CERN Large Hadron Collider (LHC) have advanced our understanding of particle physics, probing the shortest distances ever explored.
In addition to the miletone discovery of the Higgs boson, which is responsible for the generation of mass of elementary particles, the LHC has produced a large variety of quantum states with an unprecedented amount of data.  Although searching for new particles beyond the Standard Model (SM) of particle physics remains a high priority for the LHC program at the energy frontier, precision measurements of the physical processes predicted in the SM continue to lead us to a deeper understanding of nature at high energies.

In recent years, there has been a significant transformation in how we collect, analyze, and interpret data in collider experiments. This shift is driven by two major factors: the integration of quantum information science into high-energy physics and the rapid advances in machine learning (ML). Both fields have individually revolutionized our understanding of complex systems, and their combination has the potential to unlock new perspectives on fundamental physics. Quantum information science provides powerful tools for characterizing and quantifying quantum correlations in collider experiments, while machine learning enables the extraction of intricate patterns from vast datasets, improving both efficiency and accuracy in data analysis. Together, these approaches offer novel insights into the quantum nature of particle interactions, opening doors to previously inaccessible phenomena, and expanding the scope of high-energy physics research.

While quantum mechanics is intrinsic to the description of collisions that take place at high-energy colliders, it is only recently that quantum entanglement, predicted by quantum mechanics, has been measured explicitly at the unprecedented energies in collider experiments. In this context, the most common constituents of these quantum experiments are the spins of outgoing particles.  The first system that was used to study quantum entanglement at the LHC was the $t\bar{t}$ system, with both tops decaying leptonically~\cite{Afik:2020onf}.  This was extended to Bell nonlocality~\cite{Fabbrichesi:2021npl,Severi:2021cnj,Afik:2022kwm,Aguilar-Saavedra:2022uye,Fabbrichesi:2022ovb} and to semi-leptonic top pair decays~\cite{Dong:2023xiw,Han:2023fci}.  Beyond these first steps, beyond the SM physics applications~\cite{Aoude:2022imd,Severi:2022qjy,Aoude:2023hxv,Maltoni:2024tul,Duch:2024pwm,Fabbrichesi:2025ywl} have been studied as have different quantum information variables like quantum discord~\cite{Afik:2022dgh,Han:2024ugl}, quantum magic~\cite{White:2024nuc}, and others~\cite{Aguilar-Saavedra:2023hss,Aguilar-Saavedra:2024hwd,Aguilar-Saavedra:2024fig,Maltoni:2024csn}.  The top pair system was the focus of early work because of its theoretical simplicity as a two qubit system and its experimental feasibility  due to the established method of reconstructing the top quark kinematics. If we only consider particles that decay as qubits, there are only two bipartite qubit systems available in the LHC environment: $t\bar{t}$ and $\tau^+ \tau^-$.

A number of general advances in our theoretical understanding of quantum systems at colliders have also been made such as the role of fictitious states~\cite{Cheng:2023qmz,Cheng:2024btk} and the reconstruction of the density matrix using the kinematic method~\cite{Cheng:2024rxi}.  Fictitious states explain the dependence of quantum tomography on the choice of spin quantization basis.  This has important implications for observables that classify states into convex sets, like concurrence~\cite{Afik:2022kwm}, and into non-convex sets, like quantum discord~\cite{Han:2024ugl}. The kinematic method uses the spin dependence of the differential cross sections to perform quantum tomography using the production kinematics. The statistical uncertainties using this method are substantially smaller than those using the decay method~\cite{Cheng:2024rxi} due to the simplicity of the event reconstruction and the sensitivity to the observable in question. 

In addition to the $t\bar{t}$ system, there has been early work on a number of other final states, including $WW$~\cite{Aguilar-Saavedra:2022mpg,Fabbrichesi:2023cev,Morales:2023gow,Bi:2023uop,Grossi:2024jae}, $ZZ$~\cite{Aguilar-Saavedra:2022wam,Fabbrichesi:2023cev,Morales:2023gow,Bernal:2023ruk,Ruzi:2024cbt,Wu:2024ovc,Grossi:2024jae}, $\Lambda_b \bar{\Lambda}_b$~\cite{Afik:2024uif}, $Y\bar{Y}$~\cite{Wu:2024asu}, $\phi\phi$~\cite{Gabrielli:2024kbz}, $\mu^+ \mu^-$~\cite{Gao:2024leu}, and $q\bar{q}$~\cite{Cheng:2025cuv}.  Some of these other states go beyond two qubits and probe qutrits, multipartite systems~\cite{Sakurai:2023nsc,Aguilar-Saavedra:2024whi,Morales:2024jhj,Subba:2024mnl}, and other quantum informational concepts~\cite{Altomonte:2023mug,Barr:2024djo,Altomonte:2024upf}.  Studies of the $\tau^+ \tau^-$ system have been suggested at $e^+ e^-$ colliders, including at the FCC-ee~\cite{Altakach:2022ywa,Fabbrichesi:2024wcd}, at the CEPC~\cite{Ma:2023yvd}, at Belle-II~\cite{Ehataht:2023zzt}, and at the BEPC~\cite{Han:2025ewp}.\footnote{See also Ref.~\cite{larry}.}

Generally, $t\bar{t}$ has been considered the canonical qubit system in hadron colliders, whereas $\tau^+ \tau^-$ has been considered the canonical qubit system in lepton colliders.  Although both systems have at least two neutrinos in their final states, in the $t\bar{t}$ system there are enough kinematic constraints, save for discrete ambiguities, to solve the neutrino momenta at hadron or lepton colliders, while for $\tau^+ \tau^-$ there are only enough constraints at lepton colliders due to the well-determined collision energies. Because of this difficulty, there has not been a study of $\tau^+ \tau^-$ state at the LHC.\footnote{Ref.~\cite{Fabbrichesi:2022ovb} presented theoretical-level results for  $\tau^+ \tau^-$ at the LHC but did not provide a practical way to measure the quantum state.  See also Ref.~\cite{LoChiatto:2024dmx}.}  In this work, we leverage the power of ML to open up this notoriously difficult channel and show that it has excellent sensitivity to detecting quantum entanglement and Bell nonlocality, already with the LHC's current dataset.

Recent advances in generative ML models have introduced novel approaches to neutrino reconstruction, which address the limitations of conventional methods. In particular, diffusion models~\cite{diffusion_overview, diffusion_propose} have emerged as a powerful technique for event-level inference, especially in cases where analytical solutions are impossible. Unlike deterministic reconstruction methods, diffusion models learn the underlying probability distribution of missing kinematic components conditioned on observed event data, enabling a probabilistic treatment of neutrino momenta.

While diffusion models have recently been explored in high-energy physics for applications such as calorimeter shower generation~\cite{calodiffusion,Mikuni:2023tqg,Acosta:2023zik,Buhmann:2023bwk} and feature generation~\cite{Mikuni:2023dvk,Devlin:2023jzp,Sengupta2024:diff, PhysRevD.109.012005}, their broader potential remains largely unexplored, particularly in neutrino reconstruction for fully unconstrained systems. Traditional kinematic methods can still be applied in these scenarios, but their performance is often limited by high failure rates and poor resolution due to the lack of sufficient constraints. This work seeks to extend the applicability of diffusion models by addressing the challenge of reconstructing invisible particles in analytically unsolvable systems.

Building on these advancements, we introduce a diffusion-based approach for neutrino reconstruction in the $\tau^+ \tau^-$ system. Using the Point-Edge Transformer (PET) architecture~\cite{Mikuni:2024qsr, Mikuni:2025tar}, we train a generative model to infer neutrino momenta directly from event-level observables. By preserving event-level correlations and providing a statistical representation of the kinematic phase space, this method significantly improves reconstruction accuracy. The resulting framework enables precision studies of quantum information at hadron colliders and establishes a foundation for applying generative models to other unconstrained systems in collider physics.

Our results demonstrate that the $\tau^+ \tau^-$ system offers a robust and experimentally accessible platform for studying quantum entanglement and Bell nonlocality at the LHC. By using advanced machine learning techniques for neutrino reconstruction, we achieve precise measurements of the full spin density matrix, a critical advantage over previous studies limited by reconstruction challenges. Our analysis reveals a clear violation of Bell inequalities with high statistical significance, surpassing 5$\sigma$, establishing $\tau^+ \tau^-$ as an ideal system for quantum information studies in high-energy collisions. Given its experimental feasibility and the ease with which Bell nonlocality can be probed, we propose that $\tau^+ \tau^-$ should be regarded as the new benchmark system for quantum information studies at the LHC, complementing and extending the insights gained from the $t\bar{t}$ system.

The rest of the paper proceeds as follows.  In Sec.~\ref{sec:quantum} we introduce the relevant quantum information theory and its application to $pp \to \tau^+ \tau^- X$.  Our simulation details are provided in Sec.~\ref{sec:simulation}.  Next, in Sec.~\ref{sec:neutrino_rec} we describe the network architecture we employ to reconstruct neutrino momenta and demonstrate its efficacy.  Sec.~\ref{sec:reconstruction} discusses backgrounds, systematic uncertainties, and our statistical methods.  The results, of significance higher than $5\sigma$ for both entanglement and Bell nonlocality, are presented in Sec.~\ref{sec:results}.  We conclude in Sec.~\ref{sec:conclusions}.  Some technical details of the full density matrix and the reconstructed event kinematics are included in two appendices.

\section{Quantum Tomography for $\tau^+ \tau^-$}
\label{sec:quantum}

\subsection{Quantum Correlations}

A two qubit quantum state is described the density matrix $\rho$ which has the Fano-Bloch decomposition~\cite{Fano:1983zz}
\begin{equation}
\label{eq:fano}
\rho = \frac{1}{4} \left( \mathbb{I}_2 \otimes \mathbb{I}_2
+ \sum_i B^+_i \sigma_i \otimes \mathbb{I}_2
+ \sum_j B^-_j \mathbb{I}_2 \otimes \sigma_j
+ \sum_{ij} C_{ij} \sigma_i \otimes \sigma_j \right),
\end{equation}
where $\mathbb{I}_2$ is the two-dimensional identity matrix, $B^+_i$ characterizes the net polarization of $\tau^+$, $B^-_j$ characterizes the net polarization of $\tau^-$, and $C_{ij}$ describes the spin correlations, and $i,j$ run over $1,2,3$.  Quantum tomography is performed via the reconstruction of the  density matrix and thus we can compute the quantum observables of interest.

At the LHC the spin of the $\tau^+$ is the first qubit and the spin of the $\tau^-$ is the second qubit.  The quantum state that describes the $\tau^+ \tau^-$ state depends on the production channels and on the final state kinematics.  Different regions of kinematic phase space, parameterized by the invariant mass of the system $m_{\tau\tau}$ and by the scattering angle $\theta$ of $\tau^-$ with respect to the incoming beam, lead to different $\tau^+ \tau^-$ quantum states.

Quantum entanglement is the phenomenon of two subsystems that cannot be described independently.  This ``spooky action at a distance'' only occurs in quantum mechanical systems and is evaluated by choosing an entanglement monotone~\cite{Horodecki:2009zz}.  The concurrence $\mathcal{C}$ of a quantum state $\rho$ is a useful entanglement monotone which is $\mathcal{C} = 0$ for a separable system and $0 < \mathcal{C} \leq 1$ for an entangled system~\cite{Hill:1997pfa}.  A larger concurrence indicates a greater degree of entanglement.

For a bipartite qubit system the concurrence is given by~\cite{Wootters:1997id} 
\begin{equation}
\mathcal{C} = \max(0,\lambda_1 - \lambda_2 - \lambda_3 - \lambda_4),
\end{equation}
where $\lambda_i = \sqrt{r_i}$ ($i=1,2,3,4$).  The quantities $r_i$ are the eigenvalues, in descending magnitude, of the matrix  $\rho (\sigma_2 \otimes \sigma_2) \rho^* (\sigma_2 \otimes \sigma_2)$.  All of the $r_i$ are positive~\cite{Wootters:1997id}.  To reiterate:
\begin{equation}
\left\{
\begin{array}{l l}
\mathcal{C} =0         & \qquad\text{separable}, \\
0 < \mathcal{C} \leq 1 & \qquad\text{entangled}.
\end{array}
\right.
\end{equation}
Bell nonlocality is a stronger form of correlation than entanglement and is measured by whether or not a quantum state violates Bell's inequality~\cite{Bell:1964kc}.  For two qubits Bell's inequality is given by the Clauser-Horne-Shimony-Holt (CHSH) inequality~\cite{Clauser:1969ny}
\begin{equation} \label{eq:CHSH}
| \langle \vec{a}_1 \cdot \vec{\sigma} \otimes \vec{b}_1 \cdot \vec{\sigma} \rangle
- \langle \vec{a}_1 \cdot \vec{\sigma} \otimes \vec{b}_2 \cdot \vec{\sigma} \rangle
+ \langle \vec{a}_2 \cdot \vec{\sigma} \otimes \vec{b}_1 \cdot \vec{\sigma} \rangle
+\langle \vec{a}_2 \cdot \vec{\sigma} \otimes \vec{b}_2 \cdot \vec{\sigma} \rangle |
\leq 2.
\end{equation}
where $\vec{a}_1$ and $\vec{a}_2$ are measurement axes for the first qubit and $\vec{b}_1$ and $\vec{b}_2$ are measurement axes for the second qubit.  Rearranging Eq.~\eqref{eq:CHSH}, adjusting the normalization, we can write the Bell variable as
\begin{equation}
\mathcal{B}(\vec{a}_1, \vec{a}_2, \vec{b}_1, \vec{b}_2) = \frac{1}{\sqrt{2}} | \langle \vec{a}_1 \cdot \vec{\sigma} \otimes \vec{b}_1 \cdot \vec{\sigma} \rangle
- \langle \vec{a}_1 \cdot \vec{\sigma} \otimes \vec{b}_2 \cdot \vec{\sigma} \rangle
+ \langle \vec{a}_2 \cdot \vec{\sigma} \otimes \vec{b}_1 \cdot \vec{\sigma} \rangle
+\langle \vec{a}_2 \cdot \vec{\sigma} \otimes \vec{b}_2 \cdot \vec{\sigma} \rangle | - \sqrt{2}.
\end{equation}
While there is a known maximization over $\vec{a}_1$, $\vec{a}_2$, $\vec{b}_1$, and $\vec{b}_2$, in the collider setting it is often preferred to use the approximately maximal form of~\cite{Severi:2021cnj}
\begin{equation}
\label{eq:Bell}
\mathcal{B} = \max_{ij} | C_{ii} \pm C_{jj} | - \sqrt{2}. 
\end{equation}
Using this normalization,
\begin{equation}
\left\{
\begin{array}{l l}
-\sqrt{2} \leq \mathcal{B}  \leq 0 
  & \qquad\text{Bell local}, \\
0 < \mathcal{B} \leq 2-\sqrt{2} 
  & \qquad\text{Bell nonlocal},
\end{array}
\right.
\end{equation}
The numerical value of the upper limit is $0.586$.

\subsection{The Decay Method}

The density matrix $\rho$ can be reconstructed in two ways.  The first, the decay method, uses the fact that while the spin of a $\tau$ is not measured directly at the LHC, the direction of the decay products of the $\tau$ is correlated with the $\tau$'s spin.  In particular, if we choose one of the $\tau$'s decay products, which we call the spin analyzer $A$, then how correlated its direction is with the $\tau$'s spin is described by the spin analyzing power $\kappa_A$.

The spin analyzing power varies from $-1$ to $+1$ corresponding to maximally anti-correlated and maximally correlated, respectively.  A spin analyzing power of $0$ indicates no correlation between the spin and the spin analyzer direction.

The density matrix can be reconstructed starting with the double-differential cross section
\begin{equation} \label{eq:diff2_xsec}
\frac{1}{\sigma}\frac{d^2\sigma}{d\cos\theta_{A,i} d\cos\theta_{B,j}}
= \frac{1}{4}\left( 1 
+ \kappa_A B^+_i \cos\theta_{A,i}
+ \kappa_B B^-_j \cos\theta_{B,j}
+ \kappa_A \kappa_B C_{ij} \cos\theta_{A,i} \cos\theta_{B,j}
\right). 
\end{equation}
The values $\kappa_A$ and $\kappa_B$ are the spin analyzing powers of the given spin analyzers of $\tau^+$ and $\tau^-$.  The $B^+_i$, $B^-_j$, and $C_{ij}$ coefficients are the Fano coefficients from Eq.~\eqref{eq:fano}.  The angle $\cos\theta_{A,i}$ is the angle between the momentum direction of the spin analyzer $A$ from $\tau^+$ and the axis $\hat{i}$ and the angle $\cos\theta_{B,j}$ is the angle between the momentum direction of the spin analyzer $B$ from $\tau^-$ and the axis $\hat{j}$.

Integrating Eq.~\eqref{eq:diff2_xsec} lets us extract the $\tau^+$ polarization $B_i^+$
\begin{equation}
\label{eq:xsec_b+}
 \frac{1}{\sigma}\frac{d \sigma}{d\cos\theta_{A,i}}
= \frac{1}{2} \left( 1 + \kappa_A B^+_i \cos\theta_{A,i} \right), 
\end{equation}
the $\tau^-$ polarization $B_j^-$
\begin{equation}
\label{eq:xsec_b-}
\frac{1}{\sigma}\frac{d \sigma}{d\cos\theta_{B,j}}
 = \frac{1}{2} \left( 1 + \kappa_B B^-_j \cos\theta_{B,j} \right),
\end{equation}
and the spin correlation matrix $C_{ij}$
\begin{equation}
\label{eq:xsec_cij}
\frac{1}{\sigma}\frac{d \sigma}{d\cos\theta_{A,i} \cos\theta_{B,j}}
 =- \frac{1}{2} \left( 1 + \kappa_A \kappa_B C_{ij} \cos\theta_{A,i} \cos\theta_{B,j} \right) \log | \cos\theta_{A,i} \cos\theta_{B,j} |.
\end{equation}
The parameters can be extracted by computing the mean, computing the asymmetry of the distributions, or by fitting to the distribution.

The decay channels that we use in this study, along with their spin analyzing powers and branching ratios, are shown in Table~\ref{tab:decays}.  While the hadronic decays have the largest spin analyzing powers and sizable branching ratios, the leptonic decays are still important for the overall sensitivity because of trigger requirements, as will be discussed in Sec.~\ref{sec:trigger}.

\begin{table}
\centering
\caption{Relevant decay channels of $\tau$ with the corresponding spin analyzing power and branching ratio.  The spin analyzer is underlined.}
\label{tab:decays}
\begin{tabular}{ccc}
\hline
Decay & Spin Analyzing Power & Branching Ratio \\
\hline
$\underline{\pi}\nu_\tau$ & 1.00 & 10.8\% \\
$\underline{\rho}(\pi \pi^0) \nu_\tau$ & 0.41 & 25.5\% \\
$\underline{e}\nu_e\nu_\tau$ & $-0.33$ & 17.8\% \\
$\underline{\mu}\nu_\mu\nu_\tau$ & $-0.34$ & 17.4\% \\
\hline
\end{tabular}
\end{table}

\subsection{The Kinematic Method}

An alternative to the decay method is the kinematic method which uses the dependence of the differential cross section on spins of the $\tau^+ \tau^-$ state.  For this method, rather than measuring spin correlations in the rest frames of the $\tau^+$ and $\tau^-$, one only needs to measure the velocity $\beta$ and the scattering angle $\theta$ of the $\tau^+$ relative to the beam in the center-of-mass frame.

At the LHC, the leading processes that lead to the $\tau^+ \tau^-$ final state are $q\bar{q} \to \gamma \to \tau^+ \tau^-$ and $q\bar{q} \to Z \to \tau^+ \tau^-$.  For illustration, in this section we show results considering only the dominant contribution from the on-shell $Z$.  We work in the helicity basis specified by $\{ \hat{k}, \hat{n}, \hat{r} \}$ where $\hat{k}$ is the momentum direction of the $\tau$, $\hat{r} = (\hat{z} - \hat{k} \cos\theta)/\sin\theta$, and $\hat{n} = \hat{r} \times \hat{k}$, with all the directions defined in the center-of-mass frame.

For this case, the Fano coefficients, which parametrize the density matrix, are~\cite{Cheng:2024rxi} 
\begin{equation}
C_{kk} = 1, \;\;
C_{nn} = \frac{g_A^{\tau 2} - g_V^{\tau 2}}{g_A^{\tau 2} + g_V^{\tau 2}} \frac{\sin^2\theta}{1+\cos^2\theta}, \;\;
C_{rr} = \frac{g_V^{\tau 2} - g_A^{\tau 2}}{g_A^{\tau 2} + g_V^{\tau 2}} \frac{\sin^2\theta}{1+\cos^2\theta}, \;\;
B^\pm_k = \frac{2 g_A^\tau g_V^\tau}{g_A^{\tau 2} + g_V^{\tau 2}}.
\end{equation}
where $g_A^\tau = - 1/4$ and $g_V^\tau = -1/4+\sin^2\theta_w$ where $\theta_w$ is the Weinberg angle.  The other spin correlation and polarization elements are zero: $C_{kr} = C_{rk} = C_{kn} = C_{nk} = C_{nr} = C_{rn} = B^\pm_r = B^\pm_n = 0$.  There is no dependence on $\beta$ because the $Z$ is assumed to be on shell.

In our signal region (SR), to be discussed in Sec.~\ref{sec:reconstruction}, the density matrix is sensitive to the off-shell $Z$ amplitude and the inteference with the photon such that the full density matrix must be used.  The complete set of Fano coefficients is given in Appendix~\ref{app:densitymatrix}.

The concurrence for the on-shell-$Z$ spin correlation matrix is
\begin{equation}
\mathcal{C} = \frac{\sin^2\theta}{1+\cos^2\theta}\frac{g_A^{\tau 2} - g_V^{\tau 2}}{g_A^{\tau 2} + g_V^{\tau 2}}.
\end{equation}
Above the concurrence reaches a peak value, at $\theta = \pi/2$, of $\mathcal{C}=0.99$ which is nearly maximally entangled.

The Bell variable for the same quantum state is
\begin{equation}
\mathcal{B} = \frac{2}{1+\cos^2\theta} \frac{g_A^{\tau 2} + g_V^{\tau 2} \cos^2\theta}{g_A^{\tau 2} + g_V^{\tau 2}} - \sqrt{2}.
\end{equation}
With our definition in Eq.~\eqref{eq:Bell}, Bell nonlocality corresponds to $\mathcal{B}>0$.  At $\theta = \pi/2$ the Bell variable reaches $\mathcal{B}=0.57$ which is nearly the maximal possible value.

\subsection{The $\tau^+ \tau^-$ State}

\begin{figure}[h]
    \centering
    \begin{subfigure}[b]{0.49\textwidth}
        \centering
        \includegraphics[width=\textwidth]
        {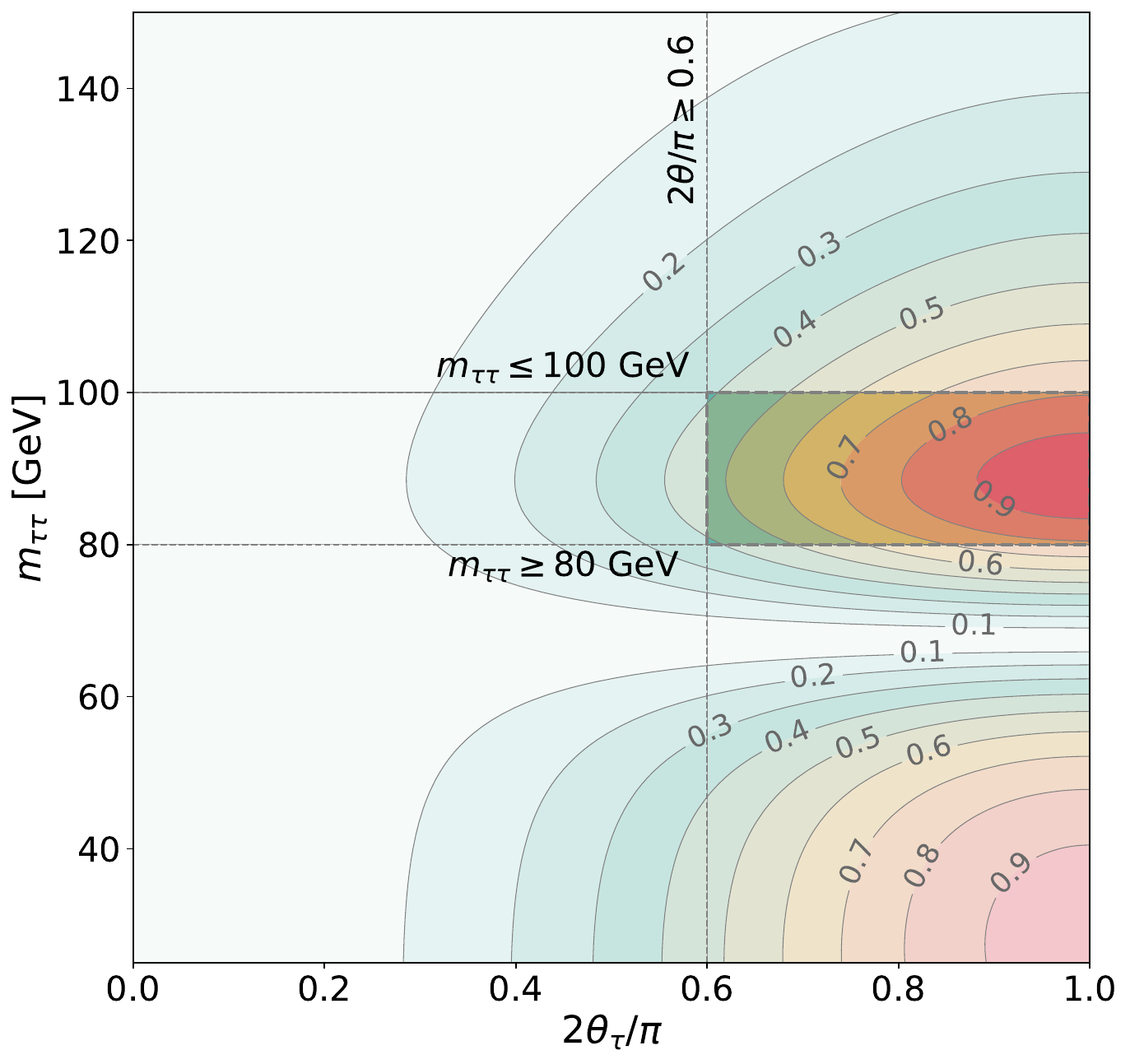}
        \caption{The concurrence $\mathcal{C}$.}
        \label{fig:kin_method_concurrence}
    \end{subfigure}
    \hfill
    \begin{subfigure}[b]{0.49\textwidth}
        \centering
        \includegraphics[width=\textwidth]
        {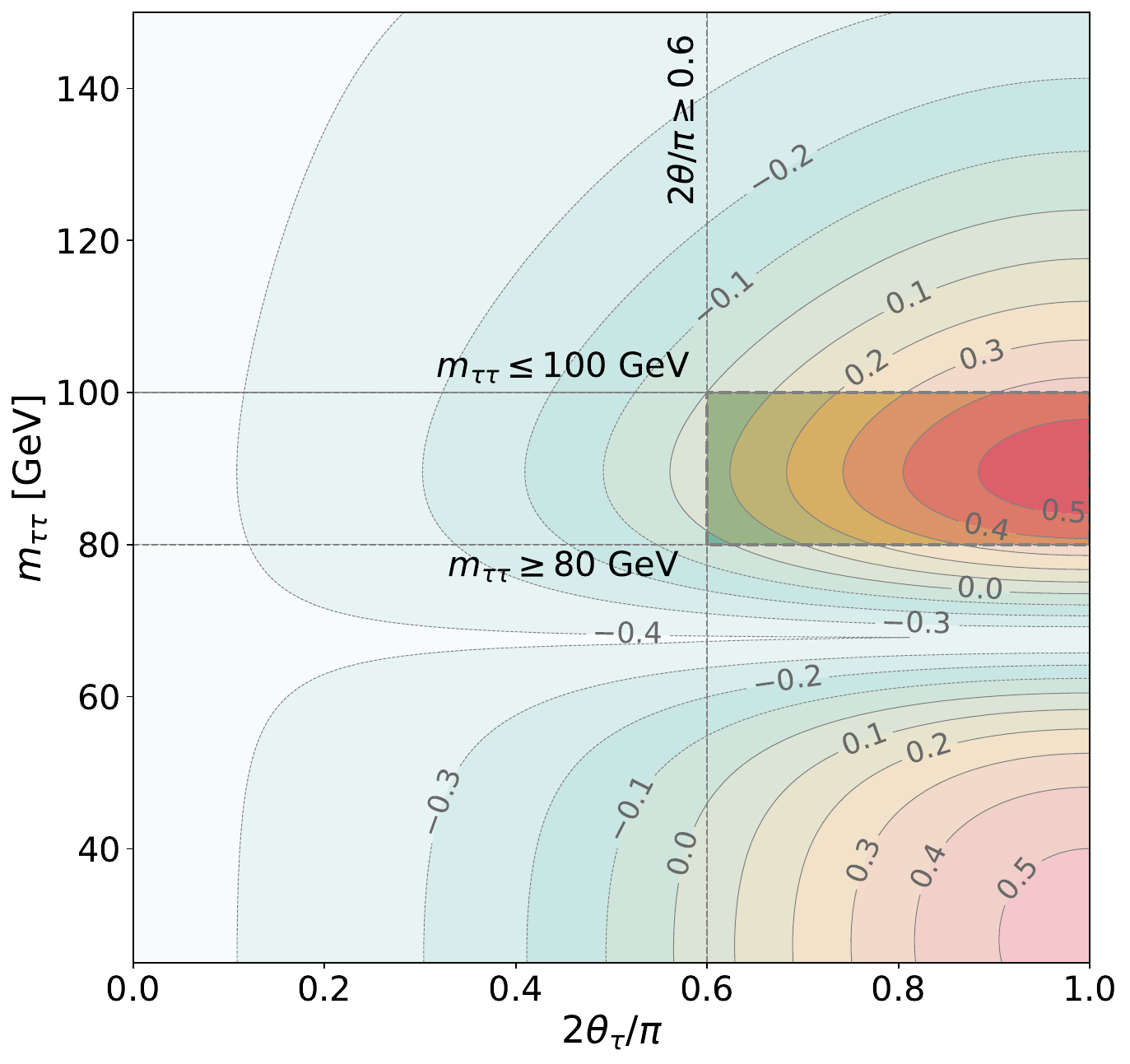}
        \caption{The Bell variable $\mathcal{B}$.}
        \label{fig:kin_method_bell}
    \end{subfigure}
    \caption{(a) The concurrence and (b) the Bell variable for $pp \to \tau^+ \tau^- X$ at the LHC in the plane of invariant mass $m_{\tau\tau}$ and scattering angle $\theta$ of $\tau^+$ relative to the beamline in the center-of-mass frame.  The black lines indicate our signal region, defined in Sec.~\ref{sec:reconstruction}.  $\mathcal{C}>0$ indicates entanglement and $\mathcal{B}>0$ indicates Bell nonlocality.}
    \label{fig:phase_space}
\end{figure}

The quantum state of $\tau^+ \tau^-$ in production $pp \to \tau^+ \tau^- X$ has been discussed in Ref.~\cite{Fabbrichesi:2022ovb} (see also~\cite{Banerjee:2023qjc}).  Below the $Z$ boson mass $m_Z$ where the $Z$ contribution is negligible, the $\tau^+ \tau^-$ state is dominantely produced through an $s$-channel photon leading to a triplet state that is nearly maximally entangled.
As the $Z$ contribution grows and the interference with the $\gamma$ contribution becomes sizable, the mixing between the $Z$ and $\gamma$ leads to a nearly separable state as seen in Fig.~\ref{fig:phase_space}.  The minimum entanglement occurs near $m_{\tau\tau} \approx 70~{\rm GeV}$.

Near the $Z$ pole the $Z$ contribution dominates and the $\tau^+ \tau^-$ system is once again in a spin triplet configuration leading to a state very close to a Bell state.  This results in nearly maximal entanglement and nearly maximal Bell nonlocality, as seen in Fig.~\ref{fig:phase_space}.  Far above the $Z$ pole, the mixing between the $Z$ and $\gamma$ again results in a nearly separable quantum state.

In this work, we focus on the $\tau^+ \tau^-$ state near the $Z$ pole.  Our signal region, to be defined in Sec.~\ref{sec:reconstruction}, is shown in Fig.~\ref{fig:phase_space} delineated by black lines.  This region results in a quantum state that exhibits a large amount of entanglement and Bell nonlocality.  The region at lower $m_{\tau\tau}$ is more difficult from a reconstruction and trigger standpoint.

\section{Simulation and Event Generation}
\label{sec:simulation}

The $pp \rightarrow \tau^+ \tau^- X$ events are generated at leading order using MadGraph 5~\cite{madgraph} at $\sqrt{s} = 13$ TeV.  The invariant mass of the final state is required to be greater than 20 GeV. $\tau$-leptons are then decayed using the \texttt{taudecay\_UFO} model in MadGraph, with the decays occurring via five orthogonal channels: $\pi\pi$, $\pi\rho$, $\rho\rho$, $\ell\pi$, and $\ell\rho$, as summarized in Table~\ref{tab:samples}.

\begin{table}[h]
\centering
\caption{Processes for the samples. Charge-conjugate processes are not included. Leptons $\ell$ include electrons ($e$) and muons ($\mu$).}
\label{tab:samples}
\begin{tabular}{cc}
\hline
Sample & Process \\
\hline
$\pi\pi$ & $p~p \rightarrow \tau^+~ \tau^- \rightarrow \pi^+ \bar{\nu_\tau}~ \pi^- \nu_\tau$ \\
$\pi\rho$ & $p~p \rightarrow \tau^+ ~\tau^- \rightarrow \pi^+ \pi^0 \bar{\nu_\tau}~ \pi^- \nu_\tau$ \\
$\rho\rho$ & $p~p \rightarrow \tau^+ ~\tau^- \rightarrow \pi^+ \pi^0 \bar{\nu_\tau}~ \pi^- \pi^0 \nu_\tau$ \\
$\ell\pi$ & $p~p \rightarrow \tau^+~ \tau^- \rightarrow \pi^+ \bar{\nu_\tau}~ \ell^- \nu_\tau \bar{\nu_\ell}$ \\
$\ell\rho$ & $p~p \rightarrow \tau^+~ \tau^- \rightarrow \pi^+ \pi^0 \bar{\nu_\tau}~ \ell^- \nu_\tau \bar{\nu_\ell}$ \\
\hline
\end{tabular}
\end{table}

The generated events are referred to as ``truth-level,'' as the kinematic information of the $\tau$-leptons and their decay products are known unambiguously, without the need for detector simulation and reconstruction. The particles in the truth-level events are input into Pythia 8~\cite{pythia}, which simulates the underlying event and performs parton showering and hadronization.  The detector response is simulated using Delphes~\cite{delphes} with the default configuration. 

Hadronically-decaying $\tau$-leptons are identified from jets with ideal efficiency if the jet's momentum exceeds 1 GeV, the absolute pseudorapidity is within 2.5, and the truth-level $\tau$ is contained within a cone of $\sqrt{\eta^2 + \phi^2} < 0.5$ around the jet.

A total of 10 million events are generated for each decay channel with inclusive polarization. Additionally, four types of polarized samples, with 10 million events each, are generated by requiring one of the $\tau$-leptons to be either right-handed or left-handed. These polarized samples are used to perform the template study.

\section{Neutrino Momentum Reconstruction Using a Generative Model}
\label{sec:neutrino_rec}

It is a well-known challenge to reconstruct the missing momenta of the two undetected neutrinos in the $\tau^+ \tau^-$ decay. Neutrino momentum reconstruction is an essential step in performing quantum tomography of the $\tau^+ \tau^-$ state.  We first consider the $\pi\pi$ sample where we have the process $pp \to \tau^+ \tau^- \to (\pi^+ \bar{\nu}_\tau) (\pi^- \nu_\tau)$.  In this system there are 8 unknowns:
\begin{equation} 
\label{eq:kin}
p_{\nu} = (E_{\nu}, \; p_{x,\nu}, \; p_{y,\nu}, \; p_{z,\nu}), 
\qquad \qquad 
p_{\bar{\nu}} = (E_{\bar{\nu}}, \; p_{x,\bar{\nu}}, \; p_{y,\bar{\nu}}, \; p_{z,\bar{\nu}}).
\end{equation}
There are 6 constraints:
\begin{align}
\label{eq:cons-nmass}
p_{\nu}^2  &= 0, &
p_{\bar{\nu}}^2  &= 0, \\
(p_{\bar{\nu}} + p_{\pi^-})^2 &= m_\tau^2, &
(p_{\nu} + p_{\pi^+})^2 & = m_\tau^2, \\
\slashed{E}_x  &= p_{x,\nu} + p_{x,\bar{\nu}} ,& \quad 
\slashed{E}_y  &= p_{y,\nu} + p_{y,\bar{\nu}} .
\end{align}
The result is an underconstrained system in which the neutrino momenta in Eq.~\eqref{eq:kin} cannot be solved analytically.  At a lepton collider there are two additional constraints for the total momentum along the $z$ direction and the total energy which allows for an analytic solution, with discrete ambiguities, for the $\pi\pi$ decay channel.

For the leptonic $\tau$ decays the constraint of Eq.~\eqref{eq:cons-nmass} cannot be used rendering even the situation at a lepton collider underconstrained.  The reconstruction of the $\tau^+ \tau^-$ is thus fundamentally challenging.

In this section, we present a diffusion-based approach tailored for the $\tau^+ \tau^-$ system, using the PET architecture to model the conditional probability distribution of neutrino momenta. 
This method generates a statistical representation of the full kinematic phase space, preserving event-level correlations while enhancing reconstruction accuracy. We describe the network architecture, training strategy, and optimization techniques designed to improve performance. Finally, we assess the model’s effectiveness in neutrino momentum reconstruction and its implications for quantum entanglement measurements at the LHC.

\begin{figure}[h!]
    \centering
    \includegraphics[width=\linewidth]{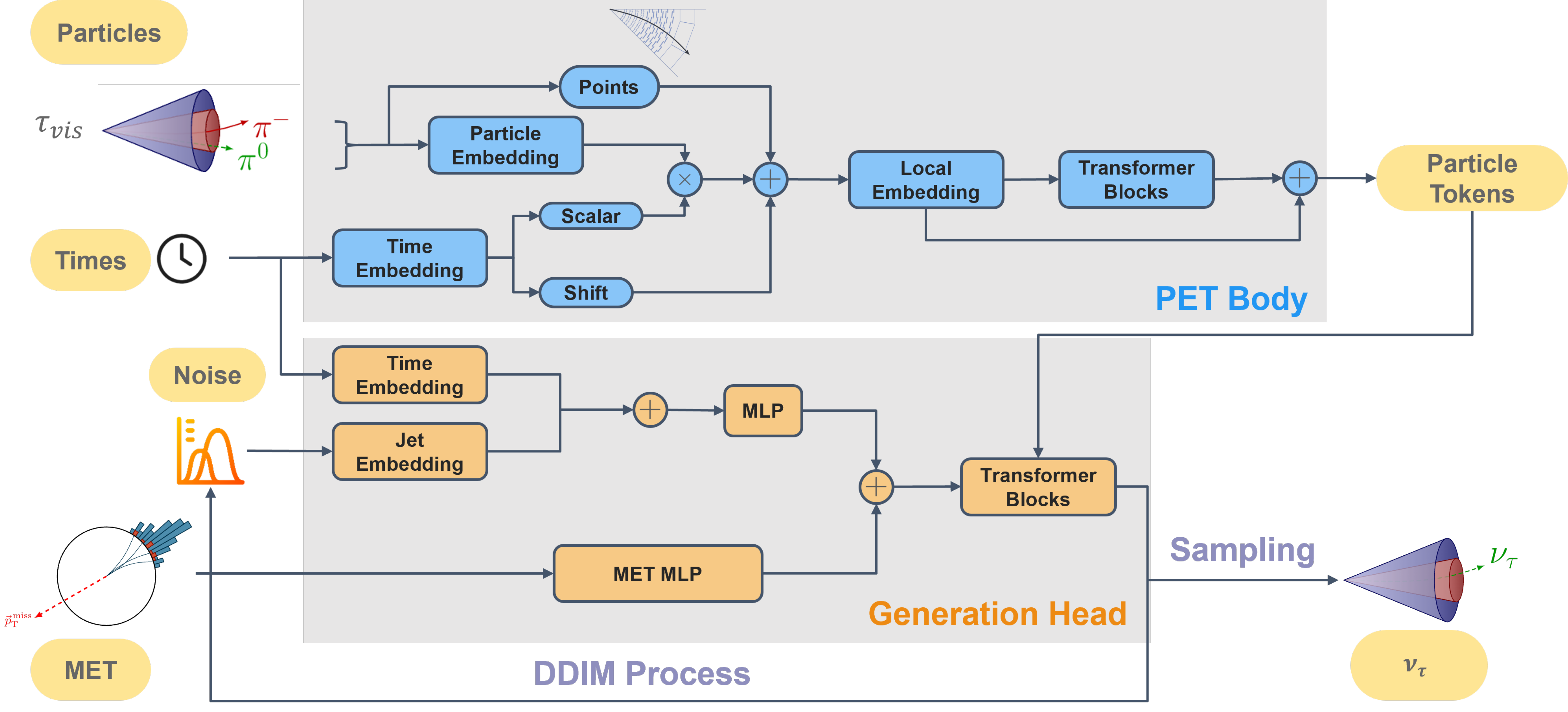}
    \caption{The architecture details of the PET body and generation head. For a given process, the visible component of the $\tau$ and the time information are input into the PET body to generate the particle token. This token, along with the noise and MET information, is then fed into the generation head for the DDIM sampling process.}
    \label{fig:PET_architecture}
\end{figure}

Compared with existing approaches, our method presents several notable advancements. Firstly, we introduce an additional Transformer layer within the generation head, thereby significantly enhancing the representational capability of the model. Secondly, we optimize the sampling procedure by replacing the modified Denoising Diffusion Probabilistic Model (DDPM) used in OmniLearn with the simpler Denoising Diffusion Implicit Model (DDIM), enabling increased sampling steps without incurring substantial computational costs. Thirdly, we incorporate random rotations into the training data augmentation process, greatly enlarging the dataset and enabling larger batch sizes, thus improving the training efficiency of the Transformer model while explicitly leveraging the rotational symmetry inherent in physical processes. Lastly, the nearest-neighbor calculations are dynamically tailored according to the number of visible components resulting from $\tau$ decays; specifically, we select the six nearest neighbors for decay channels involving $\rho$ particles and only two nearest neighbors for other channels, effectively capturing the distinct physical characteristics associated with each decay channel.

\subsection{Model Architecture and Training Strategy}

The model consists of the PET body for feature extraction and a generative head for three-momentum reconstruction. The input to the network comprises two components: missing transverse energy ($E_T^{\text{miss}}$) and reconstructed object features, summarized in Table~\ref{tab:input-vari}. The PET body processes the structured event-level features to generate diffusion tokens, which are then concatenated with $E_T^{\text{miss}}$ to guide the generative process. The architecture details of these models are shown in Fig. \ref{fig:PET_architecture}.

\begin{table}[h]
\centering
\caption{Summary of input features used for training the transformer-based diffusion model.}
\label{tab:input-vari}
\renewcommand{\arraystretch}{1.2}
\begin{tabular}{llp{6cm}}
\toprule
Category & Variables & Description \\
\midrule
$E_T^{\text{miss}}$ & ($p_T^{\text{miss}}, \phi^{\text{miss}}$) & Missing transverse momentum vector \\
\midrule
\multirow{4}{*}{$\tau$ Visible Components} & ($p_T, \eta, \phi, E$) & Four-momentum \\
& Charge & Electric charge of $\tau$-visible parts \\
& PID & Electron, muon, or pion identification \\
\midrule
\multirow{3}{*}{Small-R Jets} & ($p_T, \eta, \phi, E$) & Four-momentum \\
& Charge & Electric charge of the jet \\
& PID & Particle identification \\
\bottomrule
\end{tabular}
\end{table}

For training, separate models are optimized for each signal subchannel, ensuring that each event contains at least two neutrinos: one from $\tau^+$ decay and one from $\tau^-$. In channels with electron or muon final states, additional neutrinos from electroweak decays introduce further complexity. Instead of reconstructing each neutrino individually, we aggregate all neutrinos from the same $\tau$ decay into a single effective neutrino, preserving kinematic correlations while reducing model complexity. The network is trained to predict the three-momentum $(p_x, p_y, p_z)$ of two representative neutrinos, one from $\tau^+$ and one from $\tau^-$. 

A key limitation of generative approaches in collider physics is their reliance on large Monte Carlo datasets. To mitigate this, we employ an efficient data augmentation strategy, where each event is randomly rotated in the transverse plane relative to the center-of-mass frame. This technique enhances generalization without requiring additional MC event generation, significantly reducing computational costs. By applying this augmentation, we increase the training statistics by approximately 50 times, reaching a range of $1$M to $10$M effective events across all subchannels.

The training dataset is separately generated as an additional dataset, while the final evaluation dataset serves as the default dataset in this study.
Following the procedure outlined in Sec.~\ref{sec:simulation}, we generate 10 million raw data samples per channel. After applying the event selection detailed in Sec.~\ref{sec:reconstruction}, an additional dataset is produced specifically for training, following the same selection criteria as the final evaluation dataset. Within this training dataset, 80\% of the selected samples are used for training, while the remaining 20\% are reserved for monitoring the training process.
Model training is conducted on 16 NVIDIA A100 GPUs using Horovod \cite{horovod} on the Perlmutter Supercomputer \cite{nersc_perlmutter}. 
The learning rate follows a warm-up and cosine decay schedule, starting at $1.2 \times 10^{-4}$ and adapting dynamically to optimize convergence. The optimizer of choice is the Lion optimizer, with hyperparameters $\beta_1 = 0.95$ and $\beta_2 = 0.99$.

\subsection{Network Training Evaluation}

To mitigate potential biases in the generation process, we generate 10 candidate reconstructions per event and randomly select one for the final evaluation. This approach ensures robust estimation of neutrino kinematics while effectively preventing overfitting to any single generated sample. By transitioning from the previously employed modified DDPM method to Denoising Diffusion Implicit Models (DDIM) \cite{DDIM_paper}, we simplify the sampling process. Employing DDIM with 200 sampling steps and $\eta = 1.0$—settings equivalent to traditional Denoising Diffusion Probabilistic Models (DDPM) \cite{DDPM_paper}—allows stable and accurate predictions without significantly increasing computational demands, even as sampling steps increase. This adjustment reduces the overall computational burden, thus enhancing the model's efficiency and effectiveness. The detailed sampling procedure for each event is illustrated in Fig. \ref{fig:PET_architecture}.

To evaluate the robustness of the network and enhance the credibility of the results, results are performed on the final evaluation dataset. Evaluation is conducted at two levels: the particle level, which assesses the reconstructed three-momentum components $(p_x, p_y, p_z)$ of the neutrinos and the $\tau$ leptons, and the event level, which focuses on the mass resolution of the $\tau\tau$ system. 

\begin{figure}[h]
    \centering
    \begin{subfigure}[b]{0.48\textwidth}
        \centering
        \includegraphics[width=\textwidth]
        {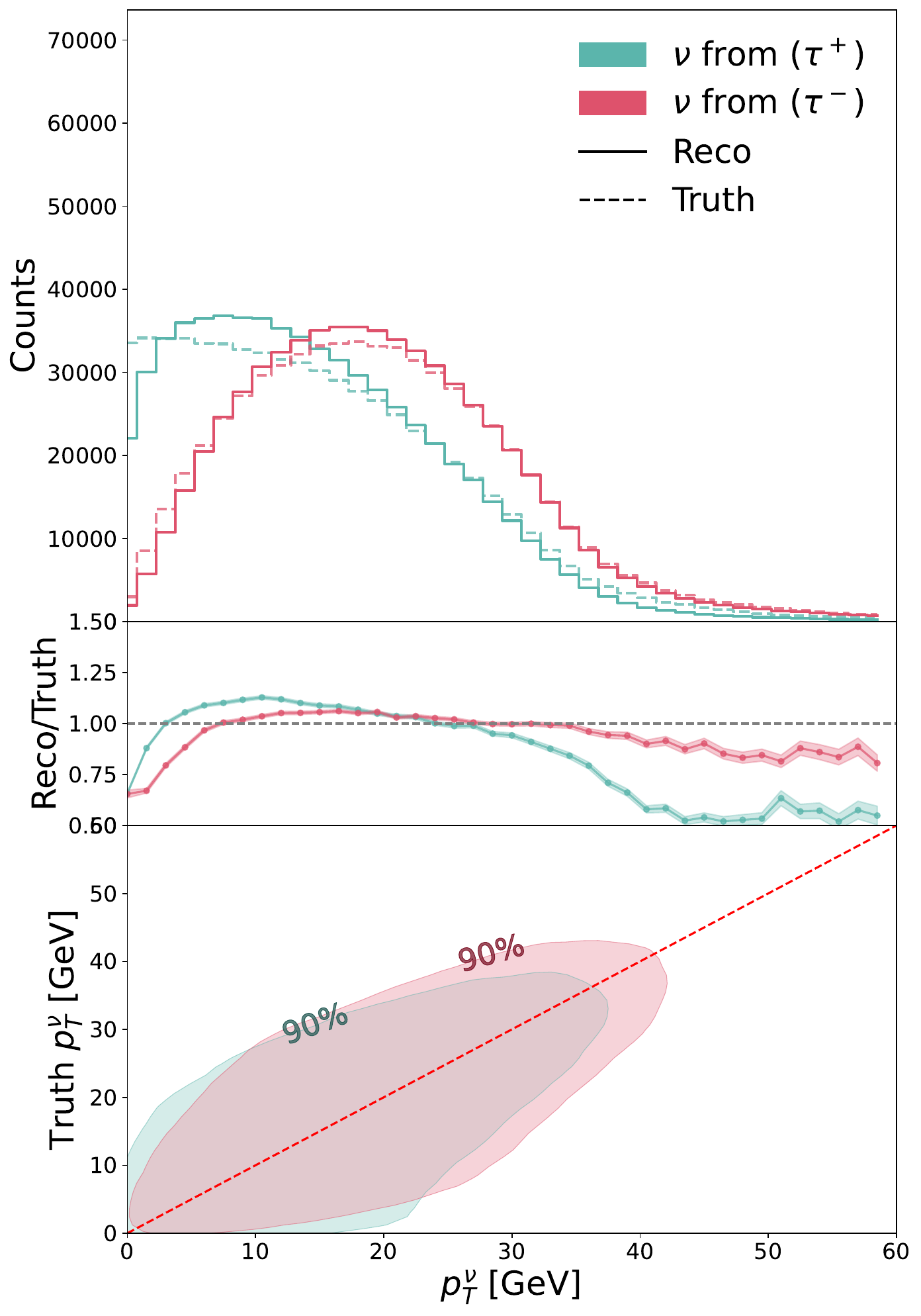}
        \caption{Diffusion-based.}
    \end{subfigure}
    \begin{subfigure}[b]{0.48\textwidth}
        \centering
        \includegraphics[width=\textwidth]
        {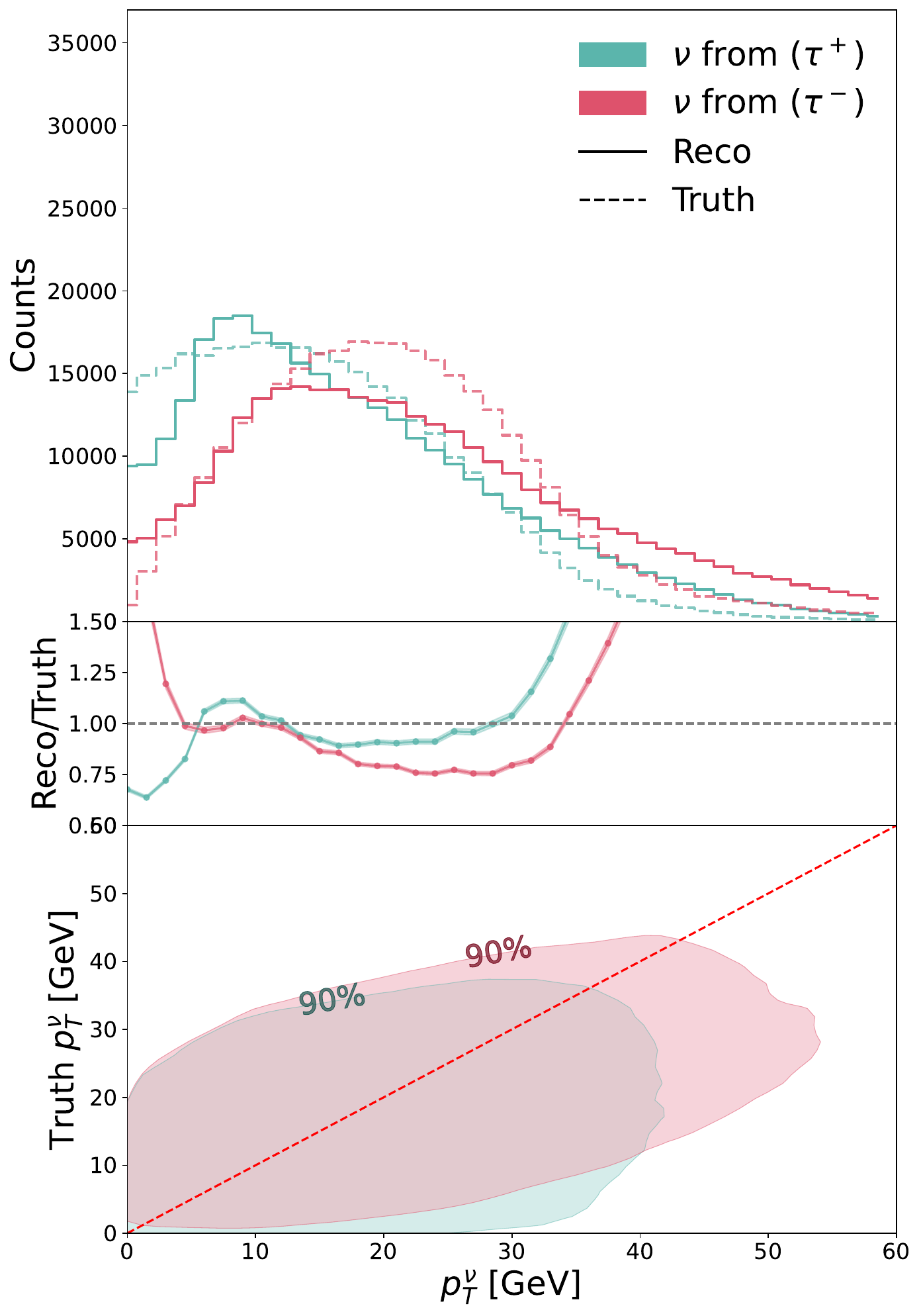}
        \caption{Missing Mass Calculator.}
    \end{subfigure}
    \caption{
     Comparison of reconstructed and truth-level neutrino transverse momentum ($p_T$) distributions for the $\ell\rho$ channels. 
    The left plot shows the result from the diffusion-based model, while the right plot corresponds to MMC.
    Each plot includes the reconstructed (solid) and truth-level (dashed) distributions for neutrinos originating from $\tau^+$ (teal) and $\tau^-$ (rose). The middle panels show the ratio of reconstructed to truth values, while the bottom panels contain two-dimensional correlation contour plots illustrating the linear dependency between reconstructed and truth values. The contour lines represent density levels, with numerical labels indicating the percentage of total data enclosed within each contour.}
    \label{fig:neutrino_kinematics}
\end{figure}

We evaluate the resolution of the reconstructed distributions using the half-width at half-maximum (HWHM) metric. This metric measures the spread of the relative error distribution near its peak and serves as our primary performance indicator. Since the relative error distribution peaks near zero, a smaller HWHM indicates more precise reconstructions with smaller relative errors.  At the particle level, the model achieved an average momentum resolution of approximately 17\% for $\tau$-leptons across different sub-channels. Among these, the best momentum resolution was observed in the $e \rho$ and $\mu \rho$ channels. Generally, channels containing $\rho$ exhibit better resolution than those without $\rho$. However, the $\tau$-lepton reconstruction results remain relatively consistent across all sub-channels, with the maximum difference not exceeding 5\%. At the event level, the invariant mass resolution of the reconstructed $\tau\tau$ system is evaluated, achieving a resolution of approximately 6\% to 8\%.

Since the model is trained to predict momenta in Cartesian coordinates ($p_x$, $p_y$, $p_z$), these components are directly used to assess its intrinsic performance, as presented in Table~\ref{tab:HFHM_merged}, where we also compare with the missing mass calculator (MMC) method~\cite{MMC_Paper}. 
However, in experimental analyses, kinematic variables such as transverse momentum ($p_T$), pseudorapidity ($\eta$), and azimuthal angle ($\phi$) are more commonly used. To bridge this gap, we transform the reconstructed momenta into ($p_T$, $\eta$, $\phi$) coordinates and present the $p_T$ distributions in Fig.~\ref{fig:neutrino_kinematics} for the $\ell\rho$ channels, which are among the most sensitive. Both the diffusion-based reconstruction and the MMC results are shown for comparison.
From the correlation plots, a strong linear correlation is observed for the diffusion-based method, demonstrating its ability to capture the underlying kinematics accurately. In contrast, the MMC results appear sparse and exhibit no clear linear pattern, indicating significantly reduced reconstruction fidelity. These comparisons highlight the superior performance of the diffusion-based approach in modeling neutrino kinematics.

To incorporate uncertainties from the diffusion model’s sampling process, predictions are derived from multiple samples per event, with a single candidate randomly selected, described in Section~\ref{sec:syst}.

\begin{table}[!h]
\centering
\caption{Half-width at half-maximum for reconstructed  $\tau$ momenta, as well as $\tau\tau$ mass, across different subchannels.}
\label{tab:HFHM_merged}
\resizebox{0.95\linewidth}{!}{%
\renewcommand{\arraystretch}{1.1}
\begin{tabular}{l*{7}{cc}}
\toprule
 & \multicolumn{2}{c}{$\pi \pi$ (\%)} & \multicolumn{2}{c}{$e \pi$ (\%)} & \multicolumn{2}{c}{$\mu \pi$ (\%)} & \multicolumn{2}{c}{$\pi \rho$ (\%)} & \multicolumn{2}{c}{$e \rho$ (\%)} & \multicolumn{2}{c}{$\mu \rho$ (\%)} & \multicolumn{2}{c}{$\rho \rho$ (\%)} \\
\cmidrule(lr){2-3} \cmidrule(lr){4-5} \cmidrule(lr){6-7} \cmidrule(lr){8-9} \cmidrule(lr){10-11} \cmidrule(lr){12-13} \cmidrule(lr){14-15}
 & ML & MMC & ML & MMC & ML & MMC & ML & MMC & ML & MMC & ML & MMC & ML & MMC \\
\midrule
$\Delta p^x_{\tau^+}$ & \textbf{18.97} & 25.99 & \textbf{18.34} & 27.91 & \textbf{19.30} & 28.56 & \textbf{16.19} & 25.45 & \textbf{15.72} & 26.72 & \textbf{16.03} & 26.62 & \textbf{16.38} & 25.65 \\[1mm]
$\Delta p^y_{\tau^+}$ & \textbf{19.01} & 26.02 & \textbf{18.54} & 27.26 & \textbf{19.33} & 28.15 & \textbf{15.96} & 25.26 & \textbf{15.50} & 26.00 & \textbf{16.04} & 27.83 & \textbf{16.38} & 25.28 \\[1mm]
$\Delta p^z_{\tau^+}$ & \textbf{19.47} & 25.48 & \textbf{19.52} & 27.46 & \textbf{20.00} & 27.19 & \textbf{16.31} & 24.85 & \textbf{15.70} & 25.65 & \textbf{16.12} & 26.69 & \textbf{16.49} & 25.02 \\[1mm]
$\Delta p^x_{\tau^-}$ & \textbf{18.77} & 25.78 & \textbf{17.06} & 28.38 & \textbf{17.69} & 27.43 & \textbf{18.11} & 26.13 & \textbf{15.69} & 26.47 & \textbf{16.02} & 26.85 & \textbf{16.34} & 25.17 \\[1mm]
$\Delta p^y_{\tau^-}$ & \textbf{18.71} & 25.96 & \textbf{16.62} & 26.22 & \textbf{17.33} & 28.13 & \textbf{17.89} & 25.79 & \textbf{15.34} & 26.87 & \textbf{16.15} & 27.39 & \textbf{16.36} & 25.40 \\[1mm]
$\Delta p^z_{\tau^-}$ & \textbf{19.69} & 25.43 & \textbf{17.06} & 26.27 & \textbf{17.75} & 27.17 & \textbf{18.44} & 25.49 & \textbf{15.81} & 26.26 & \textbf{16.26} & 25.67 & \textbf{16.72} & 24.23 \\[1mm]
\midrule
$\Delta m_{\tau\tau}$ & \textbf{7.94} & 23.27 & \textbf{6.18} & 24.91 & \textbf{6.41} & 24.31 & \textbf{7.24} & 22.72 & \textbf{5.81} & 22.70 & \textbf{5.76} & 22.81 & \textbf{6.27} & 21.89 \\  
\bottomrule
\end{tabular}%
}

\end{table}

\section{Reconstruction and Analysis Strategy}
\label{sec:reconstruction}

\subsection{Event Selection and Categorization}

This study focuses on events where $\tau$ decays result in a single charged track, as these modes have a higher branching ratio and greater spin analyzing power compared to multi-track decays. Since $\tau$ decays involve neutrinos and various hadronic final states, the spin correlations between the two $\tau$s are encoded in their decay products. To preserve this information, events are categorized into distinct subchannels based on their reconstructed final states, each exhibiting a different spin analyzing power. 

At the reconstruction level, events are categorized into seven subchannels: three hadronic-hadronic subchannels, where both $\tau$s decay hadronically, and four leptonic-hadronic subchannels, where one $\tau$ decays leptonically while the other undergoes a hadronic decay. The classification of hadronic $\tau$ decays is based on prongness, which is inherently a reconstruction-level concept rather than a direct representation of the true decay products. In practice, the detailed composition of the $\tau$ decay cannot be fully reconstructed; instead, hadronic $\tau$s are identified based on the number of detected charged tracks and associated neutral energy deposits in the calorimeter. A one-prong (1p) $\tau$ corresponds to a single reconstructed charged track. The neutral component is inferred from calorimeter clusters associated with neutral pions. Consequently, the notation “XpYn” refers to the reconstructed prongness, where X denotes the number of charged tracks, and Y represents the number of neutral clusters attributed to the $\tau$ decay. The classification of subchannels, along with their corresponding spin-analyzing power, is summarized in Table~\ref{tab:final_yields_prongness}.

To ensure a clean selection of $pp \to Z/\gamma^* \to \tau^+ \tau^-$ events, a set of preselection criteria is applied. 
The analysis is performed using $\tau$ candidates, which represent the visible components of $\tau$ decays, while the invisible component, consisting of neutrinos, is reconstructed in a later stage.
Events containing \(b\)-tagged jets are removed to suppress backgrounds from top-quark decays. For hadronic-hadronic subchannels, events with reconstructed electrons or muons are vetoed to avoid contamination from leptonic $\tau$ decays or misidentified leptons. Exactly two $\tau$ candidates are required, with opposite electric charges, ensuring a well-defined event topology. The classification of each event is then determined based on the reconstructed decay mode of the $\tau$s. Additional kinematic requirements are imposed, requiring transverse momentum \( p_T > 10 \) GeV and pseudorapidity \( |\eta| < 2.5 \) for the $\tau$ candidates.
The full reconstruction of the $\tau$s, including the neutrino kinematics, is addressed in Sec.~\ref{sec:neutrino_rec}.
For the final signal region selection, events must satisfy the invariant mass and center-of-mass $\tau$ angle constraints of
\begin{equation}
\label{eq:sr_selection}
80~{\rm GeV} < m_{\tau\tau} < 100~{\rm GeV},
\qquad\qquad
0.6 < 2\theta_\tau/\pi < 1.0 .
\end{equation}
The event yields are normalized to an integrated luminosity of 1 fb\(^{-1}\). Table~\ref{tab:final_yields_prongness} presents the final event yields after selection, categorized by the truth-labeled signal samples used in the simulation. Additionally, the table includes the spin-analyzing power for each subchannel, computed as $|\kappa_A \times \kappa_B|$, where \(\kappa_A\) and \(\kappa_B\) represent the individual spin-analyzing power for the two taus.

\begin{table}[h!]
\centering
\caption{Final signal event yields after selection for each reconstructed subchannel, normalized to an integrated luminosity of 1 fb\(^{-1}\). 
The prongness labels indicate the reconstructed decay mode classification, with "XpYn" denoting \(X\) charged and \(Y\) neutral pions in the hadronic decay.
Each entry represents the number of selected events passing all criteria in the corresponding truth sample. The spin-analyzing power $|\kappa_A \times \kappa_B|$ is also listed for each subchannel.
}
\label{tab:final_yields_prongness}
\resizebox{0.95\linewidth}{!}{%
\begin{tabular}{llcccccc}
\toprule
Subchannel & Prongness & $|\kappa_A \times \kappa_B|$ & $\tau\tau\to\ell \pi$ & $\tau\tau\to\ell\rho$ & $\tau\tau\to\pi \pi$ & $\tau\tau\to\pi \rho$ & $\tau\tau\to\rho \rho$ \\
\midrule
$\pi\pi$ & 1p0n - 1p0n & 1.0 & $ 196.14 \pm 1.29 $ & $ 12.86 \pm 0.49 $ & $ 783.83 \pm 1.42 $ & $ 61.51 \pm 0.59 $ & $ 6.12 \pm 0.28 $ \\
$\pi\rho$ & 1p0n - 1p1n & 0.41 & $ 3.92 \pm 0.18 $ & $ 572.07 \pm 3.28 $ & $ 12.93 \pm 0.18 $ & $ 1934.19 \pm 3.32 $ & $ 138.23 \pm 1.33 $ \\
$\rho\rho$ & 1p1n - 1p1n & 0.17 & $ 0.13 \pm 0.03 $ & $ 10.49 \pm 0.44 $ & $ 0.48 \pm 0.04 $ & $ 40.18 \pm 0.48 $ & $ 5420.33 \pm 8.31 $ \\
$e\pi$ & e - 1p0n & 0.33 & $ 1245.85 \pm 3.25 $ & $ 80.33 \pm 1.23 $ & $ 0.09 \pm 0.02 $ & $ 0.14 \pm 0.03 $ & $ 0.01 \pm 0.01 $ \\
$\mu\pi$ & $\mu$ - 1p0n & 0.34 & $ 1640.89 \pm 3.73 $ & $ 106.80 \pm 1.42 $ & $ 0.04 \pm 0.01 $ & $ 0.05 \pm 0.02 $ & $< 0.01$ \\
$e\rho$ & e - 1p1n & 0.14 & $ 25.32 \pm 0.46 $ & $ 3603.19 \pm 8.23 $ & $< 0.01$ & $ 0.13 \pm 0.03 $ & $ 0.47 \pm 0.08 $ \\
$\mu\rho$ & $\mu$ - 1p1n & 0.15 & $ 36.45 \pm 0.56 $ & $ 4869.69 \pm 9.57 $ & $< 0.01$ & $ 0.09 \pm 0.02 $ & $ 0.31 \pm 0.06 $ \\
\midrule
\multicolumn{8}{c}{ Signal Region ($80~{\rm GeV} < m_{\tau\tau} < 100~{\rm GeV}$ \&
$0.6 < 2\theta_\tau/\pi < 1)$
} \\
\midrule
$\pi \pi$ &  &  & $ 23.09 \pm 0.44 $ & $ 1.13 \pm 0.15 $ & $ 428.96 \pm 1.05 $ & $ 26.69 \pm 0.39 $ & $ 1.64 \pm 0.14 $ \\
$\pi \rho$ &  &  & $ 1.23 \pm 0.10 $ & $ 225.62 \pm 2.06 $ & $ 6.76 \pm 0.13 $ & $ 1137.40 \pm 2.55 $ & $ 64.13 \pm 0.90 $ \\
$\rho \rho$ &  &  & $ 0.06 \pm 0.02 $ & $ 5.72 \pm 0.33 $ & $ 0.22 \pm 0.02 $ & $ 22.86 \pm 0.36 $ & $ 3310.16 \pm 6.49 $ \\
$e \pi$ &  &  & $ 806.92 \pm 2.62 $ & $ 45.61 \pm 0.93 $ & $< 0.01$ & $< 0.01$ & $ 0.01 \pm 0.01 $ \\
$\mu \pi$ &  &  & $ 1053.87 \pm 2.99 $ & $ 59.92 \pm 1.06 $ & $ 0.01 \pm 0.01 $ & $< 0.01$ & $< 0.01$ \\
$e \rho$ &  &  & $ 15.98 \pm 0.37 $ & $ 2371.58 \pm 6.68 $ & $< 0.01$ & $ 0.05 \pm 0.02 $ & $ 0.17 \pm 0.05 $ \\
$\mu \rho$ &  &  & $ 22.55 \pm 0.44 $ & $ 3172.37 \pm 7.72 $ & $< 0.01$ & $ 0.02 \pm 0.01 $ & $ 0.05 \pm 0.03 $ \\
\midrule
\multicolumn{8}{c}{SR \& di-$\tau$ Trigger 
($p_T^{\tau_1}> 35$ GeV \& $p_T^{\tau_2}> 25$ GeV )
} \\
\midrule
$\pi \pi$ &  &  & $ 3.92 \pm 0.18 $ & $ 0.09 \pm 0.04 $ & $ 89.43 \pm 0.48 $ & $ 2.90 \pm 0.13 $ & $ 0.14 \pm 0.04 $ \\
$\pi \rho$ &  &  & $ 0.12 \pm 0.03 $ & $ 22.71 \pm 0.65 $ & $ 1.61 \pm 0.06 $ & $ 206.39 \pm 1.09 $ & $ 6.29 \pm 0.28 $ \\
$\rho \rho$ &  &  & $< 0.01$ & $ 0.56 \pm 0.10 $ & $ 0.06 \pm 0.01 $ & $ 4.51 \pm 0.16 $ & $ 629.99 \pm 2.83 $ \\
\midrule
\multicolumn{8}{c}{SR \& $e + \tau$ Trigger 
($p_T^{e}> 14$ GeV \& $p_T^{\tau}> 25$ GeV ) or 
single-$e$ Trigger ($p_T^{e}> 26$ GeV)
} \\
\midrule
$e \pi$ &  &  & $ 378.90 \pm 1.79 $ & $ 17.52 \pm 0.57 $ & $< 0.01$ & $ 0.01 \pm 0.01 $ & $< 0.01$ \\
$e \rho$ &  &  & $ 8.33 \pm 0.27 $ & $ 1233.90 \pm 4.82 $ & $< 0.01$ & $ 0.03 \pm 0.01 $ & $ 0.15 \pm 0.04 $ \\
\midrule
\multicolumn{8}{c}{SR \& $\mu + \tau$ Trigger 
($p_T^{\mu}> 17$ GeV \& $p_T^{\tau}> 25$ GeV )
or 
single-$\mu$ Trigger ($p_T^{\mu}> 26$ GeV)
} \\
\midrule
$\mu \pi$ &  &  & $ 565.94 \pm 2.19 $ & $ 25.21 \pm 0.69 $ & $< 0.01$ & $< 0.01$ & $< 0.01$ \\
$\mu \rho$ &  &  & $ 12.63 \pm 0.33 $ & $ 1862.06 \pm 5.92 $ & $< 0.01$ & $< 0.01$ & $ 0.04 \pm 0.02 $ \\
\bottomrule
\end{tabular}
}

\end{table}

\subsection{Backgrounds and Systematic Uncertainties}

Backgrounds in this analysis primarily arise from Standard Model processes where jets are misidentified as hadronic tau candidates. The dominant background contributions come from electroweak processes, top-quark production, and QCD-induced events.

Electroweak backgrounds arise from \( W + \text{jets} \) and \( Z/\gamma^* \to \ell\ell \) processes, where jet misidentification leads to signal-like events. In \( W \to \ell\nu + \text{jets} \), a jet faking a hadronic $\tau$ mimics the leptonic-hadronic signal, while in \( W \to \tau\nu + \text{jets} \), an additional jet misidentified as a $\tau$ contaminates the hadronic-hadronic category. In \( Z/\gamma^* \to \ell\ell \), background events enter the selection when one lepton is mis-reconstructed or falls outside the detector acceptance, while a jet is misidentified as a $\tau$, primarily affecting leptonic-hadronic selections. Despite the absence of genuine \( \tau^+ \tau^- \) pairs, these backgrounds remain significant due to their large production rates.

Top-quark backgrounds arise primarily from \( t\bar{t} \) production, where both top quarks decay via \( t \to Wb \), and the subsequent \( W \to \tau\nu \) decay can produce prompt $\tau$s. This background includes genuine \( \tau^+ \tau^- \) pairs but is largely suppressed by requiring events to be free of \( b \)-tagged jets. 

QCD multijet production constitutes a significant background due to the large cross section of strong interactions. However, in this analysis, which focuses on one-prong hadronic taus (1p0n and 1p0n), QCD processes have a limited ability to mimic the signal due to their distinct event kinematics. The contribution from QCD backgrounds is therefore expected to be largely suppressed after the Tau Prongness requirement.

All background processes are modeled using Monte Carlo simulations and normalized to their respective theoretical cross sections. 
Detailed yields of all backgrounds in different subchannels are listed in Table~\ref{tab:background_yields}.
Their impact is further assessed in the statistical interpretation, where systematic uncertainties associated with background modeling are incorporated.

\begin{table}[h!]
\centering
\caption{Final background yields after selection for each reconstructed subchannel, normalized to an integrated luminosity of 1 fb\(^{-1}\). 
Each entry represents the number of selected background events passing all criteria. The total background contribution is also included as the sum of all sources.}
\resizebox{0.95\linewidth}{!}{%
\begin{tabular}{lcccccc}
\toprule
Subchannel & $W \to \ell\nu$ & $W \to \tau\nu$ & $Z \to \ell\ell$ & $t\bar{t}$ & QCD & Total \\
\midrule
$\pi \pi$ & $< 0.01$ & $ 0.41 \pm 0.29 $ & $< 0.01$ & $ 0.96 \pm 0.22 $ & $< 0.01$ & $ 1.37 \pm 0.37 $ \\
$\pi \rho$ & $ 0.77 \pm 0.55 $ & $ 1.45 \pm 0.55 $ & $< 0.01$ & $ 3.28 \pm 0.41 $ & $< 0.01$ & $ 5.50 \pm 0.87 $ \\
$\rho \rho$ & $< 0.01$ & $ 4.14 \pm 0.93 $ & $< 0.01$ & $ 10.13 \pm 0.71 $ & $< 0.01$ & $ 14.27 \pm 1.17 $ \\
$e \pi$ & $ 4.25 \pm 1.28 $ & $ 3.31 \pm 0.83 $ & $ 1.51 \pm 0.75 $ & $ 42.33 \pm 1.46 $ & $< 0.01$ & $ 51.40 \pm 2.24 $ \\
$e \rho$ & $ 28.20 \pm 3.30 $ & $ 12.43 \pm 1.60 $ & $ 2.64 \pm 1.00 $ & $ 120.53 \pm 2.46 $ & $< 0.01$ & $ 163.80 \pm 4.53 $ \\
$\mu \pi$ & $ 2.70 \pm 1.02 $ & $ 5.39 \pm 1.06 $ & $< 0.01$ & $ 50.89 \pm 1.60 $ & $< 0.01$ & $ 58.98 \pm 2.17 $ \\
$\mu \rho$ & $ 34.00 \pm 3.62 $ & $ 12.84 \pm 1.63 $ & $< 0.01$ & $< 0.01$ & $< 0.01$ & $ 46.84 \pm 3.97 $ \\
\midrule
\multicolumn{7}{c}{ Signal Region ($80~{\rm GeV} < m_{\tau\tau} < 100~{\rm GeV}$ \&
$0.6 < 2\theta_\tau/\pi < 1)$
} \\
\midrule
$\pi \pi$ & $< 0.01$ & $< 0.01$ & $< 0.01$ & $ 0.05 \pm 0.05 $ & $< 0.01$ & $ 0.05 \pm 0.05 $ \\
$\pi \rho$ & $< 0.01$ & $< 0.01$ & $< 0.01$ & $ 0.20 \pm 0.10 $ & $< 0.01$ & $ 0.20 \pm 0.10 $ \\
$\rho \rho$ & $< 0.01$ & $< 0.01$ & $< 0.01$ & $ 0.96 \pm 0.22 $ & $< 0.01$ & $ 0.96 \pm 0.22 $ \\
$e \pi$ & $< 0.01$ & $ 0.21 \pm 0.21 $ & $< 0.01$ & $ 3.93 \pm 0.45 $ & $< 0.01$ & $ 4.14 \pm 0.49 $ \\
$e \rho$ & $ 2.70 \pm 1.02 $ & $ 0.62 \pm 0.36 $ & $< 0.01$ & $ 12.60 \pm 0.80 $ & $< 0.01$ & $ 15.92 \pm 1.34 $ \\
$\mu \pi$ & $< 0.01$ & $ 0.83 \pm 0.41 $ & $< 0.01$ & $ 5.59 \pm 0.53 $ & $< 0.01$ & $ 6.42 \pm 0.67 $ \\
$\mu \rho$ & $ 2.32 \pm 0.95 $ & $ 1.04 \pm 0.46 $ & $< 0.01$ & $< 0.01$ & $< 0.01$ & $ 3.35 \pm 1.05 $ \\
\midrule
\multicolumn{7}{c}{SR \& di-$\tau$ Trigger 
($p_T^{\tau_1}> 35$ GeV \& $p_T^{\tau_2}> 25$ GeV )
} \\
\midrule
$\pi \pi$ & $< 0.01$ & $< 0.01$ & $< 0.01$ & $< 0.01$ & $< 0.01$ & $< 0.01$ \\
$\pi \rho$ & $< 0.01$ & $< 0.01$ & $< 0.01$ & $ 0.05 \pm 0.05 $ & $< 0.01$ & $ 0.05 \pm 0.05 $ \\
$\rho \rho$ & $< 0.01$ & $< 0.01$ & $< 0.01$ & $ 0.10 \pm 0.07 $ & $< 0.01$ & $ 0.10 \pm 0.07 $ \\
\midrule
\multicolumn{7}{c}{SR \& $e + \tau$ Trigger 
($p_T^{e}> 14$ GeV \& $p_T^{\tau}> 25$ GeV ) or 
single-$e$ Trigger ($p_T^{e}> 26$ GeV)
} \\
\midrule
$e \pi$ & $< 0.01$ & $< 0.01$ & $< 0.01$ & $ 2.87 \pm 0.38 $ & $< 0.01$ & $ 2.87 \pm 0.38 $ \\
$e \rho$ & $ 1.93 \pm 0.86 $ & $ 0.62 \pm 0.36 $ & $< 0.01$ & $ 11.19 \pm 0.75 $ & $< 0.01$ & $ 13.74 \pm 1.20 $ \\
\midrule
\multicolumn{7}{c}{SR \& $\mu + \tau$ Trigger 
($p_T^{\mu}> 17$ GeV \& $p_T^{\tau}> 25$ GeV )
or 
single-$\mu$ Trigger ($p_T^{\mu}> 26$ GeV)
} \\
\midrule
$\mu \pi$ & $< 0.01$ & $ 0.41 \pm 0.29 $ & $< 0.01$ & $ 4.13 \pm 0.46 $ & $< 0.01$ & $ 4.55 \pm 0.54 $ \\
$\mu \rho$ & $ 1.16 \pm 0.67 $ & $ 0.62 \pm 0.36 $ & $< 0.01$ & $< 0.01$ & $< 0.01$ & $ 1.78 \pm 0.76 $ \\
\bottomrule
\end{tabular}
}

\label{tab:background_yields}
\end{table}

\subsubsection{Systematics Uncertainties}
\label{sec:syst}

The systematic uncertainties in this analysis follow the methodology established in FAIR Universe~\cite{FAIR_Universe}, allowing for parameterized and configurable variations. Systematic uncertainties are categorized into two types: object-level systematics, which modify reconstructed object kinematics, and normalization systematics, which impact cross sections and integrated luminosity.

Object-level systematic uncertainties are implemented by modifying the kinematics of reconstructed objects before neutrino reconstruction, simulating realistic detector effects. Following the standard approach~\cite{Cowan:2010js}, uncertainties are varied by $\pm1\sigma$. The jet energy scale (JES) and tau energy scale (TES) are adjusted by 5\%~\cite{ATLAS-JET-PERF} and 3\%~\cite{ATLAS-TAU-PERF}, respectively, directly impacting the transverse momenta of jets and $\tau$s, which in turn modify the missing transverse energy (MET). Additionally, two independent soft MET components, each with a 1~GeV variation in the $x$ and $y$ directions, are introduced to account for potential soft energy fluctuations in the detector. 

To incorporate uncertainties from the diffusion model’s sampling process, predictions are derived from multiple samples per event, with a single candidate randomly selected. The model’s uncertainty is quantified by evaluating the interval between the 25th and 75th percentiles of predicted neutrino kinematics and applying this range as the systematic variation for each event.

Normalization uncertainties arise from uncertainties in luminosity determination and cross section estimations. The integrated luminosity is modeled with an uncertainty of $0.83\%$ from ATLAS Run 2~\cite{ATLAS_Run2_Lumi}. The normalization uncertainties for different physics processes are taken from the generator results. Given the challenges of accurately simulating background processes, a conservative 50\% uncertainty is assigned to the normalization of major background components rather than relying solely on theoretical cross section uncertainties. 

\begin{figure}[h]
    \centering
    \begin{subfigure}[b]{0.32\textwidth}
        \centering
        \includegraphics[width=\textwidth]
        {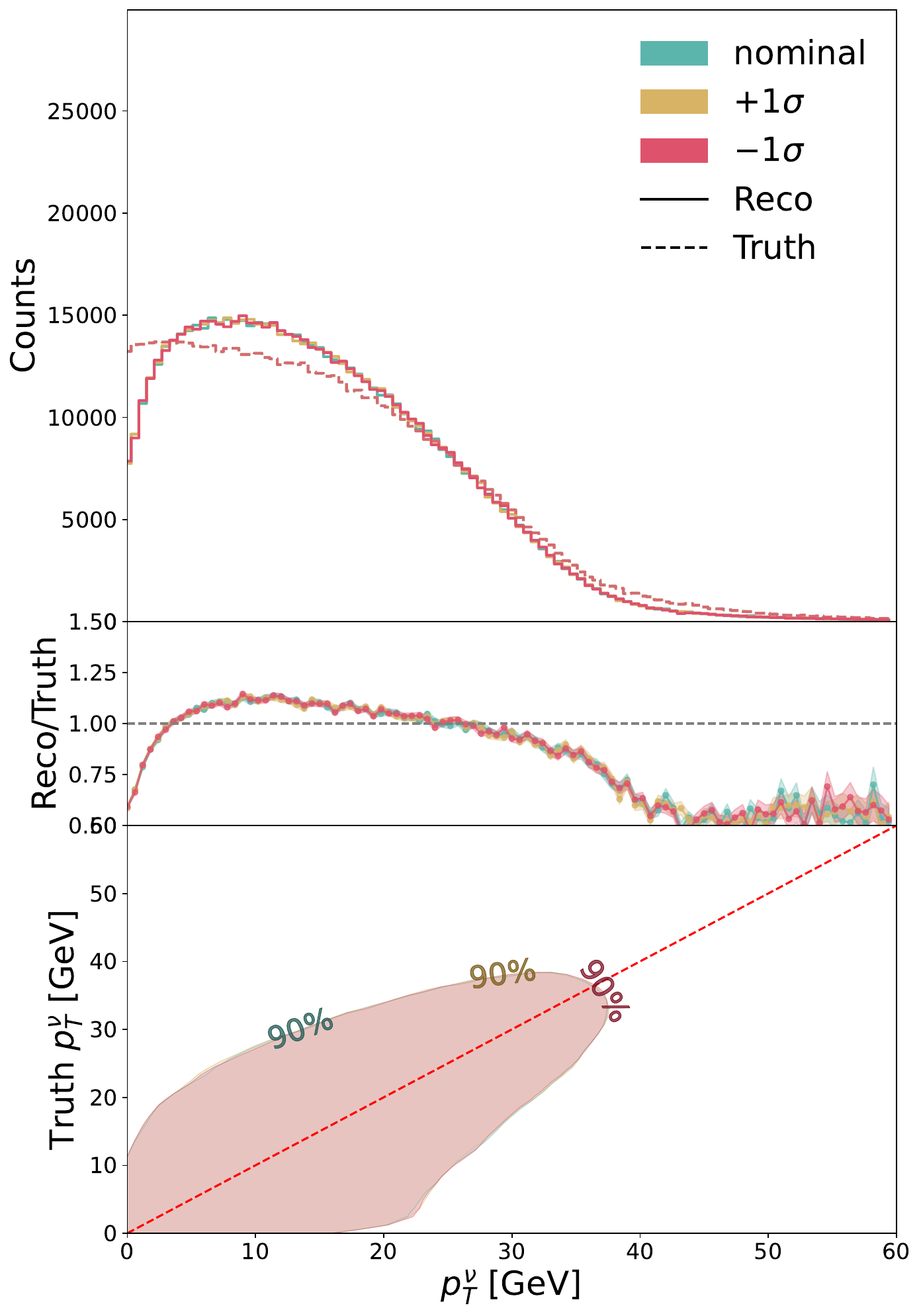}
    \end{subfigure}
    \hfill
    \begin{subfigure}[b]{0.32\textwidth}
        \centering
        \includegraphics[width=\textwidth]
        {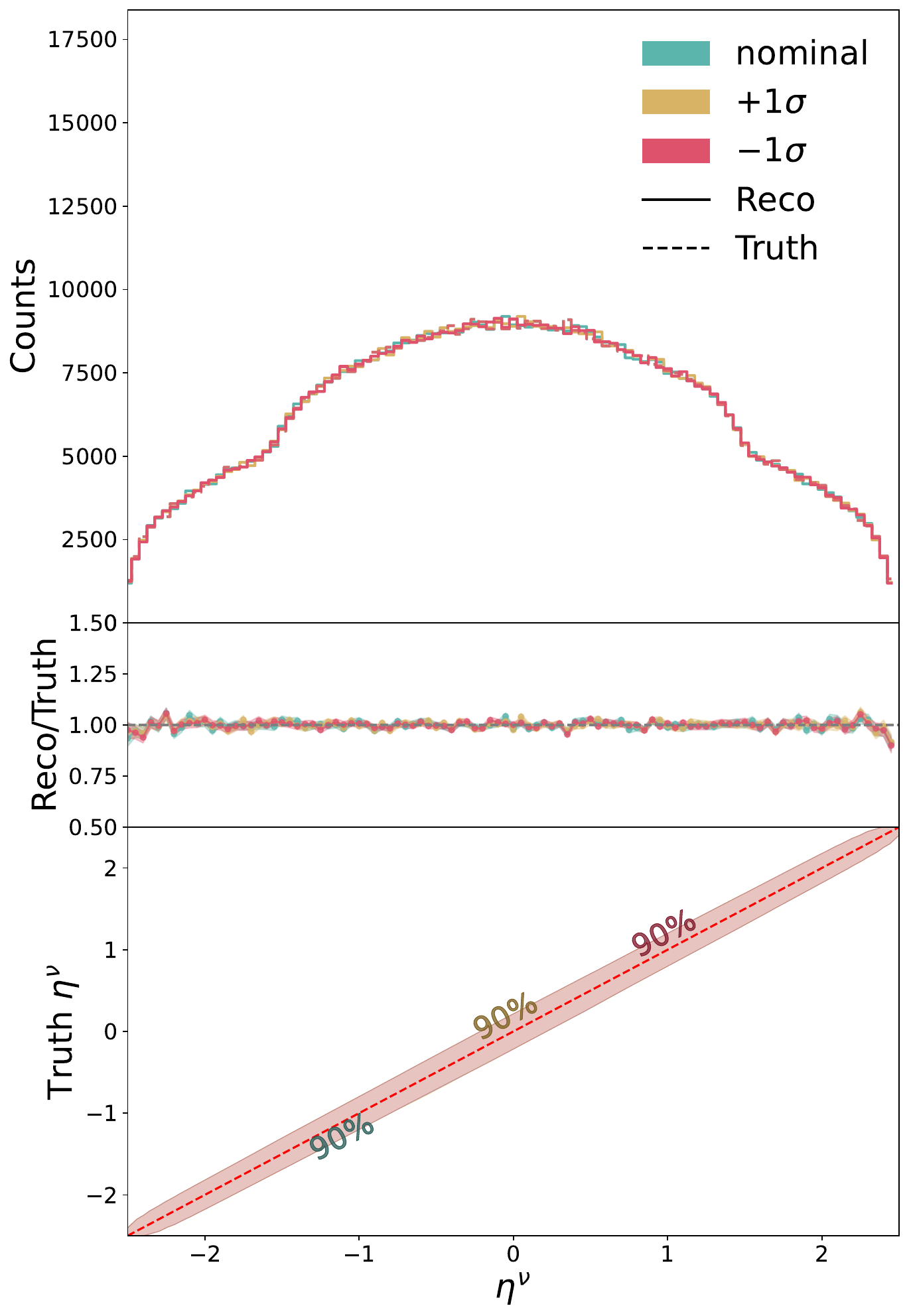}
        \end{subfigure}
    \hfill
    \begin{subfigure}[b]{0.32\textwidth}
        \centering
        \includegraphics[width=\textwidth]
        {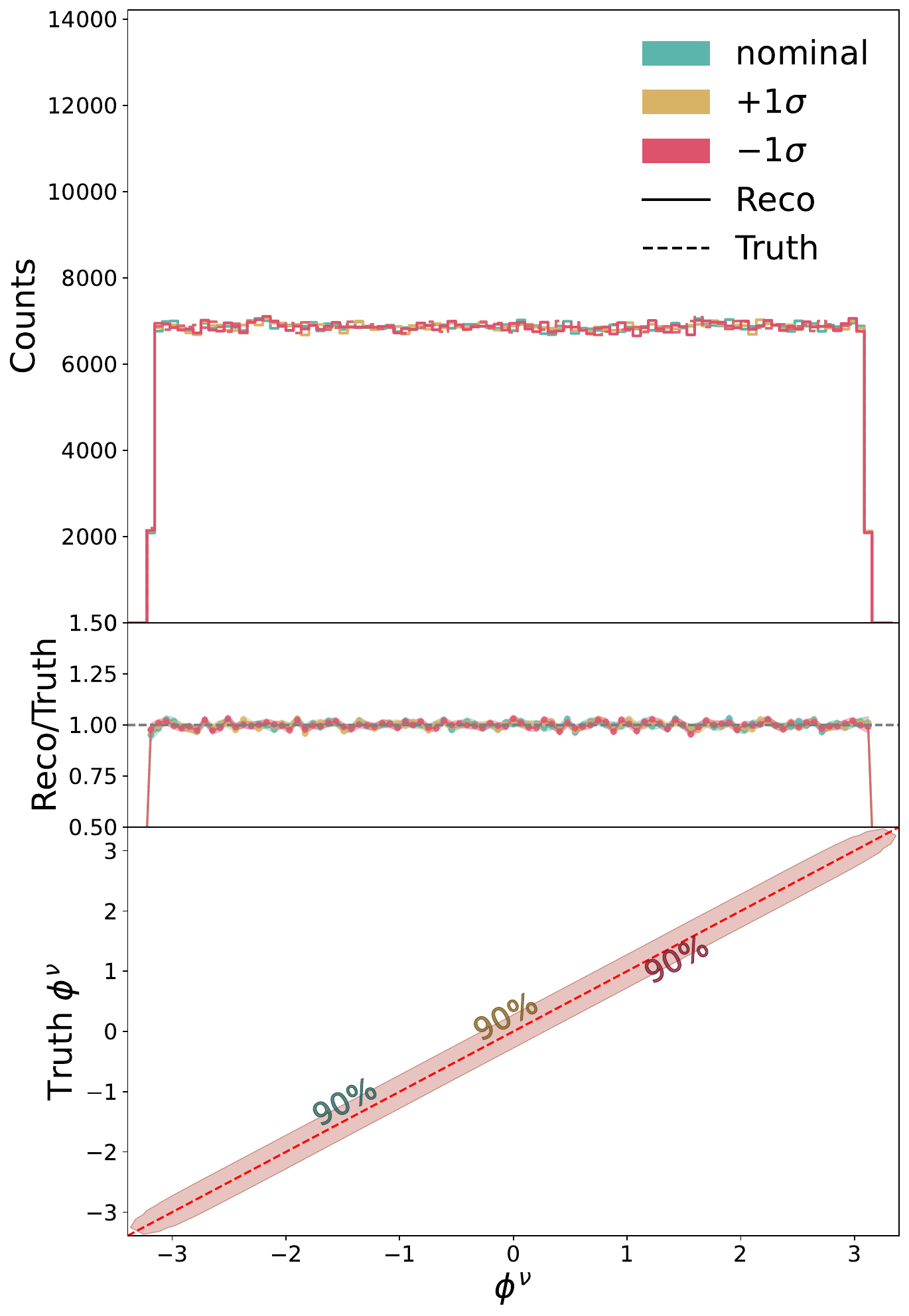}
    \end{subfigure}
    \caption{
     Comparison of reconstructed neutrino kinematic distributions under systematic variations from soft MET along the $x$ direction for $\ell\rho$ channel. The three columns correspond to transverse momentum ($p_T$), pseudorapidity ($\eta$), and azimuthal angle ($\phi$). Each plot includes the nominal reconstruction (teal) along with systematically shifted variations: $+1\sigma$ (goldenrod) and $-1\sigma$ (rose) shifts applied to neutrinos originating from $\tau^+$. The middle panels show the ratio of systematically varied reconstructions and nominal values to the truth values, while the bottom panels contain two-dimensional correlation contour plots illustrating the linear dependency between the truth and systematically shifted predictions. The contour lines represent density levels, with numerical labels indicating the percentage of total data enclosed within each contour.}
    \label{fig:neutrino_systematics}
\end{figure}

Figure~\ref{fig:neutrino_systematics} shows that the reconstructed neutrino kinematics remain consistent under systematic variations in the soft MET component along the $x$ direction in the $\ell\rho$ channels. The consistency between the systematically varied and nominal reconstructions confirms the stability of our neutrino prediction against systematic uncertainties. 

\subsection{Trigger Considerations}
\label{sec:trigger}

Triggers play a crucial role in hadron collider experiments as they determine which events are recorded for further analysis. To accurately mimic the conditions of a real experiment, it is essential to consider the impact of trigger selection on event yields. As an example, we examine the triggers used in the ATLAS experiment and evaluate their feasibility in selecting $\tau$ events. Specifically, we take two triggers from the 2018 dataset to explore their applicability to our study~\cite{ATLAS:Trigger, ATLAS:TauTrigger, ATLAS:TauTriggerPublic}.

The di-tau trigger\footnote{ \texttt{HLT\_tau35\_medium1\_tracktwo\_tau25\_medium1\_tracktwo\_L1TAU20IM\_2TAU12IM}} is designed for events with two hadronically decaying $\tau$-leptons, covering the $\pi\pi$, $\pi\rho$, and $\rho\rho$ subchannels. This trigger requires the leading $\tau$ to have $p_T > 35$ GeV and the subleading $\tau$ to have $p_T > 25$ GeV. 
The tau + muon trigger\footnote{ \texttt{HLT\_mu14\_tau25\_medium1\_tracktwo}} is applied to events in the $\mu\pi$ and $\mu\rho$ channels, requiring a $\tau$ with $p_T > 25$ GeV and a muon with $p_T > 14$ GeV. 
Additionally, the single-muon trigger\footnote{ \texttt{HLT\_mu26\_ivarmedium}} is used, requiring a muon with $p_T > 26$ GeV.
For events involving electrons, the tau + electron trigger\footnote{ \texttt{HLT\_e17\_lhmedium\_nod0\_ivarloose\_tau25\_medium1\_tracktwo}} is applied, requiring a $\tau$ with $p_T > 25$ GeV and an electron with $p_T > 17$ GeV. The single electron trigger\footnote{ \texttt{HLT\_e26\_lhtight\_nod0\_ivarloose}} is also utilized, selecting events with an electron satisfying $p_T > 26$ GeV.

To emulate the trigger effect in our study, we impose these trigger conditions in addition to the signal region selection criteria. The di-tau trigger imposes a significant efficiency penalty on events with low-$p_T$ visible $\tau$s, particularly for single-pion decays. However, both the $\rho\rho$ and $\pi\rho$ channels remain viable under the di-tau trigger.
In contrast, the tau + muon trigger accommodates both the $\mu\pi$ and $\mu\rho$ channels, leading to a higher event rate. Similarly, the tau + electron trigger allows for the $e\pi$ and $e\rho$ channels, expanding the accessible phase space. Additionally, the single-lepton triggers for muons and electrons provide an independent selection pathway for events containing a high-$p_T$ isolated lepton. 

The impact of trigger selections on both signal and background processes is summarized in Tables~\ref{tab:final_yields_prongness} and~\ref{tab:background_yields}.  
These selections, which include stringent $p_T$ thresholds on $\tau$ and lepton candidates, are designed to suppress large backgrounds, particularly from QCD multijet and electroweak processes.
Among the dominant backgrounds, $t\bar{t}$ is suppressed by factors ranging from 2 to 9 across all subchannels, while $W \to \ell\nu$ and $W \to \tau\nu$ are reduced by approximately a factor of 2. QCD and $Z \to \ell\ell$ backgrounds are rendered negligible under the full selection. This suppression establishes a low-background environment in both hadronic and leptonic channels, with residual contributions primarily from $t\bar{t}$ in the leptonic final states. 
As a result, the overall background contribution is significantly reduced, leading to a clean analysis region.

\subsection{Statistical Methodology}

This analysis employs a template fit method to extract the full spin density matrix of the $\tau^+\tau^-$ system, including both polarization and spin correlation terms. The primary objective is to measure the polarization components, $B^{\pm}_i$, where $i$ represents the helicity basis ($n, r, k$), as well as the spin correlation terms, $C_{ij}$, with both indices spanning the helicity basis.  
The conventional approach relies on reconstructing $B^{\pm}_i$ and $C_{ij}$ from angular distributions following theoretical predictions. However, during the reconstruction, kinematic selection criteria, such as a transverse momentum threshold of $p_T > 10$ GeV, significantly distort these distributions, making an unfolding procedure challenging. To mitigate this issue, we adopt a template fit approach, which provides a more robust method for extracting the spin density matrix elements.  

Template construction begins with the generation of signal samples with fixed polarization states. The observed distribution is modeled as a weighted sum of an SM sample and additional polarized samples, with a weight parameter, $x$, controlling the relative contribution of the SM sample. The polarized samples are generated with different polarization values other than the SM one, with a spin correlation matrix that is not entangled. By varying $x$, different polarization states and spin correlation matrices can be obtained, forming the basis of the template fit.

To facilitate the extraction of $B^{\pm}_i$ and $C_{ij}$, 15 distinct measurement regions are defined, each divided into eight bins. Within each bin, a linear interpolator is constructed as a function of $x$, allowing for a continuous variation of $B^{\pm}_i$ and $C_{ij}$. These interpolators are then used to compute the expected event yields in the negative log-likelihood (NLL) function~\cite{Cowan:2010js}. The likelihood function follows a Poisson distribution for binned event counts, incorporating statistical uncertainties from Monte Carlo samples and systematic uncertainties constrained by Gaussian priors. The 15 parameters are simultaneously fitted across all regions, treating them as the primary observables of interest. This simultaneous fit accounts for correlations among systematic uncertainties across different regions. 

Minimization of the likelihood function is performed using the Minuit algorithm~\cite{iminuit}, with the Hesse matrix providing estimates of parameter uncertainties. The fit results yield $x \pm \sigma_x$, from which the underlying $B^{\pm}_i$ and $C_{ij}$ are determined. Due to the non-linear dependence of these terms on $x$, uncertainty propagation introduces asymmetric errors in $B^{\pm}_i$ and $C_{ij}$.  
With the extracted spin density matrix elements, further observables related to quantum entanglement are computed. The concurrence is determined by solving the eigenvalues of the full spin density matrix, with uncertainty propagation handled through a first-order approximation in the differentiation of eigenvalues~\cite{MATRIX_ERROR}. In contrast, Bell nonlocality involves a linear combination of spin correlation terms, allowing for a direct propagation of uncertainties.

\section{Results}
\label{sec:results}

\begin{figure}[!htp]
    \centering
    \includegraphics[width=0.95\textwidth]{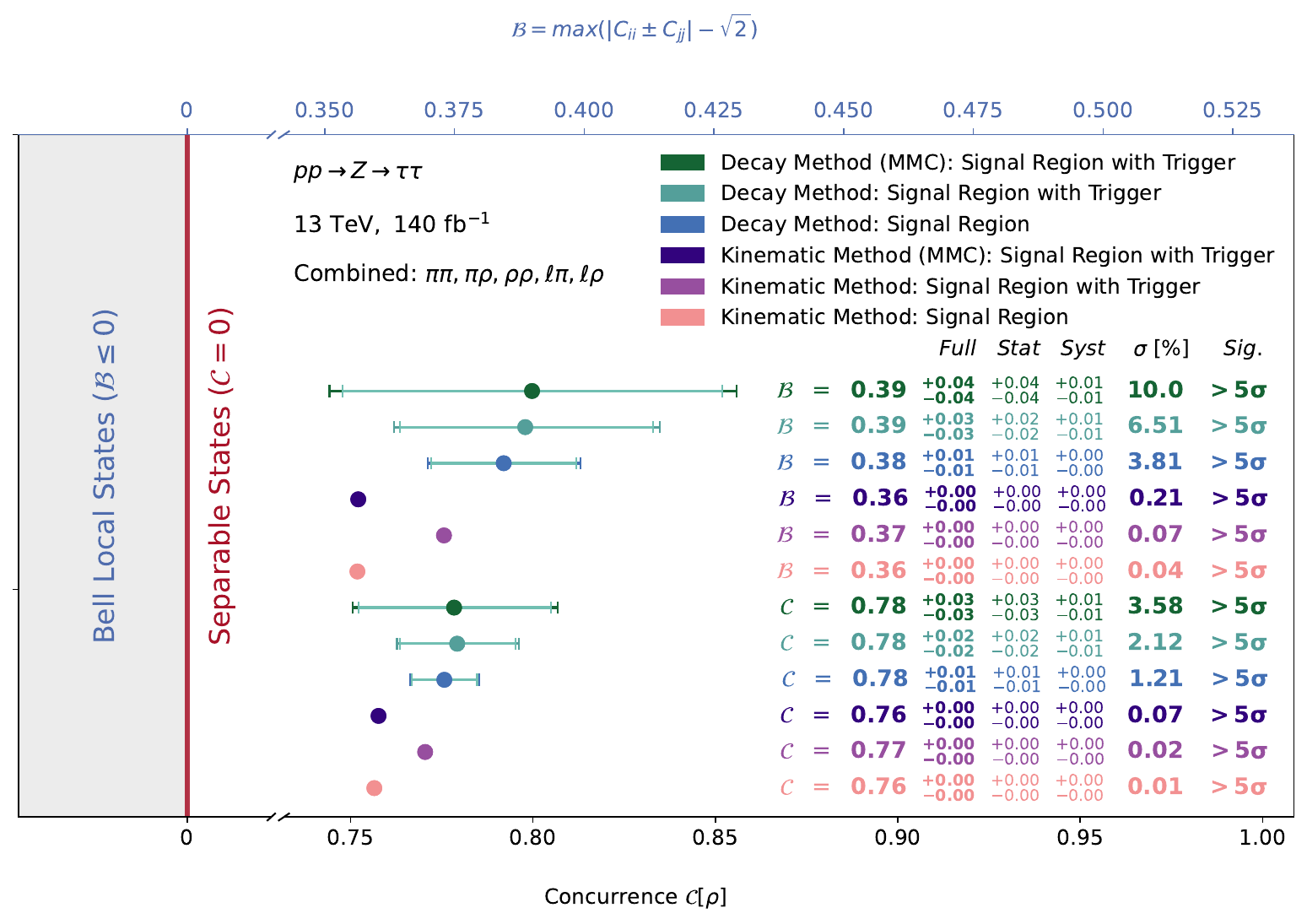}
    \caption{Results for concurrence (bottom axis) and Bell nonlocality (top axis). 
    Results are shown for an integrated luminosity of $140~\text{fb}^{-1}$, with a combined result that incorporates all seven studied channels.}
    \label{fig:poi_results}
\end{figure}

With the extracted central values for concurrence and Bell nonlocality, along with their respective uncertainties, we present the full set of results across different analysis methods and selection criteria. Assuming an integrated luminosity of $140~\text{fb}^{-1}$ (ATLAS Full Run 2~\cite{ATLAS_Run2_Lumi}), we provide the central values along with their upper and lower uncertainties as well as the precision in percentage.

Figure~\ref{fig:poi_results} summarizes all results from this paper, presenting the measured central values of concurrence and Bell nonlocality. The figure includes results from multiple analysis approaches: the decay method, the kinematic method, and the decay method with trigger Selection. Each measurement is shown with statistical-only (stat-only) uncertainties as well as combined statistical and systematic (stat+syst) uncertainties, providing a complete picture of the impact of different sources of uncertainty. The left panel features truncated axes, where the $\mathcal{C} = 0$ line indicates no entanglement, and $\mathcal{B} \leq 0$ corresponds to Bell local states.   

We see that both entanglement and Bell nonlocality can be observed with a significance well above $5\sigma$.  The precision, accounting for statistical and systematic uncertainties, on the concurrence is $6.5\%$ with the decay method and $<0.1\%$ with the kinematic method.  The measurement of Bell nonlocality has a precision of $2.1\%$ with the decay method and  $<0.1\%$ with the kinematic method.  These precise measurements, already with the current dataset, highlight the $\tau^+ \tau^-$ channel as likely the first two qubit state at the LHC to observe Bell nonlocality.

\begin{figure}[!htb]
    \centering
    \begin{subfigure}[b]{0.95\textwidth}
        \centering
        \includegraphics[width=\textwidth]
        {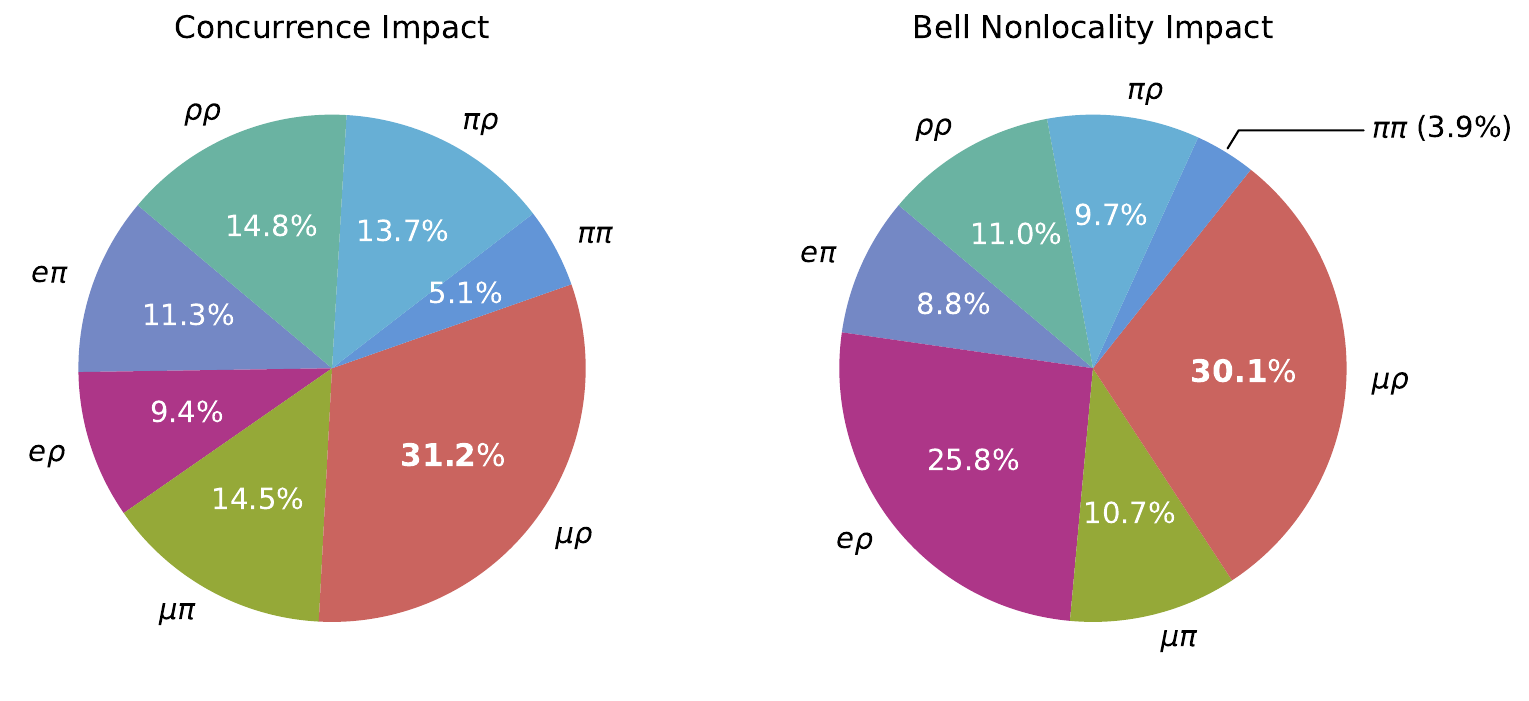}

        \end{subfigure}
    \caption{Impact of each decay channel on the concurrence (left) and on Bell nonlocality (right) using the decay method where the signal region includes the trigger requirement.  The impact is measured by the inverse variance.}
    \label{fig:pie_chart_TFTO}
\end{figure}

We take the combined results from the decay method in the signal region with trigger requirements as our primary results, providing the most reliable determination. Figure~\ref{fig:pie_chart_TFTO} presents a pie chart illustrating the contribution of each subchannel to the final combined results, where the impact is assessed based on the inverse variance.
Additionally, we highlight the most sensitive individual channels: $\rho\rho$ from the $\tau_\text{had}\tau_\text{had}$ category and $\mu\rho$ from the $\tau_\text{lep}\tau_\text{had}$ category. These selections represent the most precise measurements within their respective decay modes and offer a clear benchmark for channel-by-channel comparisons. 
The detailed values of the polarization terms ($B^\pm_i$) and spin correlation terms ($C_{ij}$) can be found in Table~\ref{tab:full_spin_density_result_comp}.

We show results for both the decay method and for the kinematic method.  At theory-level these two methods lead to the same mean value of concurrence $\mathcal{C}$ and the Bell variable $\mathcal{B}$, however, in the detailed simulation we perform, they can differ by a small amount.  The source of this difference is mainly due to the fact that template fitting is required for the decay method while for the kinematic method it is not used because the kinematics are already measured well.  One advantage of the kinematic method is that the analysis is simpler since template fitting is not required.

\begin{table}[h]
    \centering
        \caption{
    Summary of the fitted values for the polarization terms ($B^\pm_i$) and spin correlation terms ($C_{ij}$), with individual measurements provided for the $\rho\rho$ and $\mu\rho$ channels, as well as a combined result that incorporates all seven studied channels. Each entry represents the central fitted value with its corresponding upper and lower uncertainties. 
    Uncertainties are multiplied by 100 for better readability due to their small magnitudes.}
    \label{tab:full_spin_density_result_comp}
    \renewcommand{\arraystretch}{1.15}  
    \resizebox{1.0\linewidth}{!}{%
        \begin{tabular}{c | c c c | c c c | c c c | c c c}
            \toprule

 & \multicolumn{3}{c}{Decay Method (SR Only)} & \multicolumn{3}{c}{Decay Method (SR \& Trigger)} & \multicolumn{3}{c}{Kinematic Method (SR Only)} & \multicolumn{3}{c}{Kinematic Method (SR \& Trigger)} \\ \midrule
 & $\mu \rho$ & $\rho \rho$ & $\text{Combined}$ & $\mu \rho$ & $\rho \rho$ & $\text{Combined}$ & $\mu \rho$ & $\rho \rho$ & $\text{Combined}$ & $\mu \rho$ & $\rho \rho$ & $\text{Combined}$ \\ \midrule
$B^{-}_{k}$ & $\mathbf{-0.21}^{+\mathbf{ 0.02}}_{-\mathbf{ 0.02}}$ & $\mathbf{-0.21}^{+\mathbf{ 0.01}}_{-\mathbf{ 0.01}}$ & $\mathbf{-0.21}^{+\mathbf{ 0.00}}_{-\mathbf{ 0.00}}$ & $\mathbf{-0.21}^{+\mathbf{ 0.04}}_{-\mathbf{ 0.04}}$ & $\mathbf{-0.21}^{+\mathbf{ 0.01}}_{-\mathbf{ 0.01}}$ & $\mathbf{-0.21}^{+\mathbf{ 0.01}}_{-\mathbf{ 0.01}}$ & $\mathbf{-0.20}^{+\mathbf{ 0.06}}_{-\mathbf{ 0.06}}$ & $\mathbf{-0.20}^{+\mathbf{ 0.06}}_{-\mathbf{ 0.06}}$ & $\mathbf{-0.19}^{+\mathbf{ 0.02}}_{-\mathbf{ 0.02}}$ & $\mathbf{-0.21}^{+\mathbf{ 0.06}}_{-\mathbf{ 0.06}}$ & $\mathbf{-0.21}^{+\mathbf{ 0.09}}_{-\mathbf{ 0.09}}$ & $\mathbf{-0.21}^{+\mathbf{ 0.03}}_{-\mathbf{ 0.03}}$ \\
$B^{-}_{n}$ & $-0.01^{+ 0.00}_{- 0.00}$ & $\phantom{-} 0.01^{+ 0.00}_{- 0.00}$ & $\phantom{-} 0.00^{+ 0.00}_{- 0.00}$ & $-0.01^{+ 0.00}_{- 0.00}$ & $\phantom{-} 0.01^{+ 0.00}_{- 0.00}$ & $\phantom{-} 0.00^{+ 0.00}_{- 0.00}$ & $\phantom{-} 0.00^{+ 0.00}_{- 0.00}$ & $\phantom{-} 0.00^{+ 0.00}_{- 0.00}$ & $\phantom{-} 0.00^{+ 0.00}_{- 0.00}$ & $\phantom{-} 0.00^{+ 0.00}_{- 0.00}$ & $\phantom{-} 0.00^{+ 0.00}_{- 0.00}$ & $\phantom{-} 0.00^{+ 0.00}_{- 0.00}$ \\
$B^{-}_{r}$ & $\phantom{-} 0.00^{+ 0.00}_{- 0.00}$ & $-0.01^{+ 0.00}_{- 0.00}$ & $-0.00^{+ 0.00}_{- 0.00}$ & $\phantom{-} 0.00^{+ 0.00}_{- 0.00}$ & $-0.01^{+ 0.00}_{- 0.00}$ & $-0.00^{+ 0.00}_{- 0.00}$ & $-0.00^{+ 0.00}_{- 0.00}$ & $-0.00^{+ 0.00}_{- 0.00}$ & $-0.00^{+ 0.00}_{- 0.00}$ & $-0.00^{+ 0.00}_{- 0.00}$ & $-0.00^{+ 0.00}_{- 0.00}$ & $-0.00^{+ 0.00}_{- 0.00}$ \\
$B^{+}_{k}$ & $\mathbf{-0.21}^{+\mathbf{ 0.02}}_{-\mathbf{ 0.02}}$ & $\mathbf{-0.21}^{+\mathbf{ 0.02}}_{-\mathbf{ 0.02}}$ & $\mathbf{-0.21}^{+\mathbf{ 0.01}}_{-\mathbf{ 0.01}}$ & $\mathbf{-0.21}^{+\mathbf{ 0.06}}_{-\mathbf{ 0.06}}$ & $\mathbf{-0.21}^{+\mathbf{ 0.02}}_{-\mathbf{ 0.02}}$ & $\mathbf{-0.21}^{+\mathbf{ 0.01}}_{-\mathbf{ 0.01}}$ & $\mathbf{-0.20}^{+\mathbf{ 0.06}}_{-\mathbf{ 0.06}}$ & $\mathbf{-0.20}^{+\mathbf{ 0.06}}_{-\mathbf{ 0.06}}$ & $\mathbf{-0.19}^{+\mathbf{ 0.02}}_{-\mathbf{ 0.02}}$ & $\mathbf{-0.21}^{+\mathbf{ 0.06}}_{-\mathbf{ 0.06}}$ & $\mathbf{-0.21}^{+\mathbf{ 0.09}}_{-\mathbf{ 0.09}}$ & $\mathbf{-0.21}^{+\mathbf{ 0.03}}_{-\mathbf{ 0.03}}$ \\
$B^{+}_{n}$ & $\phantom{-} 0.00^{+ 0.00}_{- 0.00}$ & $\phantom{-} 0.00^{+ 0.00}_{- 0.00}$ & $\phantom{-} 0.00^{+ 0.00}_{- 0.00}$ & $\phantom{-} 0.00^{+ 0.00}_{- 0.00}$ & $\phantom{-} 0.00^{+ 0.00}_{- 0.00}$ & $\phantom{-} 0.00^{+ 0.00}_{- 0.00}$ & $\phantom{-} 0.00^{+ 0.00}_{- 0.00}$ & $\phantom{-} 0.00^{+ 0.00}_{- 0.00}$ & $\phantom{-} 0.00^{+ 0.00}_{- 0.00}$ & $\phantom{-} 0.00^{+ 0.00}_{- 0.00}$ & $\phantom{-} 0.00^{+ 0.00}_{- 0.00}$ & $\phantom{-} 0.00^{+ 0.00}_{- 0.00}$ \\
$B^{+}_{r}$ & $\phantom{-} 0.00^{+ 0.00}_{- 0.00}$ & $\phantom{-} 0.01^{+ 0.00}_{- 0.00}$ & $\phantom{-} 0.00^{+ 0.00}_{- 0.00}$ & $\phantom{-} 0.00^{+ 0.00}_{- 0.00}$ & $\phantom{-} 0.01^{+ 0.00}_{- 0.00}$ & $\phantom{-} 0.00^{+ 0.00}_{- 0.00}$ & $-0.00^{+ 0.00}_{- 0.00}$ & $-0.00^{+ 0.00}_{- 0.00}$ & $-0.00^{+ 0.00}_{- 0.00}$ & $-0.00^{+ 0.00}_{- 0.00}$ & $-0.00^{+ 0.00}_{- 0.00}$ & $-0.00^{+ 0.00}_{- 0.00}$ \\\midrule
$C_{kk}$ & $\mathbf{\phantom{-} 1.02}^{+\mathbf{ 0.00}}_{-\mathbf{ 0.00}}$ & $\mathbf{\phantom{-} 1.01}^{+\mathbf{ 0.00}}_{-\mathbf{ 0.00}}$ & $\mathbf{\phantom{-} 1.01}^{+\mathbf{ 0.00}}_{-\mathbf{ 0.00}}$ & $\mathbf{\phantom{-} 1.02}^{+\mathbf{ 0.00}}_{-\mathbf{ 0.00}}$ & $\mathbf{\phantom{-} 1.01}^{+\mathbf{ 0.00}}_{-\mathbf{ 0.00}}$ & $\mathbf{\phantom{-} 1.01}^{+\mathbf{ 0.00}}_{-\mathbf{ 0.00}}$ & $\mathbf{\phantom{-} 1.00}^{+\mathbf{ 0.00}}_{-\mathbf{ 0.00}}$ & $\mathbf{\phantom{-} 1.00}^{+\mathbf{ 0.00}}_{-\mathbf{ 0.00}}$ & $\mathbf{\phantom{-} 1.00}^{+\mathbf{ 0.00}}_{-\mathbf{ 0.00}}$ & $\mathbf{\phantom{-} 1.00}^{+\mathbf{ 0.00}}_{-\mathbf{ 0.00}}$ & $\mathbf{\phantom{-} 1.00}^{+\mathbf{ 0.00}}_{-\mathbf{ 0.00}}$ & $\mathbf{\phantom{-} 1.00}^{+\mathbf{ 0.00}}_{-\mathbf{ 0.00}}$ \\
$C_{nn}$ & $\mathbf{\phantom{-} 0.80}^{+\mathbf{ 0.03}}_{-\mathbf{ 0.03}}$ & $\mathbf{\phantom{-} 0.77}^{+\mathbf{ 0.05}}_{-\mathbf{ 0.05}}$ & $\mathbf{\phantom{-} 0.79}^{+\mathbf{ 0.01}}_{-\mathbf{ 0.01}}$ & $\mathbf{\phantom{-} 0.80}^{+\mathbf{ 0.07}}_{-\mathbf{ 0.07}}$ & $\mathbf{\phantom{-} 0.77}^{+\mathbf{ 0.06}}_{-\mathbf{ 0.06}}$ & $\mathbf{\phantom{-} 0.79}^{+\mathbf{ 0.03}}_{-\mathbf{ 0.03}}$ & $\mathbf{\phantom{-} 0.78}^{+\mathbf{ 0.03}}_{-\mathbf{ 0.03}}$ & $\mathbf{\phantom{-} 0.77}^{+\mathbf{ 0.03}}_{-\mathbf{ 0.03}}$ & $\mathbf{\phantom{-} 0.77}^{+\mathbf{ 0.01}}_{-\mathbf{ 0.01}}$ & $\mathbf{\phantom{-} 0.79}^{+\mathbf{ 0.03}}_{-\mathbf{ 0.03}}$ & $\mathbf{\phantom{-} 0.81}^{+\mathbf{ 0.23}}_{-\mathbf{ 0.23}}$ & $\mathbf{\phantom{-} 0.79}^{+\mathbf{ 0.03}}_{-\mathbf{ 0.03}}$ \\
$C_{rr}$ & $\mathbf{-0.80}^{+\mathbf{ 0.05}}_{-\mathbf{ 0.05}}$ & $\mathbf{-0.80}^{+\mathbf{ 0.03}}_{-\mathbf{ 0.03}}$ & $\mathbf{-0.79}^{+\mathbf{ 0.02}}_{-\mathbf{ 0.02}}$ & $\mathbf{-0.80}^{+\mathbf{ 0.08}}_{-\mathbf{ 0.08}}$ & $\mathbf{-0.80}^{+\mathbf{ 0.04}}_{-\mathbf{ 0.04}}$ & $\mathbf{-0.79}^{+\mathbf{ 0.03}}_{-\mathbf{ 0.03}}$ & $\mathbf{-0.78}^{+\mathbf{ 0.03}}_{-\mathbf{ 0.03}}$ & $\mathbf{-0.77}^{+\mathbf{ 0.03}}_{-\mathbf{ 0.03}}$ & $\mathbf{-0.77}^{+\mathbf{ 0.01}}_{-\mathbf{ 0.01}}$ & $\mathbf{-0.79}^{+\mathbf{ 0.03}}_{-\mathbf{ 0.03}}$ & $\mathbf{-0.81}^{+\mathbf{ 0.23}}_{-\mathbf{ 0.23}}$ & $\mathbf{-0.79}^{+\mathbf{ 0.03}}_{-\mathbf{ 0.03}}$ \\
$C_{kn}$ & $-0.01^{+ 0.00}_{- 0.00}$ & $\phantom{-} 0.03^{+ 0.00}_{- 0.00}$ & $\phantom{-} 0.00^{+ 0.00}_{- 0.00}$ & $-0.01^{+ 0.00}_{- 0.00}$ & $\phantom{-} 0.03^{+ 0.00}_{- 0.00}$ & $-0.00^{+ 0.00}_{- 0.00}$ & $\phantom{-} 0.00^{+ 0.00}_{- 0.00}$ & $\phantom{-} 0.00^{+ 0.00}_{- 0.00}$ & $\phantom{-} 0.00^{+ 0.00}_{- 0.00}$ & $\phantom{-} 0.00^{+ 0.00}_{- 0.00}$ & $\phantom{-} 0.00^{+ 0.00}_{- 0.00}$ & $\phantom{-} 0.00^{+ 0.00}_{- 0.00}$ \\
$C_{kr}$ & $\phantom{-} 0.01^{+ 0.00}_{- 0.00}$ & $-0.00^{+ 0.00}_{- 0.00}$ & $-0.00^{+ 0.00}_{- 0.00}$ & $\phantom{-} 0.01^{+ 0.00}_{- 0.00}$ & $-0.00^{+ 0.00}_{- 0.00}$ & $-0.01^{+ 0.00}_{- 0.00}$ & $\phantom{-} 0.00^{+ 0.00}_{- 0.00}$ & $\phantom{-} 0.00^{+ 0.00}_{- 0.00}$ & $\phantom{-} 0.00^{+ 0.00}_{- 0.00}$ & $\phantom{-} 0.01^{+ 0.00}_{- 0.00}$ & $\phantom{-} 0.01^{+ 0.00}_{- 0.00}$ & $\phantom{-} 0.01^{+ 0.00}_{- 0.00}$ \\
$C_{nr}$ & $\phantom{-} 0.03^{+ 0.00}_{- 0.00}$ & $\phantom{-} 0.03^{+ 0.00}_{- 0.00}$ & $\phantom{-} 0.02^{+ 0.00}_{- 0.00}$ & $\phantom{-} 0.03^{+ 0.00}_{- 0.00}$ & $\phantom{-} 0.03^{+ 0.00}_{- 0.00}$ & $\phantom{-} 0.02^{+ 0.00}_{- 0.00}$ & $\phantom{-} 0.00^{+ 0.00}_{- 0.00}$ & $\phantom{-} 0.00^{+ 0.00}_{- 0.00}$ & $\phantom{-} 0.00^{+ 0.00}_{- 0.00}$ & $\phantom{-} 0.00^{+ 0.00}_{- 0.00}$ & $\phantom{-} 0.00^{+ 0.00}_{- 0.00}$ & $\phantom{-} 0.00^{+ 0.00}_{- 0.00}$ \\
$C_{rk}$ & $\phantom{-} 0.01^{+ 0.00}_{- 0.00}$ & $\phantom{-} 0.00^{+ 0.00}_{- 0.00}$ & $\phantom{-} 0.00^{+ 0.00}_{- 0.00}$ & $\phantom{-} 0.01^{+ 0.00}_{- 0.00}$ & $\phantom{-} 0.00^{+ 0.00}_{- 0.00}$ & $\phantom{-} 0.00^{+ 0.00}_{- 0.00}$ & $\phantom{-} 0.00^{+ 0.00}_{- 0.00}$ & $\phantom{-} 0.00^{+ 0.00}_{- 0.00}$ & $\phantom{-} 0.00^{+ 0.00}_{- 0.00}$ & $\phantom{-} 0.01^{+ 0.00}_{- 0.00}$ & $\phantom{-} 0.01^{+ 0.00}_{- 0.00}$ & $\phantom{-} 0.01^{+ 0.00}_{- 0.00}$ \\
$C_{rn}$ & $\phantom{-} 0.02^{+ 0.00}_{- 0.00}$ & $\phantom{-} 0.00^{+ 0.00}_{- 0.00}$ & $\phantom{-} 0.02^{+ 0.00}_{- 0.00}$ & $\phantom{-} 0.02^{+ 0.00}_{- 0.00}$ & $\phantom{-} 0.00^{+ 0.00}_{- 0.00}$ & $\phantom{-} 0.02^{+ 0.00}_{- 0.00}$ & $\phantom{-} 0.00^{+ 0.00}_{- 0.00}$ & $\phantom{-} 0.00^{+ 0.00}_{- 0.00}$ & $\phantom{-} 0.00^{+ 0.00}_{- 0.00}$ & $\phantom{-} 0.00^{+ 0.00}_{- 0.00}$ & $\phantom{-} 0.00^{+ 0.00}_{- 0.00}$ & $\phantom{-} 0.00^{+ 0.00}_{- 0.00}$ \\
$C_{nk}$ & $\phantom{-} 0.00^{+ 0.00}_{- 0.00}$ & $-0.00^{+ 0.00}_{- 0.00}$ & $-0.00^{+ 0.00}_{- 0.00}$ & $\phantom{-} 0.00^{+ 0.01}_{- 0.01}$ & $-0.00^{+ 0.00}_{- 0.00}$ & $-0.00^{+ 0.00}_{- 0.00}$ & $\phantom{-} 0.00^{+ 0.00}_{- 0.00}$ & $\phantom{-} 0.00^{+ 0.00}_{- 0.00}$ & $\phantom{-} 0.00^{+ 0.00}_{- 0.00}$ & $\phantom{-} 0.00^{+ 0.00}_{- 0.00}$ & $\phantom{-} 0.00^{+ 0.00}_{- 0.00}$ & $\phantom{-} 0.00^{+ 0.00}_{- 0.00}$ \\

            \bottomrule
        \end{tabular}%
    }
\end{table}

\subsection{Systematics Impact}
\label{sec:syst_impact}

The impact of systematic uncertainties on measuring concurrence and Bell nonlocality is evaluated using a covariance matrix-based approach~\cite{Pinto:2023yob}, which is applicable exclusively to the template fit method.

The impact of each nuisance parameter (NP) on a given parameter of interest (POI) is determined from the covariance matrix as  $I_n = \sigma_a C_{a,n} \sigma_n $, where  $\sigma_a$  is the total uncertainty of the POI,  $C_{a,n}$  is the correlation coefficient between the POI and NP, and  $\sigma_n$  is the post-fit uncertainty of the NP. The total systematic uncertainty is obtained by summing the individual impacts in quadrature,  $\sigma_{\text{sys}} = \sqrt{\sum_n I_n^2}$, while the statistical uncertainty is extracted by subtracting the systematic contribution from the total uncertainty in quadrature.

Systematic uncertainties are categorized into groups corresponding to MC statistics, neutrino sampling, JES, TES, Soft MET, luminosity, signal cross sections, and background cross sections. This method is exclusive to the decay-based approach as it relies on template fitting.
The results are shown in Table~\ref{tab:systematics_impact}.

\begin{table}[h]
    \centering
        \caption{
    Summary of systematic uncertainty impacts on the measurement of concurrence and Bell nonlocality, ranked by their relative effects to the total uncertainty. 
    The table presents results for the two most sensitive channels, $\mu\rho$ and $\rho\rho$, as well as the combined result of all 7 subchannels. 
    The uncertainties are evaluated in the signal region (SR Only) and with trigger effects included (SR \& Trigger). 
    The total systematic uncertainty is computed as the quadrature sum of individual contributions.
    }
    \label{tab:systematics_impact}
    \renewcommand{\arraystretch}{1.05} 
    \resizebox{0.95\linewidth}{!}{
        \begin{tabular}{l | c c c | c c c }
            \toprule

 & \multicolumn{3}{c}{SR Only} & \multicolumn{3}{c}{SR \& Trigger} \\ \midrule 
 & $\rho\rho$ & $\mu\rho$ & $\text{Combined}$ & $\rho\rho$ & $\mu\rho$ & $\text{Combined}$ \\ \midrule 
\textbf{All Systematics} & \textbf{29.81\%} & \textbf{29.76\%} & \textbf{31.29\%} & \textbf{31.00\%} & \textbf{29.82\%} & \textbf{31.35\%} \\
MC Statistics & 29.31\% & 29.56\% & 30.05\% & 28.93\% & 29.55\% & 28.66\% \\
Luminosity & 0.12\% & 0.08\% & 0.90\% & 7.73\% & 0.74\% & 6.10\% \\
Background Cross-Section & 3.39\% & 0.07\% & 2.01\% & 1.51\% & 0.05\% & 1.06\% \\
Signal Cross-Section & 0.23\% & 0.23\% & 1.71\% & 3.12\% & 0.52\% & 2.41\% \\
Tau Energy Scale & 1.47\% & 2.50\% & 2.12\% & 1.20\% & 0.89\% & 1.47\% \\
Jet Enery Scale & 1.67\% & 1.50\% & 4.49\% & 2.41\% & 1.72\% & 8.05\% \\
Soft MET ($p_x$, $p_y$) & 3.66\% & 1.90\% & 6.57\% & 6.68\% & 3.42\% & 7.11\% \\
$\nu$ Sampling & 0.02\% & 0.02\% & 0.03\% & 0.06\% & 0.03\% & 0.07\% \\

            \bottomrule
        \end{tabular}
    }
\end{table}

\section{Discussion and Conclusions}
\label{sec:conclusions}

This study explores the potential of the $pp \to \tau^+\tau^- X$ process as a novel probe for quantum entanglement and Bell nonlocality at the LHC. The results highlight both methodological advantages and experimental considerations, offering a path toward precision measurements of fundamental quantum correlations in hadronic environments.  Some key observations are summarized as follows.

\begin{itemize}

\item \textbf{Comparison Between $\tau^+\tau^-$ and $t\bar{t}$ Channels} \\
A comparison with $t\bar{t}$ further underscores the advantages of $\tau\tau$ for quantum entanglement and Bell nonlocality. 
Experimentally, the inclusive $Z\to\tau\tau$ cross section, without any $\tau$ branching ratio included, is measured to be 1848~pb~\cite{ZtautauXsec:2018}, while the inclusive $t\bar{t}$ production cross section, without any $t$ branching ratio included, is 833.9~pb~\cite{TtbarNNLO}.
While the $t\bar{t}$ threshold region achieves a comparable event yield for entanglement studies~\cite{Han:2023fci}, the $\tau\tau$ process benefits from lower theoretical uncertainties~\cite{Fabbrichesi:2022ovb} and a cleaner final state. 
For Bell nonlocality, the advantage of $\tau\tau$ is even more pronounced. 
The event yield in the boosted $t\bar{t}$ region, where Bell nonlocality becomes accessible, is 20.93 (13.76) events per fb$^{-1}$ under the weak (strong) scenario. In contrast, the $\tau\tau$ process yields two orders of magnitude higher across multiple subchannels, enabling substantially improved statistical precision. 
Moreover, in $t\bar{t}$, entanglement and Bell nonlocality emerge in distinct kinematic regions, requiring separate analysis strategies. In $\tau\tau$, both effects are accessible within the same phase space near the $Z$ pole, allowing a unified measurement. This makes $\tau^+\tau^-$ a more favorable channel both systematically and statistically.

\item \textbf{Neutrino Momentum Reconstruction: Machine Learning vs MMC} \\
Missing Mass Calculator is a traditional tool that uses known distributions between visible and invisible decay products to constrain neutrino kinematics. However, it suffers from computational inefficiencies and non-negligible failure rates due to its reliance on low-dimensional scans. In this work, a diffusion-based machine learning model is adopted to capture high-dimensional correlations. The ML method achieves a tau-pair mass resolution of approximately 6\%, significantly outperforming the 20\% resolution of MMC. 
It also correctly reproduces the angular distribution $\cos\theta$, which is heavily distorted in the MMC reconstruction. The ML model successfully reconstructs all events without convergence issues, retains more statistics in the signal region about an order of magnitude, and remains robust under key systematic variations, including tau energy scale, soft MET, and neutrino sampling, making it well-suited for precision measurements.

\item \textbf{Background Estimation and Fake Taus} \\
In real experimental conditions, significant background arises from jets misidentified as taus, which are difficult to simulate accurately. While the current study applies a tau-prongness requirement that effectively reduces these backgrounds, suppressing even QCD contributions to negligible levels, and this scenario does not reflect the full complexity of a hadron collider environment. The results here should therefore be regarded as an idealized benchmark. Future experimental applications will require dedicated strategies to model and estimate fake backgrounds, likely involving data-driven methods and in situ measurements, to ensure accurate interpretation of the results.

\item \textbf{Trigger Requirements and Event Selection} \\
At hadron colliders, trigger strategies are essential for rejecting large QCD backgrounds, typically requiring high $p_T$ thresholds for both leptons and hadronic taus. These trigger cuts reduce the overall event yield by roughly an order of magnitude. Nevertheless, the large production rate of $Z \to \tau^+\tau^-$ ensures that the remaining statistics are sufficient for performing meaningful measurements. A detailed trigger study could further enhance the measurement precision, especially if extended to future high-luminosity data-taking periods.

\item \textbf{Decay Method vs Kinematic Method} \\
The decay method and the kinematic method offer complementary approaches for probing quantum correlations in the $\tau^+\tau^-$ final state. The kinematic method is constructed using only two observables ($m_{\tau\tau}$ and $\theta_\tau$) and relies on the analytic prediction derived under Standard Model (SM) assumptions. As a result, this method remains robust even in the presence of background contamination, provided that sufficient statistics are available near the $Z$ pole. Its strength lies in leveraging the theoretical form of the angular distribution, which compensates for experimental imperfections such as signal purity. 
In contrast, the decay method is more general and does not assume any specific theoretical framework, making it model-independent by construction. It directly reconstructs the spin correlation from the full decay kinematics, offering a more reliable interpretation in scenarios beyond the SM. Although it typically yields larger uncertainties due to its greater complexity and sensitivity to detector effects, it remains the more experimentally preferred approach for precision measurements. The consistency observed between the two methods in our results is encouraging, as it supports the validity of the measurement across different theoretical and experimental treatments.

\end{itemize}

In conclusion, our results demonstrate that the $\tau^+ \tau^-$ system offers a robust and experimentally accessible platform for studying quantum entanglement and Bell nonlocality at the LHC. By using advanced machine learning techniques for neutrino reconstruction, we achieve precise measurements of the full spin density matrix, a critical advantage over previous studies limited by reconstruction challenges. Our analysis reveals a clear violation of Bell inequalities with high statistical significance, surpassing 5$\sigma$, establishing $\tau^+ \tau^-$ as an ideal system for quantum information studies in high-energy collisions. Given its experimental feasibility and sensitivity with which Bell nonlocality can be probed, we propose that $\tau^+ \tau^-$ should be regarded as the new benchmark system for quantum information studies at the LHC, complementing and extending the insights gained from the $t\bar{t}$ system.

\section{Acknowledgements}

The authors would like to thank Quentin Baut, Kun Cheng, Lawrence Lee, Navin McGinnis, and Vinicius Mikuni for helpful conversations. 
YZ and SCH are supported by the U.S. National Science Foundation under Grant Number 2209034.
SCH is partially supported by DGE-2021540.
Zhou and Li are supported by National Key R\&D Program of China (Nos. 2023YFA1605703 and 2023YFA1605700), Shanghai Pilot Program for Basic Research – Shanghai Jiao Tong University (No. 21TQ1400209), National Natural Science Foundation of China (No. 12150006). Zhou is supported by T.D. Lee scholarship.
This work was supported in part by the US Department of Energy under grant No. DE-SC0007914 and in part by Pitt PACC.  ML is also supported by the National Science Foundation under grant No. PHY-2112829.

\appendix

\section{Full Density Matrix}
\label{app:densitymatrix}

The production of $\tau^+ \tau^-$ proceeds via $q\bar{q} \to \gamma \to \tau^+ \tau^-$ and $q\bar{q} \to Z \to \tau^+ \tau^-$.  We define the production density matrix for each partonic process as
\begin{equation}
    R^{q\bar{q}}_{ab,\bar{a}\bar{b}}(\hat{s},\theta) \propto \mathcal{M}(q\bar{q}\to\tau^+_a \tau^-_b) \mathcal{M}^*(q\bar{q}\to\tau^+_{\bar{a}} \tau^-_{\bar{b}}),
\end{equation}
where $a,b$ are the spin indices for $\tau^+\tau^-$, $\sqrt{\hat{s}}$ is the partonic center of mass energy and $\theta$ is scattering angle of the $q\bar{q}\to\tau^+ \tau^-$ process. The $4\times 4$ matrix $R^{q\bar{q}}$ has a decomposition
\begin{equation}
R^{q\bar{q}}(\hat{s},\theta) = \frac{1}{4} \bigg( \tilde{A}^{q\bar{q}}\, \mathbb{I}_2 \otimes \mathbb{I}_2
+ \sum_i \tilde{B}^{+,q\bar{q}}_i \sigma_i \otimes \mathbb{I}_2
+ \sum_j \tilde{B}^{-,q\bar{q}}_j \mathbb{I}_2 \otimes \sigma_j
+ \sum_{ij} \tilde{C}_{ij}^{q\bar{q}} \sigma_i \otimes \sigma_j \bigg),
\end{equation}
which is similar to the spin density matrix but normalized differently.  The density matrix is found by appropriately weighting initial states by their parton luminosity
\begin{align}
    \rho(\hat{s},\theta) &= \frac{\mathcal{L}^{u\bar{u}}(\xi)\mathbb{R}^{u\bar{u}}(\hat{s},\theta) + \big(\mathcal{L}^{d\bar{d}}(\xi) + \mathcal{L}^{s\bar{s}}(\xi)\big)\mathbb{R}^{d\bar{d}}(\hat{s},\theta)}{\mathcal{L}^{u\bar{u}}(\xi)\,\mathrm{Tr} \mathbb{R}^{u\bar{u}}(\hat{s},\theta) + \big(\mathcal{L}^{d\bar{d}}(\xi) + \mathcal{L}^{s\bar{s}}(\xi)\big)\,\mathrm{Tr}\mathbb{R}^{d\bar{d}}(\hat{s},\theta)}, \label{eq:R2rho}
    \\
    \mbox{with}\hspace{0.7em}&\hspace{-0.2em} \mathbb{R}^{q\bar{q}}(\hat{s},\theta)=R^{q\bar{q}}(\hat{s},\theta)+R^{q\bar{q}}(\hat{s},\pi-\theta),
\end{align}
where $\xi=\hat{s}/s$ and the parton luminosity is defined in term of the parton distribution function of the corresponding quark $f^{q}(x)$, given by
\begin{align}
    \mathcal{L}^{q\bar{q}}(\xi) = 2\int_\xi^1 \frac{\dd x}{x} f^q(x) f^{\bar{q}}\big(\frac{\xi}{x}\big),
\end{align}
and $\mathbb{R}^{q\bar{q}}$ accounts for the contribution from both $q\bar{q}$ and $\bar{q}q$.
Eq.~(\ref{eq:R2rho}) implies
\begin{align}
    B_i^{\pm}(\hat{s},\theta) &= \frac{\mathcal{L}^{u\bar{u}}(\xi) \mathbb{B}_i^{\pm,u\bar{u}}(\hat{s},\theta) + \big(\mathcal{L}^{d\bar{d}}(\xi) + \mathcal{L}^{s\bar{s}}(\xi)\big) \mathbb{B}_i^{\pm,d\bar{d}}(\hat{s},\theta)}{\mathcal{L}^{u\bar{u}}(\xi)\, \mathbb{A}^{u\bar{u}}(\hat{s},\theta) + \big(\mathcal{L}^{d\bar{d}}(\xi) + \mathcal{L}^{s\bar{s}}(\xi)\big)\, \mathbb{A}^{d\bar{d}}(\hat{s},\theta)}, \\
    C_{ij}(\hat{s},\theta) &= \frac{\mathcal{L}^{u\bar{u}}(\xi)\, \mathbb{C}_{ij}^{u\bar{u}}(\hat{s},\theta) + \big(\mathcal{L}^{d\bar{d}}(\xi) + \mathcal{L}^{s\bar{s}}(\xi)\big)\, \mathbb{C}_{ij}^{d\bar{d}}(\hat{s},\theta)}{\mathcal{L}^{u\bar{u}}(\xi)\, \mathbb{A}^{u\bar{u}}(\hat{s},\theta) + \big(\mathcal{L}^{d\bar{d}}(\xi) + \mathcal{L}^{s\bar{s}}(\xi)\big)\, \mathbb{A}^{d\bar{d}}(\hat{s},\theta)},
\end{align}
with coefficients given by
\begin{align}
    \mathbb{A}^{q\bar{q}}(\hat{s},\theta)&=\tilde{A}^{q\bar{q}}(\hat{s},\theta)+\tilde{A}^{q\bar{q}}(\hat{s},\pi-\theta),\\
    \mathbb{B}_i^{\pm,q\bar{q}}(\hat{s},\theta)&=\tilde{B}_i^{\pm,q\bar{q}}(\hat{s},\theta)+\tilde{B}_i^{\pm,q\bar{q}}(\hat{s},\pi-\theta),\\
    \mathbb{C}_{ij}^{q\bar{q}}(\hat{s},\theta)&=\tilde{C}_{ij}^{q\bar{q}}(\hat{s},\theta)+\tilde{C}_{ij}^{q\bar{q}}(\hat{s},\pi-\theta).
\end{align}
For $q\bar{q} \to \tau^+ \tau^-$ with an $s$-channel $\gamma$ and $Z$, the $\tilde{A}$, $\tilde{B}^\pm_i$, and $\tilde{C}_{ij}$ coefficients are given by
\begin{align}
\scalebox{0.86}{$\tilde{A}^{q\bar{q}} =$}&\scalebox{0.86}{$ \, \frac{1}{16} \bigg\{ Q_q^2 Q_\tau^2 \big[ 2 - \beta^2 \sin^2\theta \big] 
+ 2Q_q Q_\tau \operatorname{Re} \big[ \chi(\hat{s}) \big] 
\Big[ 2\beta g_A^q g_A^\tau \cos\theta + g_V^q g_V^\tau \big( 2 - \beta^2 \sin^2\theta \big) \Big]$} \notag \\
& \scalebox{0.86}{$+ \big| \chi(\hat{s}) \big|^2 
\bigg[ \big( g_V^{q2} + g_A^{q2} \big) \Big( 2g_V^{\tau2} + 2\beta^2 g_A^{\tau2} - \beta^2 
\big( g_V^{\tau2} + g_A^{\tau2}\big) \sin^2\theta \Big) + 8\beta g_V^q g_V^\tau g_A^q g_A^\tau \cos\theta \bigg] 
\bigg\},$} \\
\scalebox{0.86}{$\tilde{B}_k^{\pm,q \bar{q}}=$} & \scalebox{0.86}{$- \frac{1}{8} \Big\{Q_q Q_\tau \operatorname{Re} \big[ \chi(\hat{s}) \big] \Big[\beta g_A^\tau g_V^q\big(1+\cos ^2 \theta\big)+2 g_A^q g_V^\tau \cos \theta\Big]$}\notag \\
& \scalebox{0.86}{$+ \big| \chi(\hat{s}) \big|^2 \Big[2 g_A^q g_V^q\big(\beta^2 g_A^{\tau 2}+g_V^{\tau 2}\big) \cos \theta+\beta g_A^\tau g_V^\tau\big(g_V^{q 2}+g_A^{q 2}\big)\big(1+\cos ^2 \theta\big)\Big]\Big\},$} \\
\scalebox{0.86}{$\tilde{B}_r^{\pm,q \bar{q}}=$} & \scalebox{0.86}{$- \frac{1}{8} \sin \theta \sqrt{1-\beta^2}\Big\{Q_q Q_\tau \operatorname{Re} \big[ \chi(\hat{s}) \big] \big[\beta g_A^\tau g_V^q \cos \theta+2 g_A^q g_V^\tau\big] \notag$} \\
& \hspace{6em} \scalebox{0.86}{$+ \big| \chi(\hat{s}) \big|^2 g_V^\tau\Big[\beta g_A^\tau\big(g_V^{q 2}+g_A^{q 2}\big) \cos \theta+2 g_A^q g_V^q g_V^\tau\Big] \Big\},$} \\
\scalebox{0.86}{$\tilde{B}_n^{\pm,q \bar{q}}=$} & \scalebox{0.86}{$\,0,$}
\\
\scalebox{0.86}{$\tilde{C}_{n n}^{q \bar{q}}=$}&\scalebox{0.86}{$-\frac{1}{16} \beta^2 \sin ^2 \theta\Big\{Q_q^2 Q_\tau^2+2 Q_q Q_\tau \operatorname{Re} \big[ \chi(\hat{s}) \big] g_V^q g_V^\tau-\big| \chi(\hat{s}) \big|^2 \big(g_V^{q 2}+g_A^{q 2}\big)\big(g_A^{\tau 2}-g_V^{\tau 2}\big)\Big\},$} \\
\scalebox{0.86}{$\tilde{C}_{r r}^{q \bar{q}} =$}& \scalebox{0.86}{$ -\frac{1}{16} \sin ^2 \theta \Big\{\big(\beta^2-2\big) Q_q^2 Q_\tau^2+2 Q_q Q_\tau \operatorname{Re} \big[ \chi(\hat{s}) \big] g_V^q g_V^\tau\big(\beta^2-2\big)$} \notag \\
& \hspace{5em} \scalebox{0.86}{$+\big| \chi(\hat{s}) \big|^2 \Big[\beta^2\big(g_A^{\tau 2}+g_V^{\tau 2}\big)-2 g_V^{\tau 2}\Big]\big(g_V^{q 2}+g_A^{q 2}\big)\Big\},$} \\
\scalebox{0.86}{$\tilde{C}_{k k}^{q \bar{q}} = $}& \scalebox{0.86}{$\, \frac{1}{16}\bigg\{Q_q^2 Q_\tau^2\Big[\big(\beta^2-2\big) \sin ^2 \theta+2\Big]$} \notag\\
& \scalebox{0.86}{$+2 Q_q Q_\tau \operatorname{Re} \big[ \chi(\hat{s}) \big] \Big[2 \beta g_A^q g_A^\tau \cos \theta+g_V^q g_V^\tau\big((\beta^2-2) \sin ^2 \theta+2\big)\Big]$} \notag\\
& \scalebox{0.86}{$+\big| \chi(\hat{s}) \big|^2 \bigg[8 \beta g_A^q g_A^\tau g_V^q g_V^\tau \cos \theta 
+ \big(g_V^{q 2}+g_A^{q 2}\big)\Big(2 g_V^{\tau 2} \cos ^2 \theta-\beta^2 \big(g_A^{\tau 2}-g_V^{\tau 2}\big) \sin ^2 \theta+2 \beta^2 g_A^{\tau 2}\Big)\!\bigg]\! \bigg\},$} \\
\scalebox{0.86}{$\tilde{C}_{k r}^{q \bar{q}} =$} &\scalebox{0.86}{$\, \tilde{C}_{r k}^{q \bar{q}}= \frac{1}{8} \sin \theta \sqrt{1-\beta^2} \Big\{Q_q^2 Q_\tau^2 \cos \theta+Q_q Q_\tau \operatorname{Re} \big[ \chi(\hat{s}) \big] \left[\beta g_A^q g_A^\tau+2 g_V^q g_V^\tau \cos \theta\right]$} \notag \\
& \scalebox{0.86}{$\hspace{3em} +\big| \chi(\hat{s}) \big|^2 \Big[2 \beta g_A^q g_A^\tau g_V^q g_V^\tau+g_V^{\tau 2}\big(g_V^{q 2}+g_A^{q 2}\big) \cos \theta\Big]\Big\},$} \\
\scalebox{0.86}{$\tilde{C}_{nr}^{q \bar{q}} =$} &\scalebox{0.86}{$\, \tilde{C}_{rn}^{q \bar{q}} = \tilde{C}_{nk}^{q \bar{q}} =\tilde{C}_{kn}^{q \bar{q}}=0,$}
\end{align}
where $\beta = \sqrt{1-4m_\tau^2/\hat{s}}$, and $Q_{\tau}=-1$, $Q_{u}=2/3$, $Q_{d}=-1/3$ are the electric charges, while $g_V^i$ and $g_A^i$ are the vector and axial-vector couplings given by
\begin{equation}
    g_V^i=\frac{1}{2}T_3^i- Q_i \sin ^2 \theta_W, \quad g_A^i=\frac{1}{2}T_3^i.
\end{equation}
Additionally, we have
\begin{align}
    \operatorname{Re} \big[ \chi(q^2) \big]&=\frac{q^2(q^2-m_Z^2)}{\sin ^2 \theta_W \cos ^2 \theta_W\big[(q^2-m_Z^2)^2+q^4 \Gamma_Z^2 / m_Z^2\big]}, \\
    \big|\chi(q^2)\big|^2&=\frac{q^4}{\sin ^4 \theta_W \cos ^4 \theta_W \big[(q^2-m_Z^2)^2+q^4 \Gamma_Z^2 / m_Z^2\big]},
\end{align}
where $\theta_W$ is the weak-mixing angle, $m_Z$ and $\Gamma_Z$ are the mass and width of $Z$, respectively.

\section{Additional Results}

\subsection{Kinematic Distributions}

\begin{figure}[h!]
    \centering
    \begin{subfigure}{\textwidth}
        \centering
        \begin{subfigure}{0.48\textwidth}
            \centering
            \includegraphics[width=\linewidth]{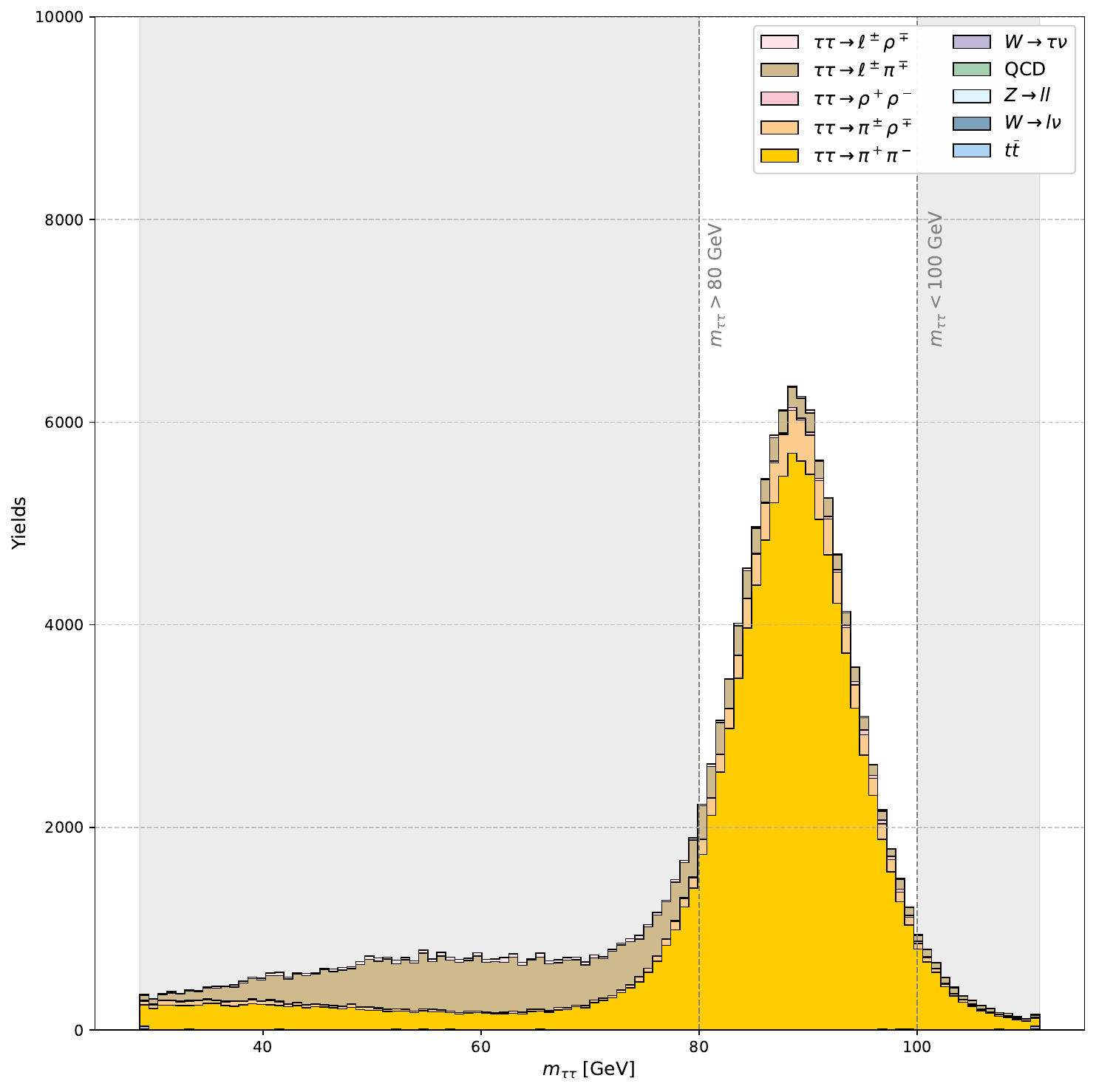}
        \end{subfigure}
        \hfill
        \begin{subfigure}{0.48\textwidth}
            \centering
            \includegraphics[width=\linewidth]{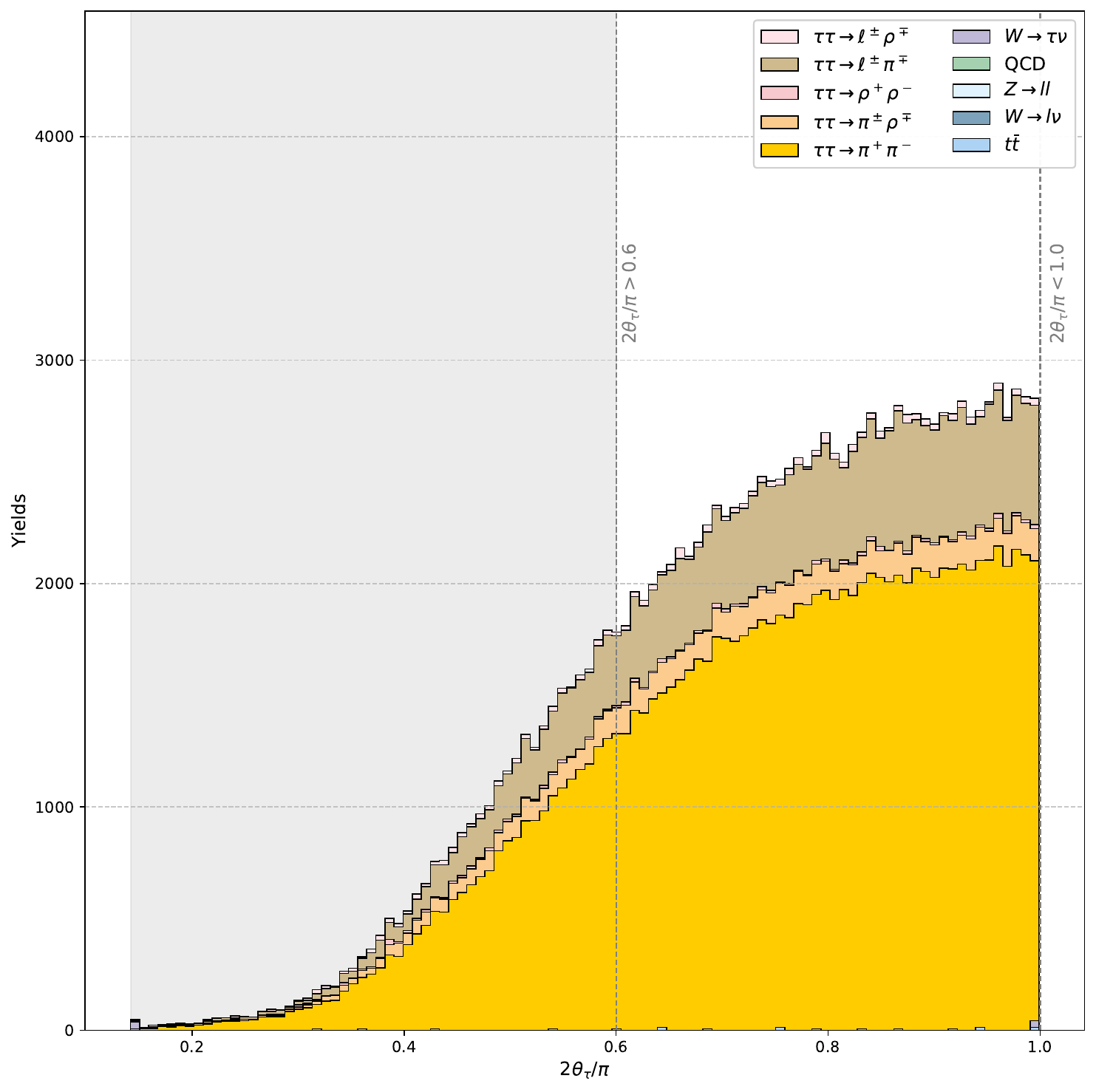}
        \end{subfigure}
    \end{subfigure}
    \caption{Distribution of $m_{\tau\tau}$ (left) and of $\theta_\tau$ (right) for the $\pi\pi$ subchannel.}
    \label{fig:extra-pipi}
\end{figure}

\begin{figure}[h!]
    \centering
    \begin{subfigure}{\textwidth}
        \centering
        \begin{subfigure}{0.48\textwidth}
            \centering
            \includegraphics[width=\linewidth]{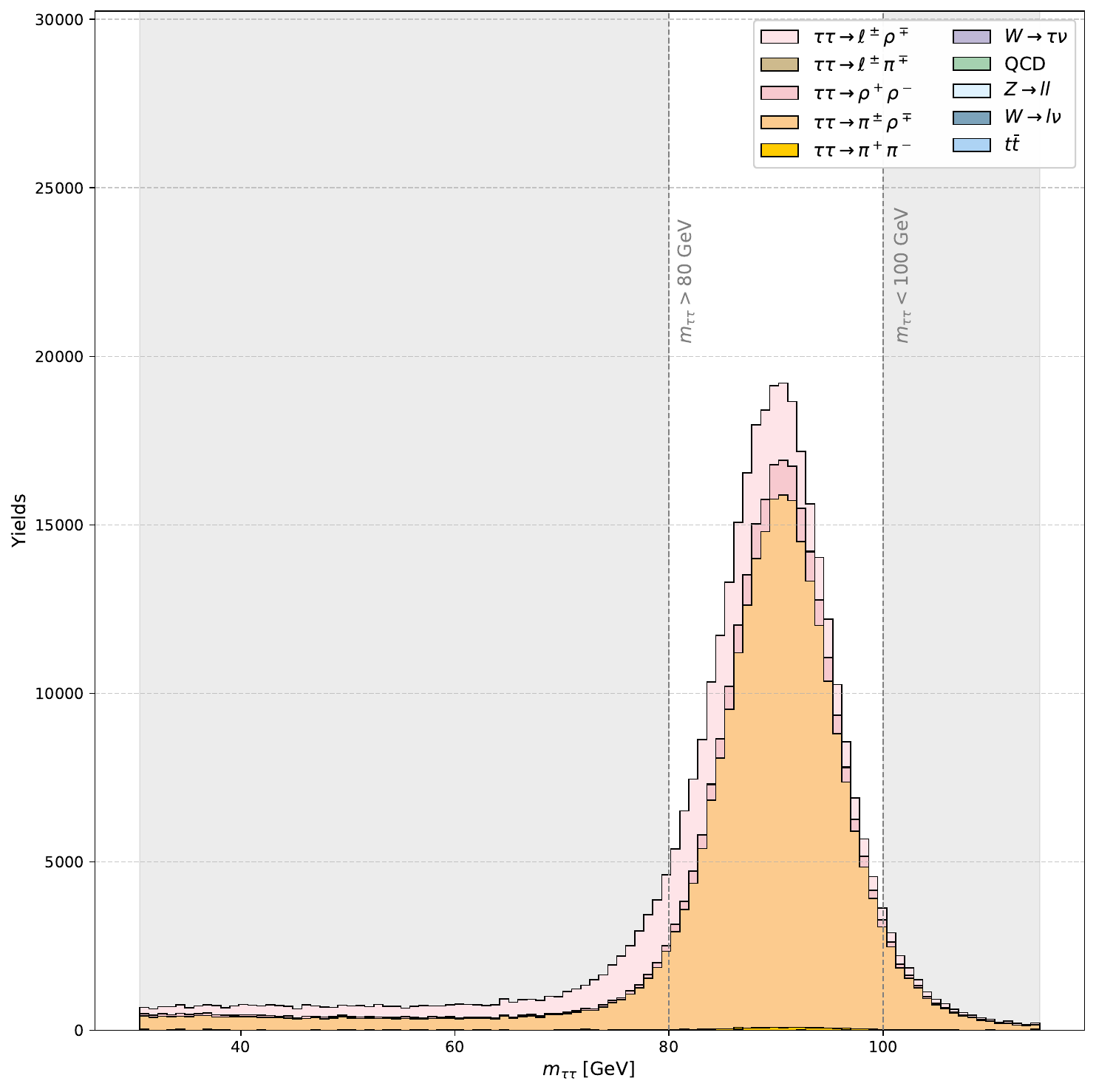}
        \end{subfigure}
        \hfill
        \begin{subfigure}{0.48\textwidth}
            \centering
            \includegraphics[width=\linewidth]{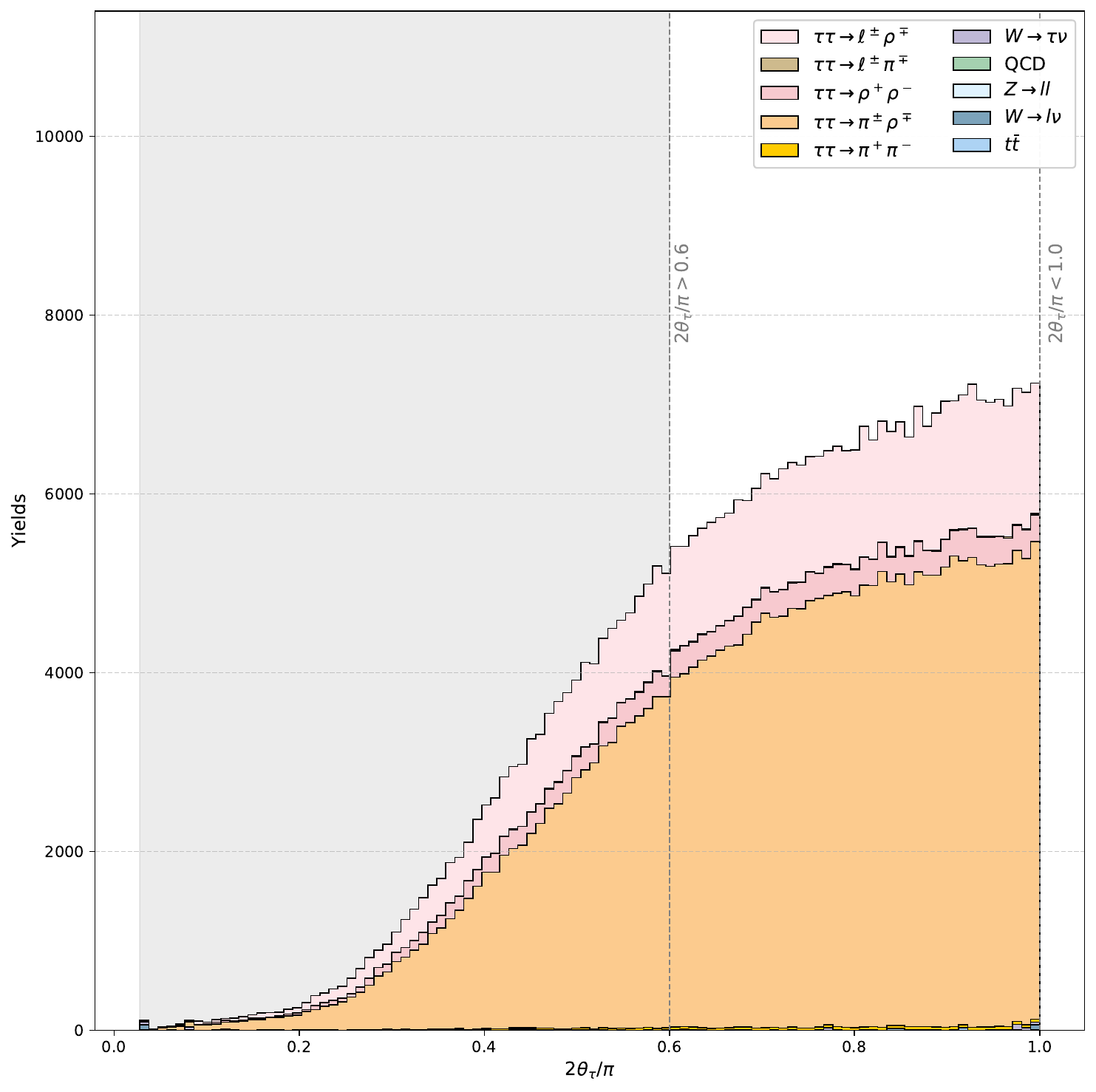}
        \end{subfigure}
    \end{subfigure}
    \caption{Distribution of $m_{\tau\tau}$ (left) and of $\theta_\tau$ (right) for the $\pi\rho$ subchannel.}
    \label{fig:extra-pirho}
\end{figure}

\begin{figure}[h!]
    \centering
    \begin{subfigure}{\textwidth}
        \centering
        \begin{subfigure}{0.48\textwidth}
            \centering
            \includegraphics[width=\linewidth]{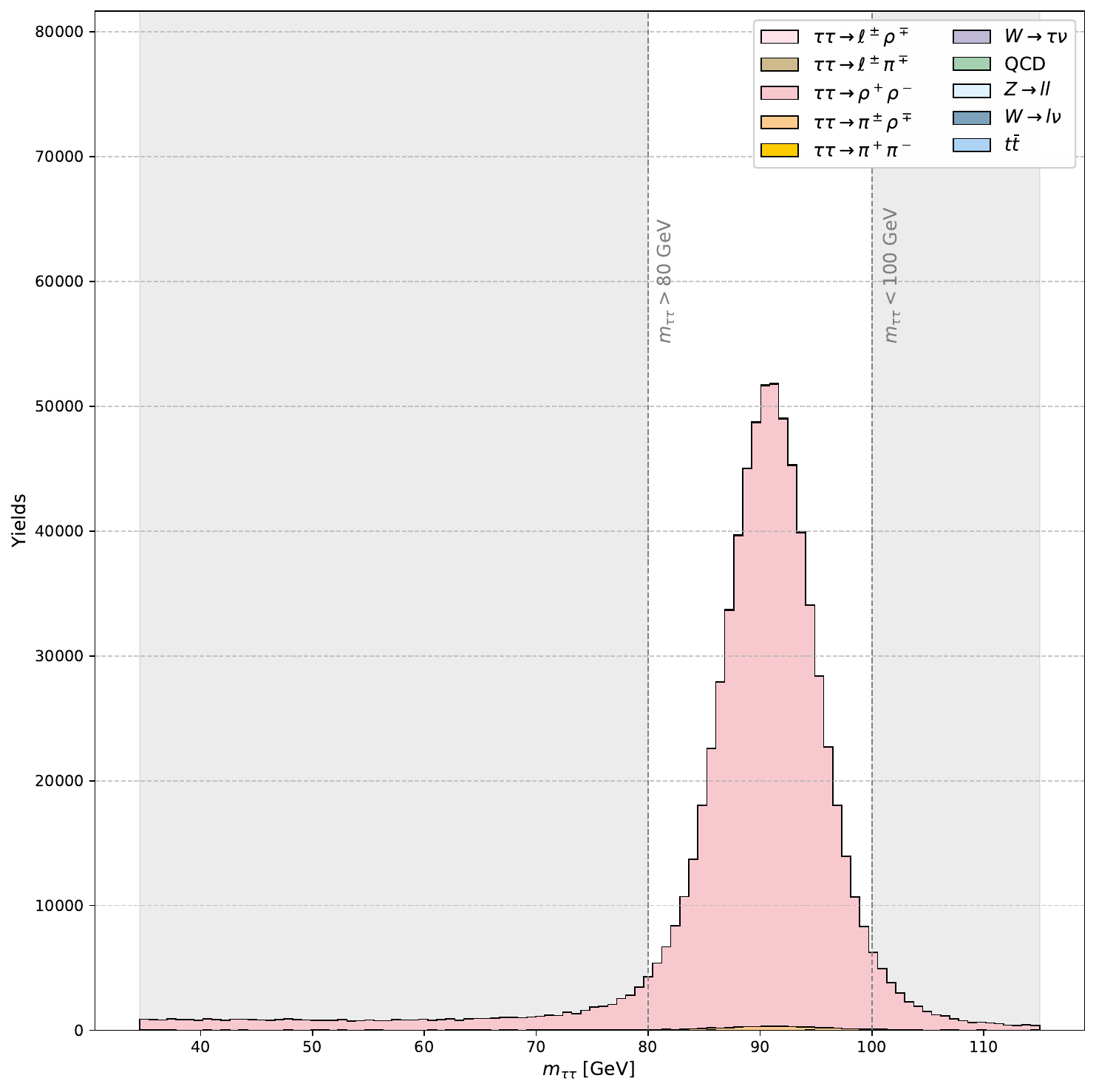}
        \end{subfigure}
        \hfill
        \begin{subfigure}{0.48\textwidth}
            \centering
            \includegraphics[width=\linewidth]{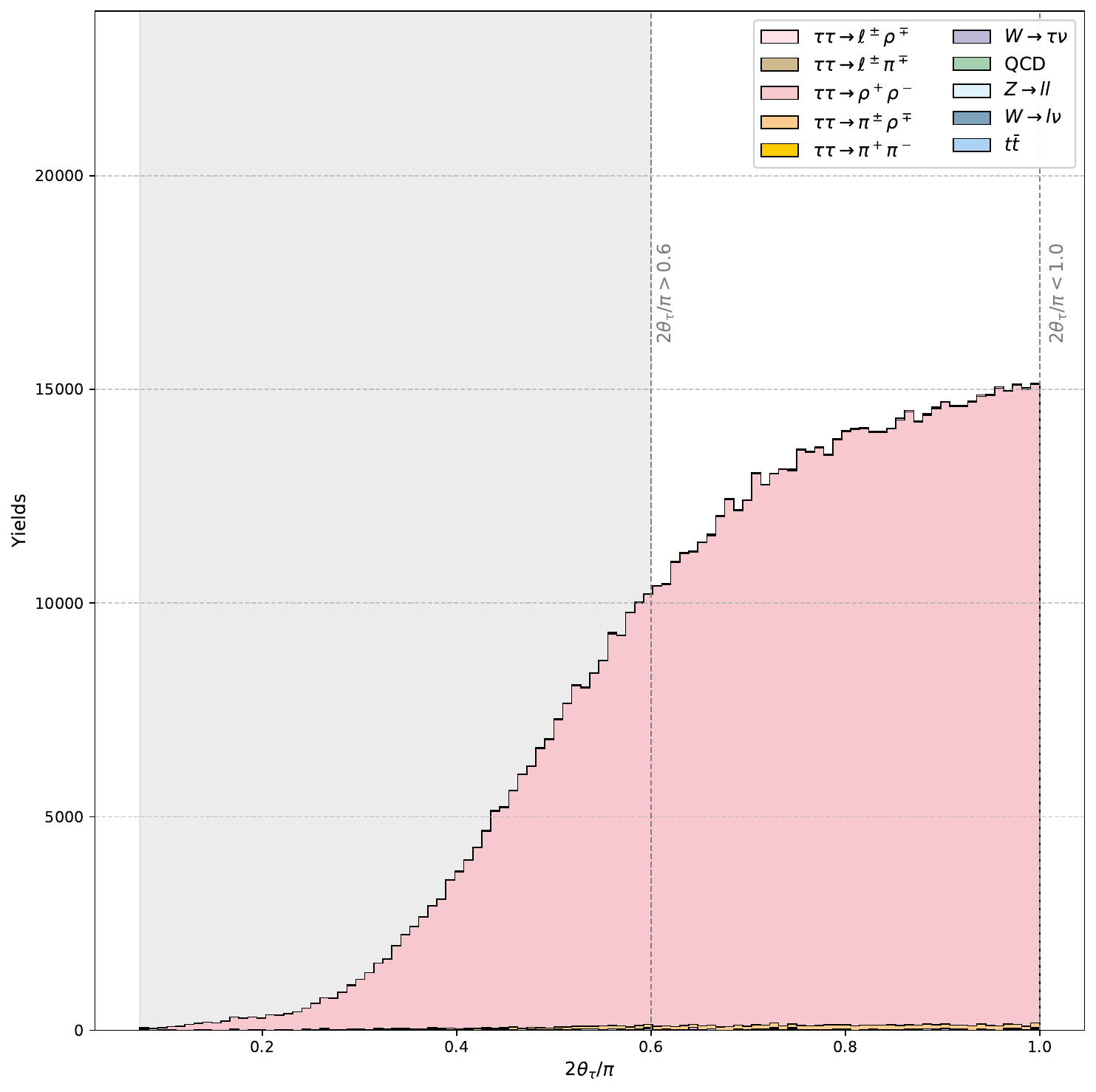}
        \end{subfigure}
    \end{subfigure}
 \caption{Distribution of $m_{\tau\tau}$ (left) and of $\theta_\tau$ (right) for the $\rho\rho$ subchannel.}
    \label{fig:extra-rhorho}
\end{figure}

\begin{figure}[h!]
    \centering
    \begin{subfigure}{\textwidth}
        \centering
        \begin{subfigure}{0.48\textwidth}
            \centering
            \includegraphics[width=\linewidth]{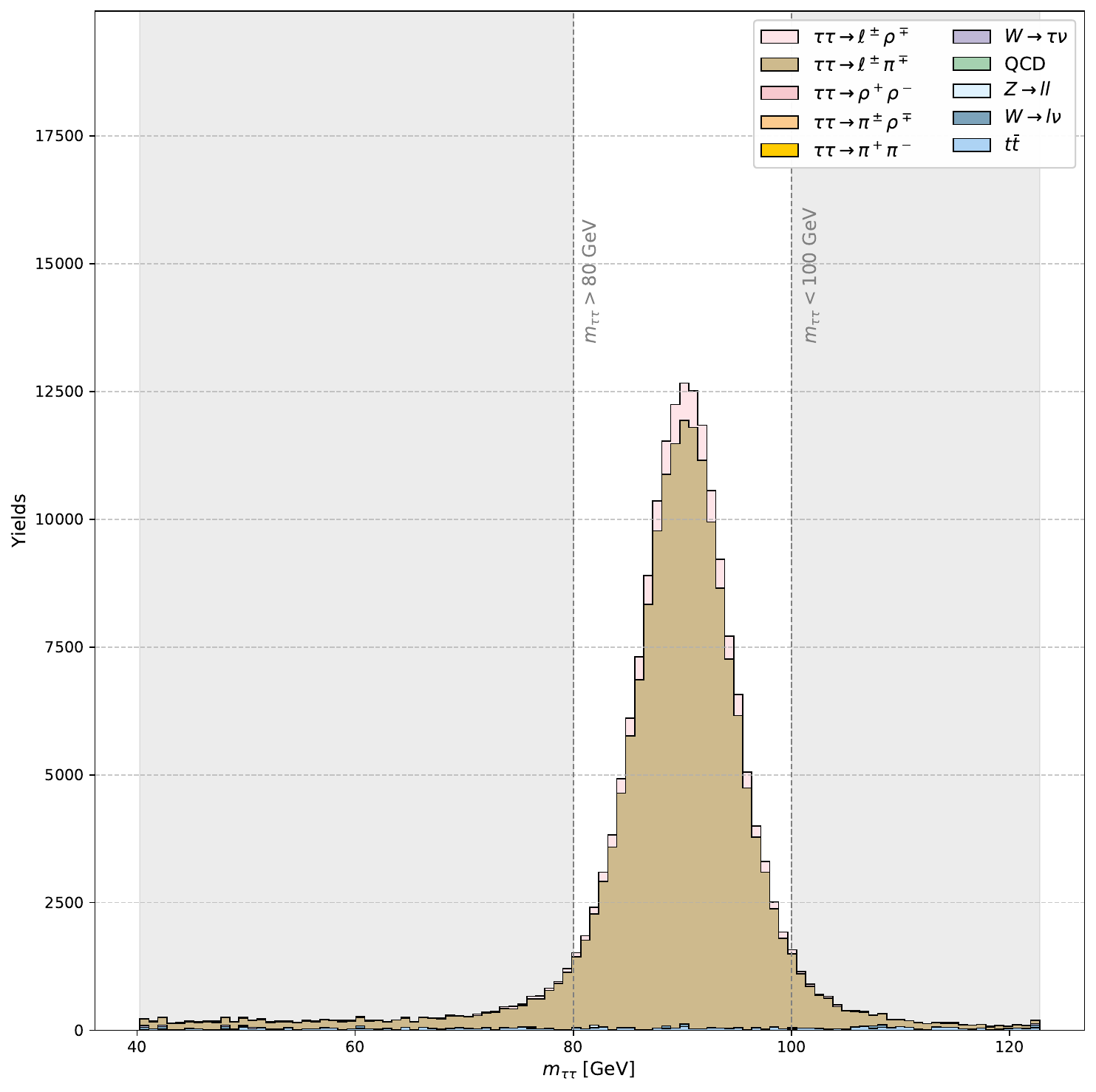}
        \end{subfigure}
        \hfill
        \begin{subfigure}{0.48\textwidth}
            \centering
            \includegraphics[width=\linewidth]{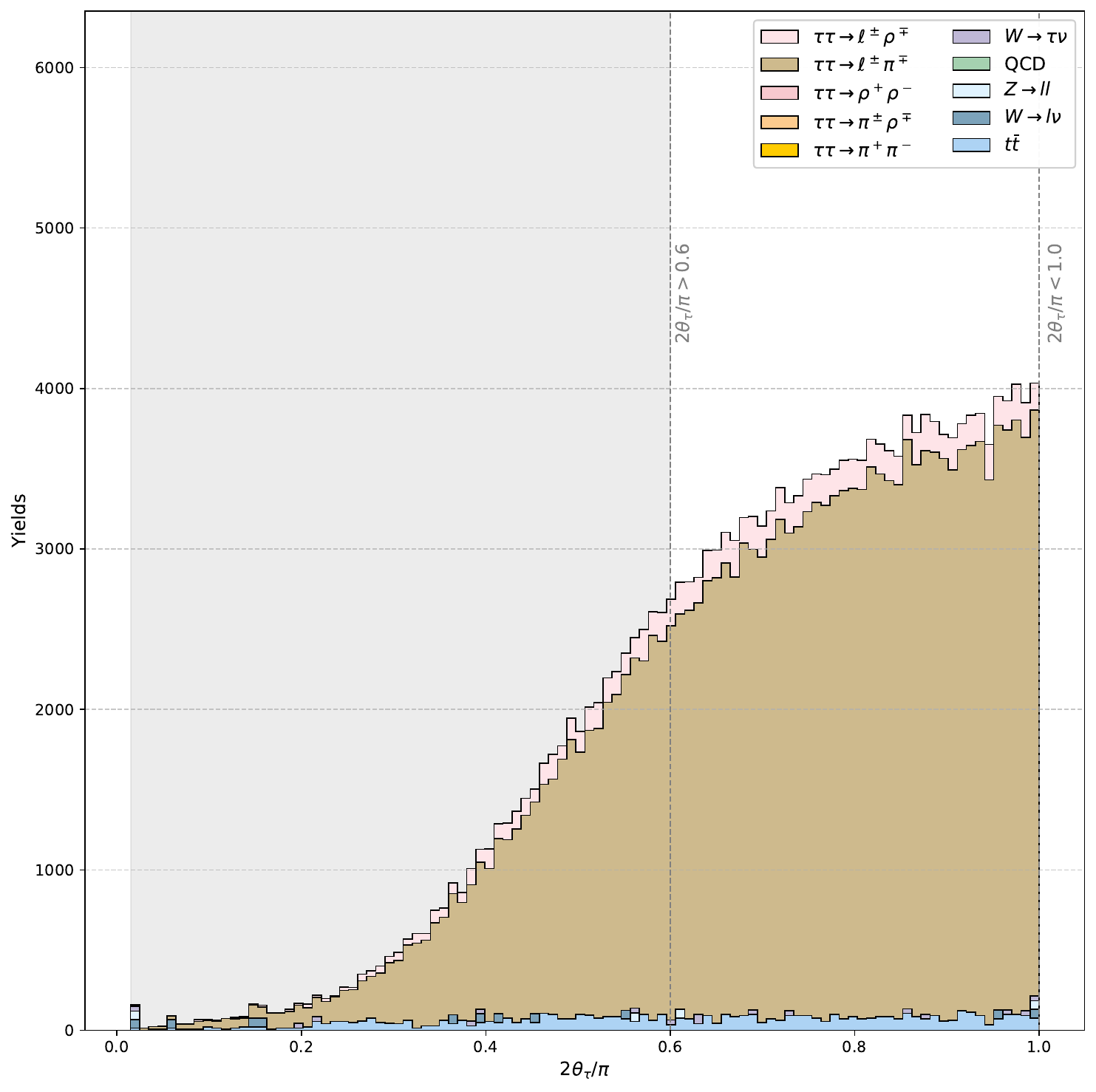}
        \end{subfigure}
    \end{subfigure}
    \caption{Distribution of $m_{\tau\tau}$ (left) and of $\theta_\tau$ (right) for the $e\pi$ subchannel.}
    \label{fig:extra-epi}
\end{figure}

\begin{figure}[h!]
    \centering
    \begin{subfigure}{\textwidth}
        \centering
        \begin{subfigure}{0.48\textwidth}
            \centering
            \includegraphics[width=\linewidth]{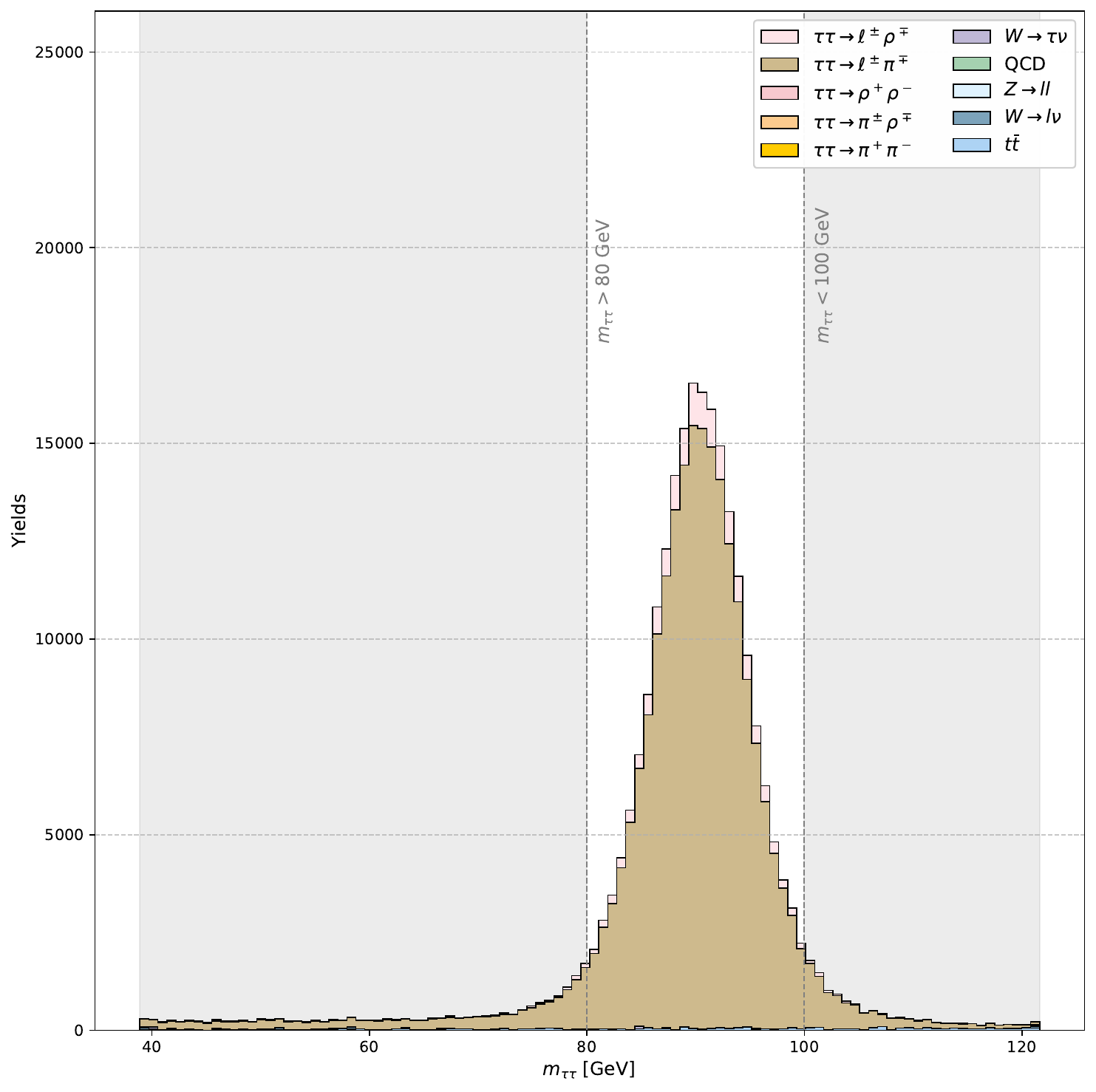}
        \end{subfigure}
        \hfill
        \begin{subfigure}{0.48\textwidth}
            \centering
            \includegraphics[width=\linewidth]{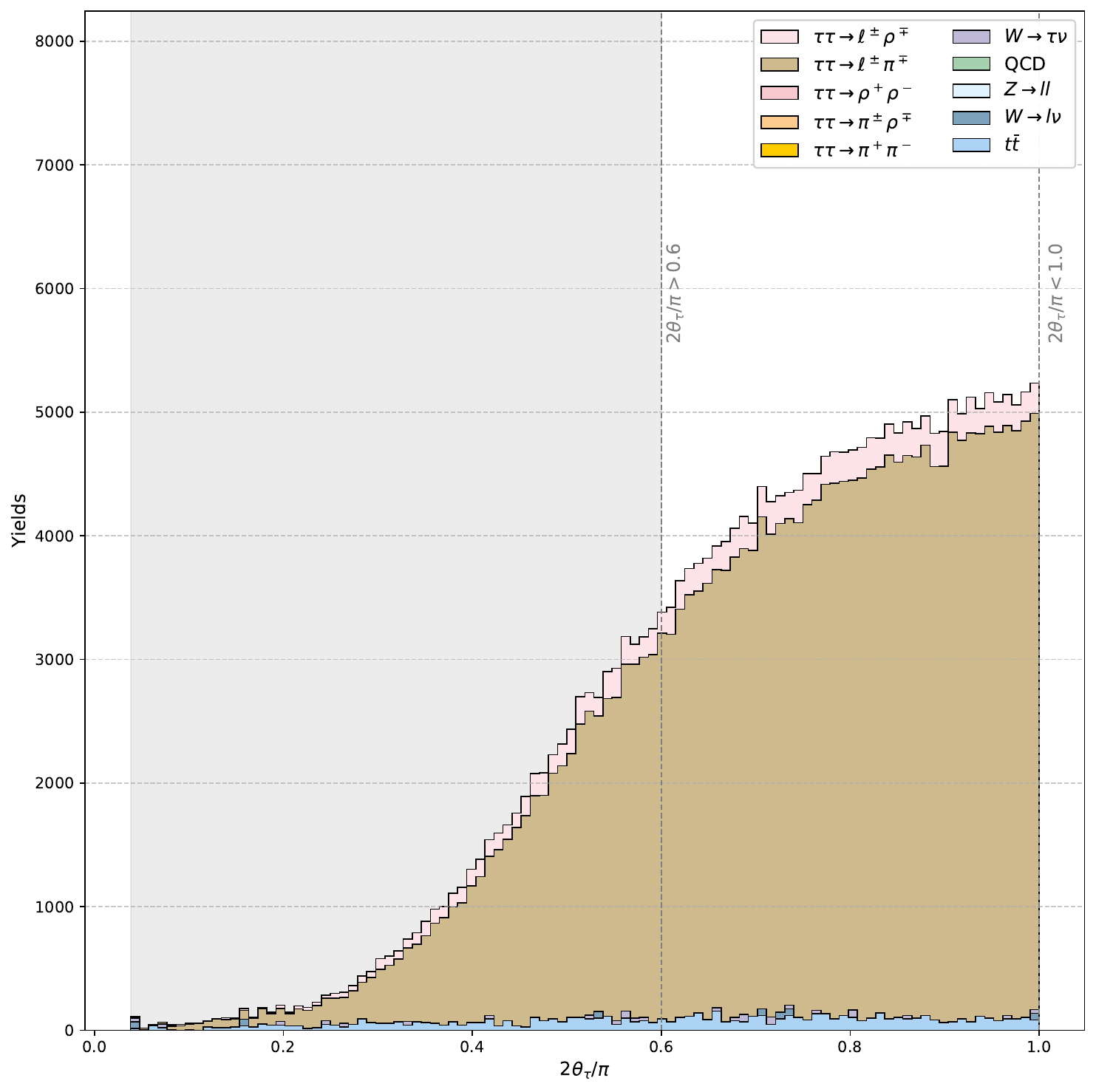}
        \end{subfigure}
    \end{subfigure}
    \caption{Distribution of $m_{\tau\tau}$ (left) and of $\theta_\tau$ (right) for the $\mu\pi$ subchannel.}
    \label{fig:extra-mupi}
\end{figure}

\begin{figure}[h!]
    \centering
    \begin{subfigure}{\textwidth}
        \centering
        \begin{subfigure}{0.48\textwidth}
            \centering
            \includegraphics[width=\linewidth]{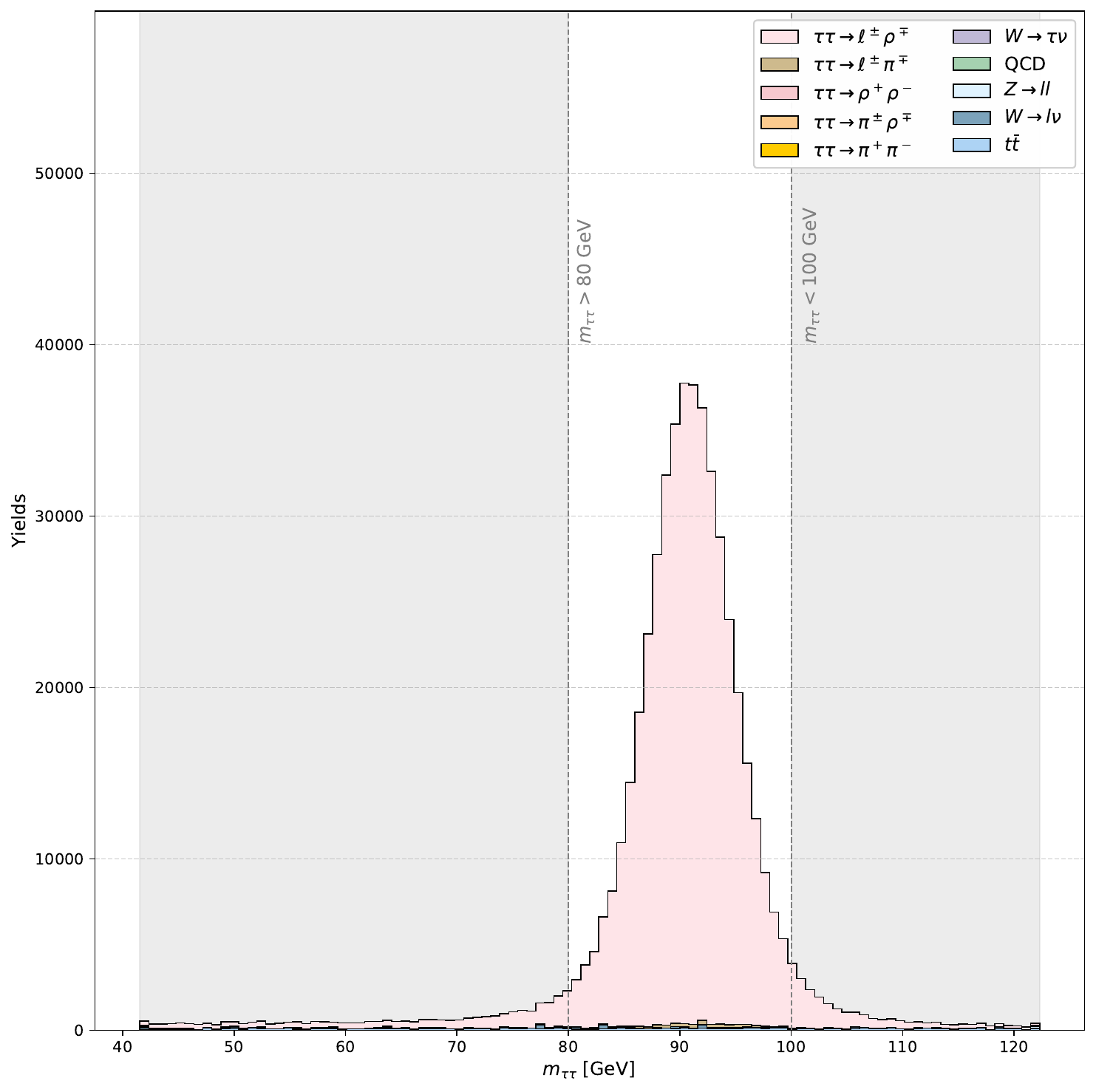}
        \end{subfigure}
        \hfill
        \begin{subfigure}{0.48\textwidth}
            \centering
            \includegraphics[width=\linewidth]{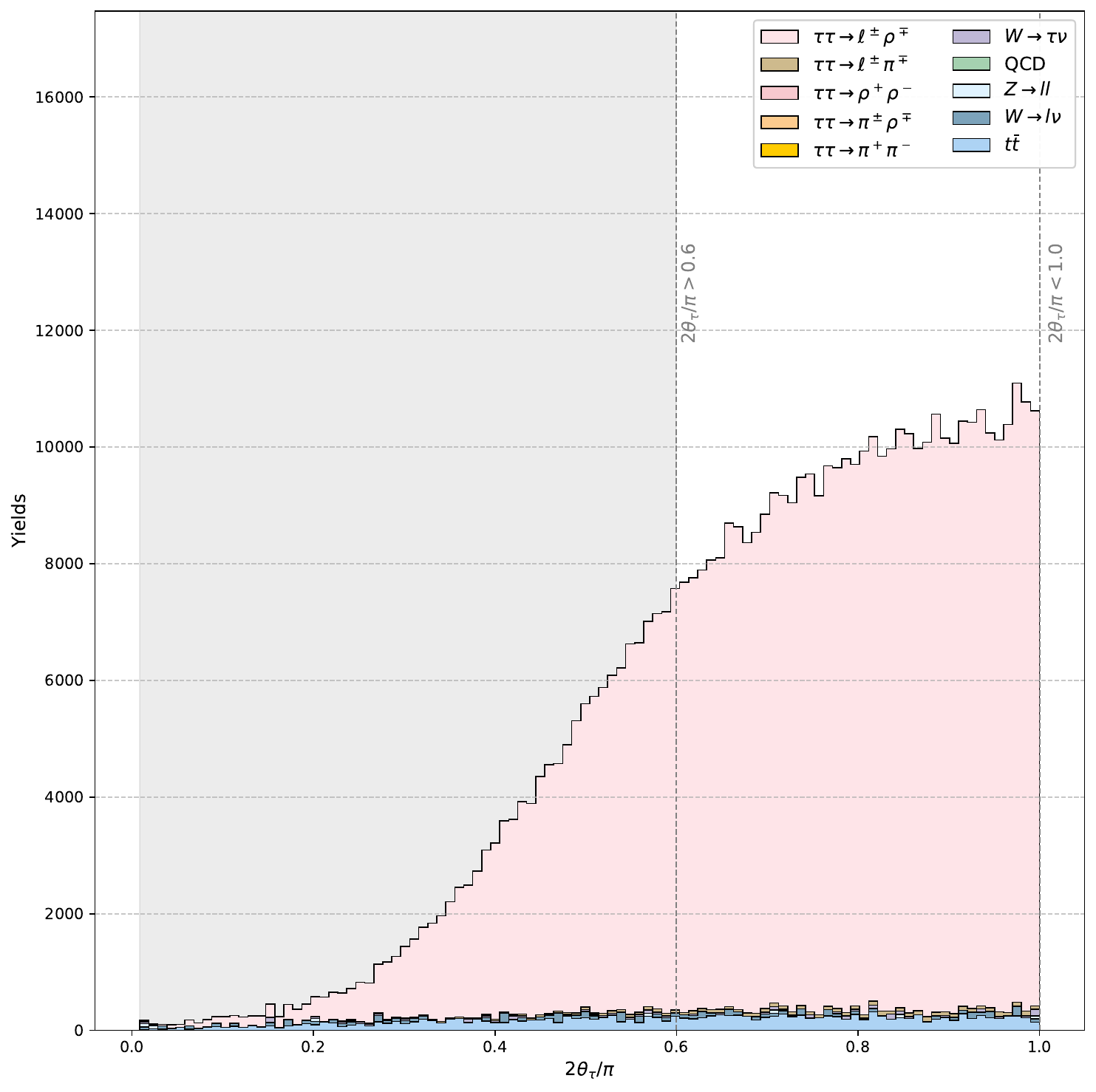}
        \end{subfigure}
    \end{subfigure}
 \caption{Distribution of $m_{\tau\tau}$ (left) and of $\theta_\tau$ (right) for the $e\rho$ subchannel.}
    \label{fig:extra-erho}
\end{figure}

\begin{figure}[h!]
    \centering
    \begin{subfigure}{\textwidth}
        \centering
        \begin{subfigure}{0.48\textwidth}
            \centering
            \includegraphics[width=\linewidth]{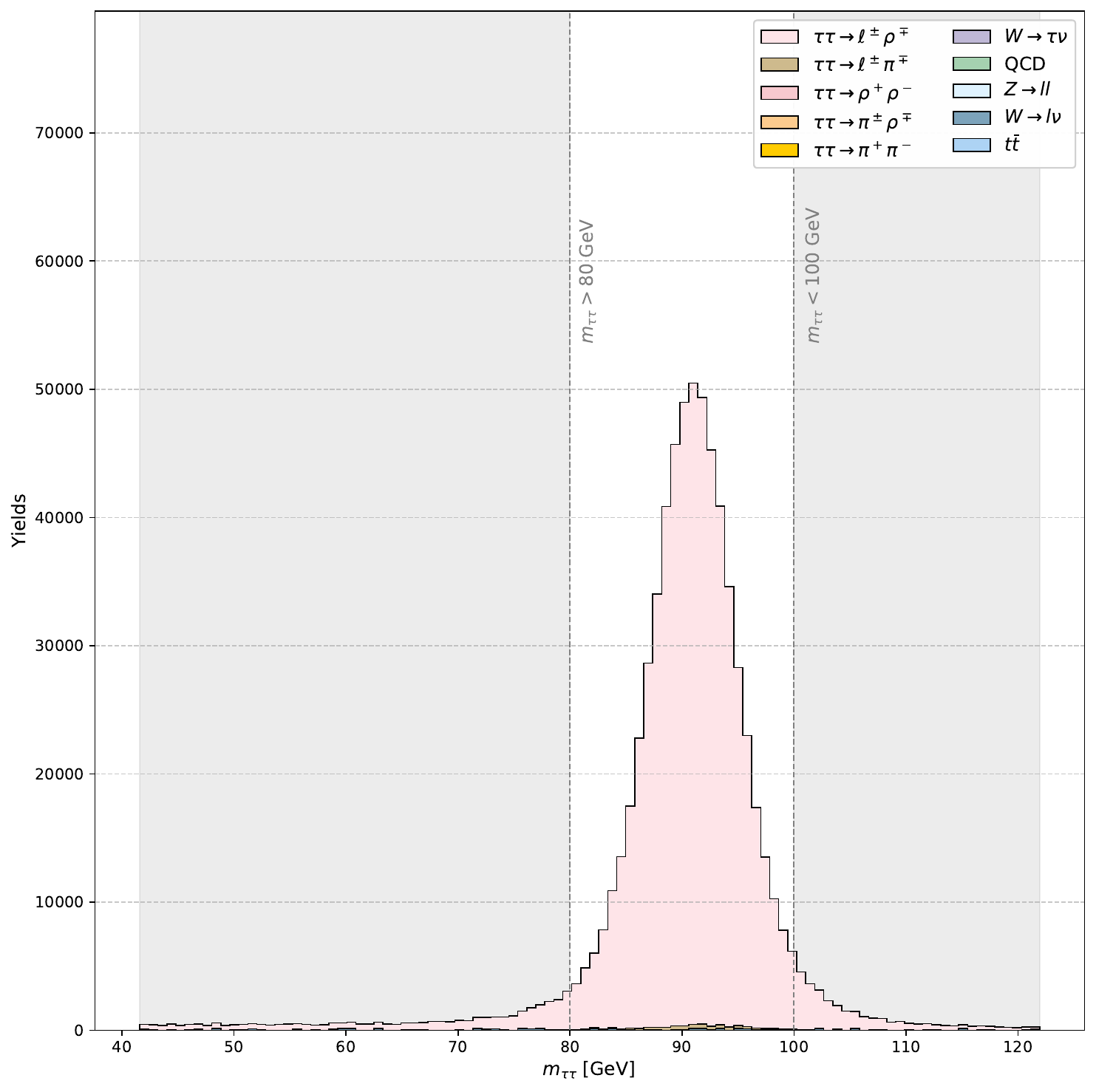}
        \end{subfigure}
        \hfill
        \begin{subfigure}{0.48\textwidth}
            \centering
            \includegraphics[width=\linewidth]{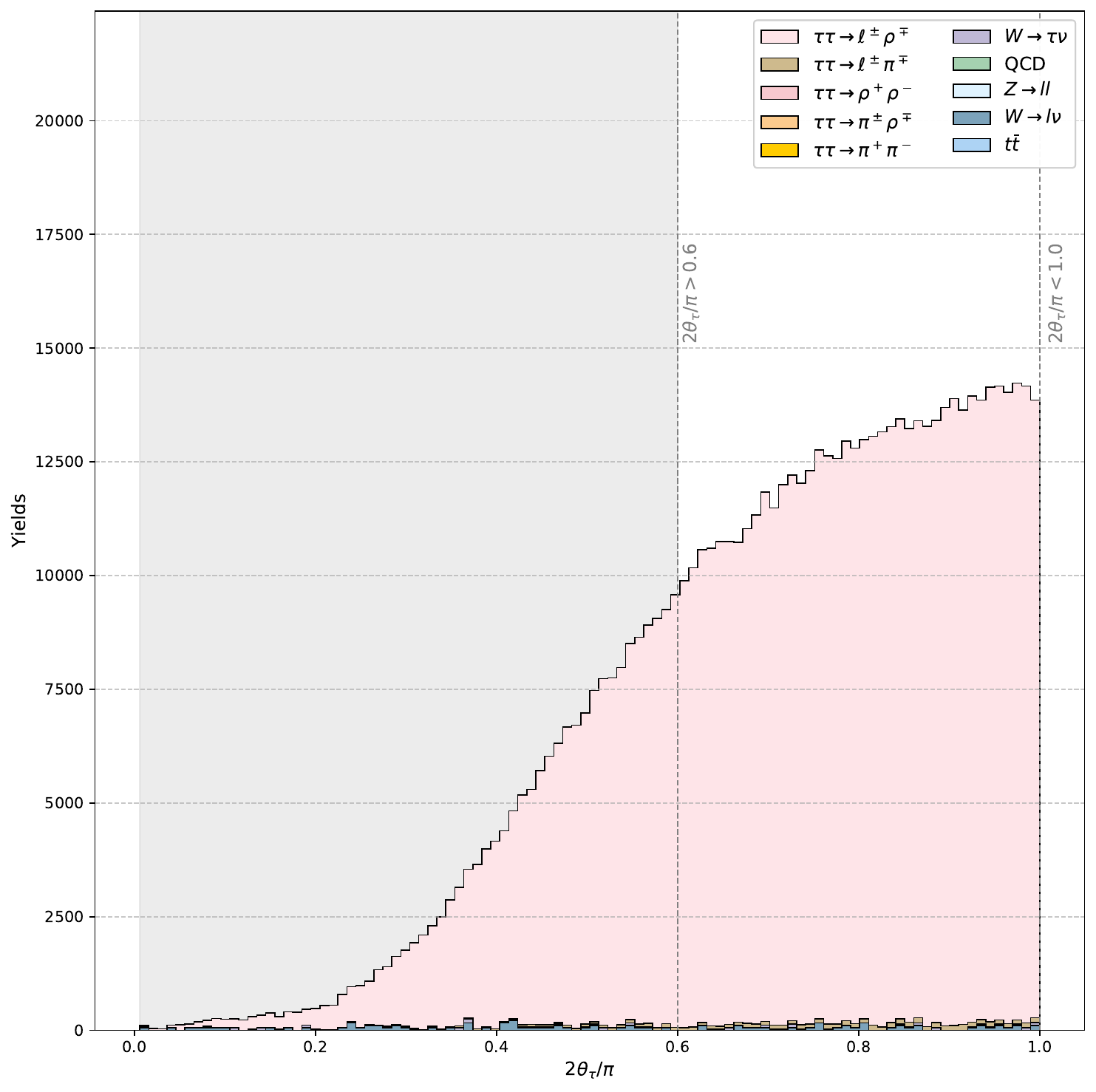}
        \end{subfigure}
    \end{subfigure}
 \caption{Distribution of $m_{\tau\tau}$ (left) and of $\theta_\tau$ (right) for the $\mu\rho$ subchannel.}
    \label{fig:extra-murho}
\end{figure}

We show distributions of $m_{\tau\tau}$ and of $\theta_\tau$ for the subchannel $\pi\pi$ in Fig.~\ref{fig:extra-pipi}, for the subchannel $\pi\rho$ in Fig.~\ref{fig:extra-pirho}, for the subchannel $\rho\rho$ in Fig.~\ref{fig:extra-rhorho}, for the subchannel $e\pi$ in Fig.~\ref{fig:extra-epi}, for the subchannel $\mu\pi$ in Fig.~\ref{fig:extra-mupi},  for the subchannel $e\rho$ in Fig.~\ref{fig:extra-erho}, and for the subchannel $\mu\rho$ in Fig.~\ref{fig:extra-murho}.

These distributions are obtained after applying the full neutrino reconstruction and all selections preceding the final signal region requirements defined in Eq.~(\ref{eq:sr_selection}). The vertical lines in the plots indicate the $m_{\tau\tau}$ and $\theta_{\tau\tau}$ cuts used in the final signal region selection.

\subsection{Neutrino Reconstruction}

\begin{figure}[h]
    \centering
    \begin{subfigure}[b]{0.32\textwidth}
        \centering
        \includegraphics[width=\textwidth]
        {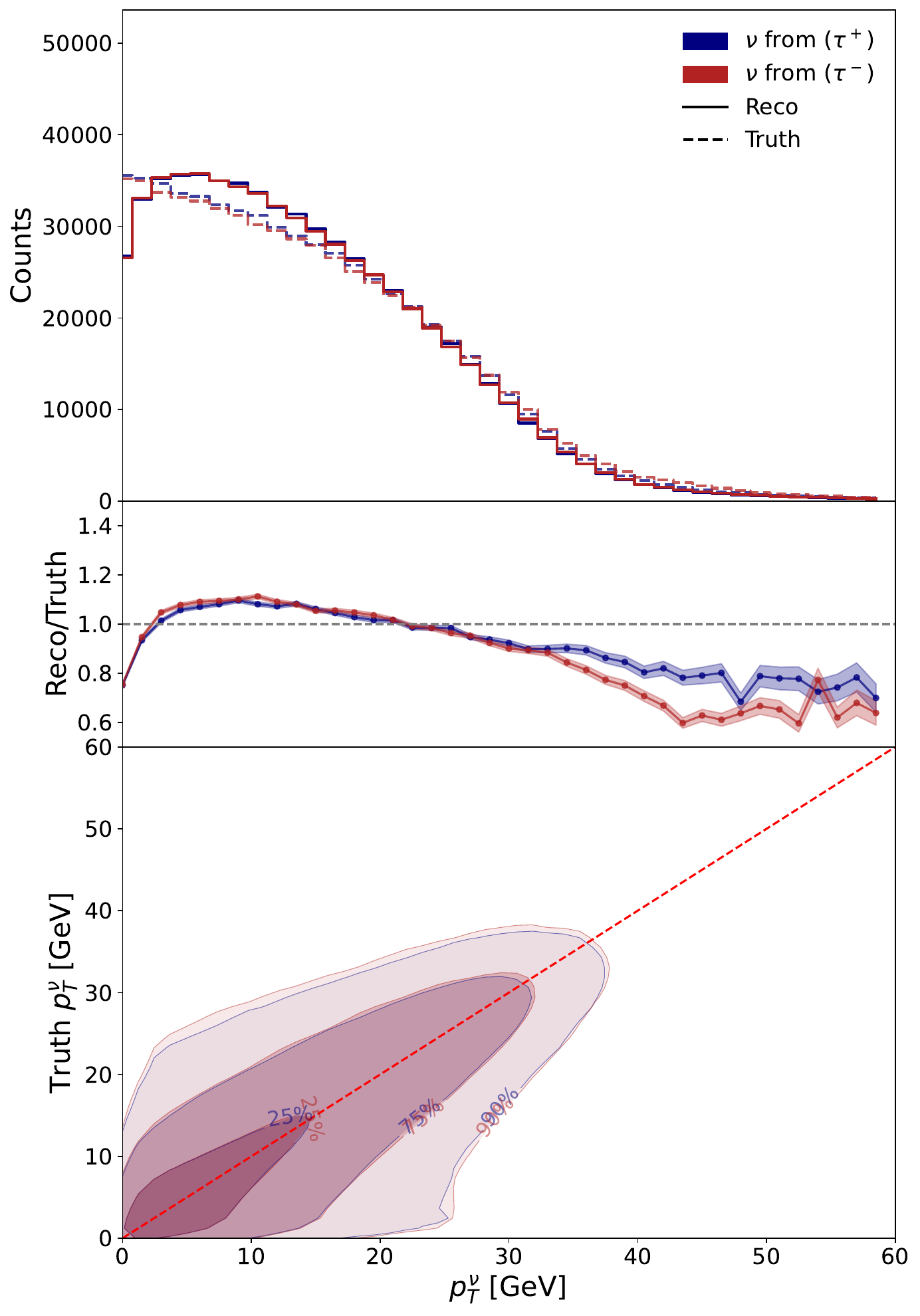}
    \end{subfigure}
    \hfill
    \begin{subfigure}[b]{0.32\textwidth}
        \centering
        \includegraphics[width=\textwidth]
        {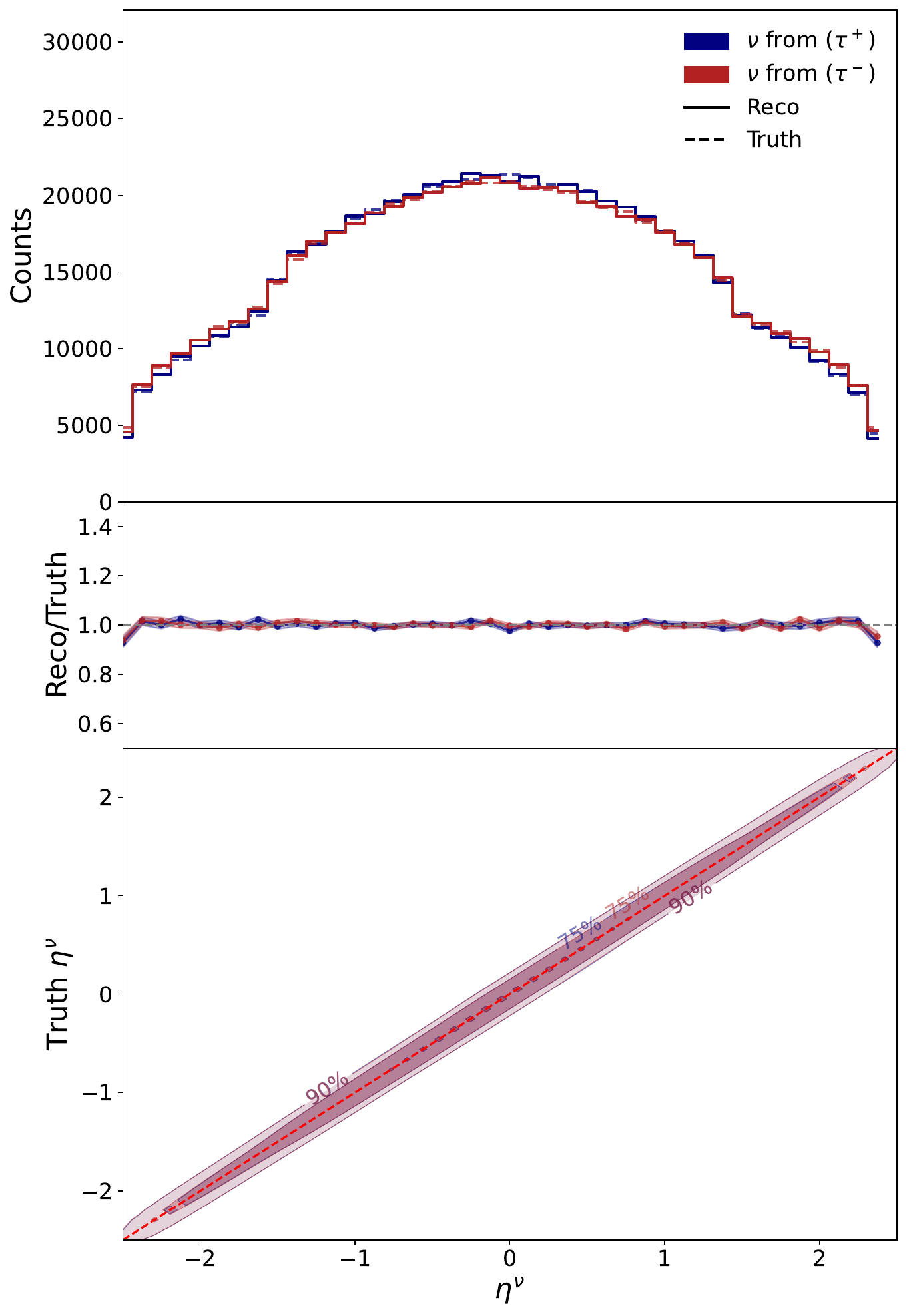}
    \end{subfigure}
    \hfill
    \begin{subfigure}[b]{0.32\textwidth}
        \centering
        \includegraphics[width=\textwidth]
        {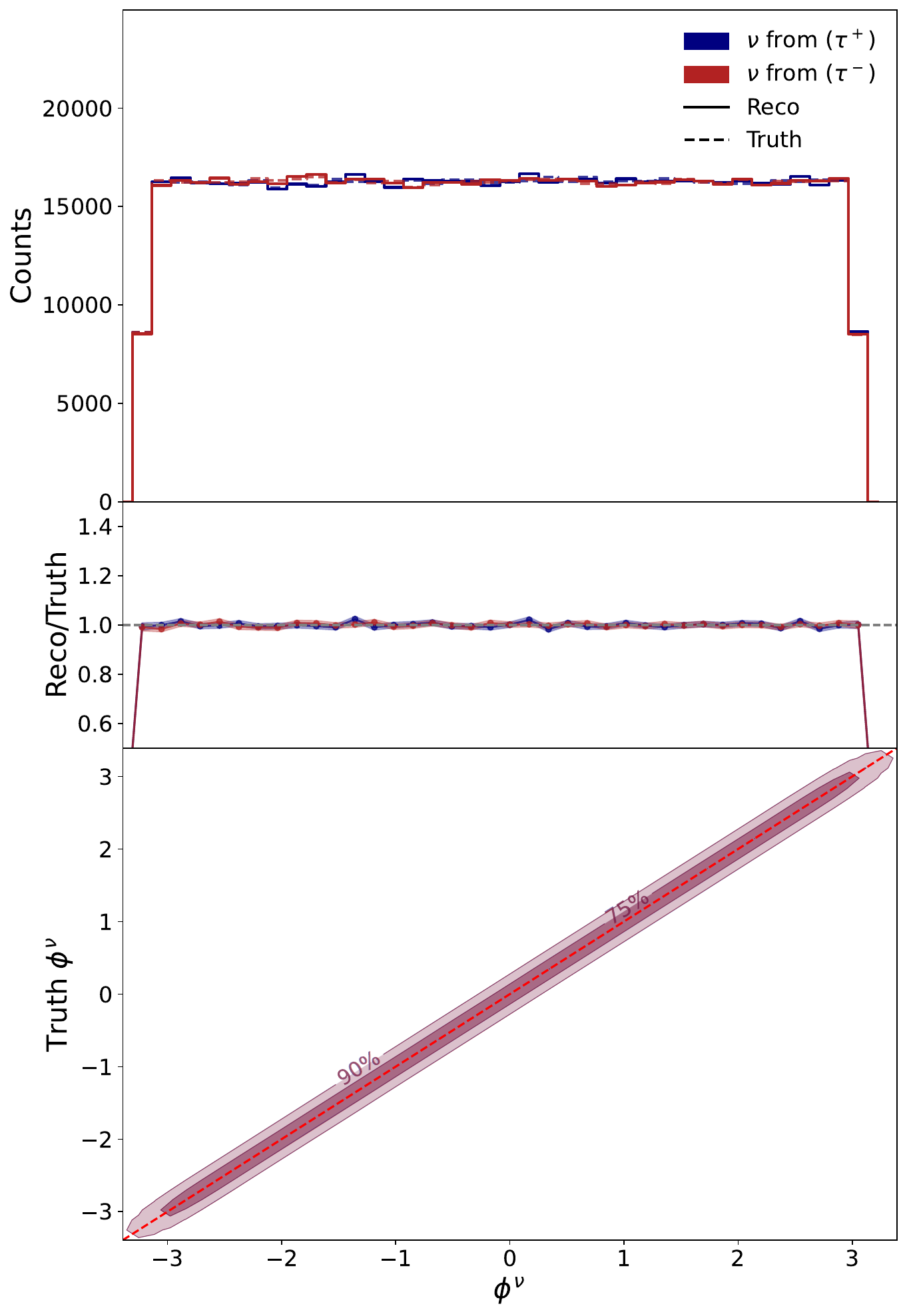}
    \end{subfigure}
    \caption{
     Comparison of reconstructed and truth-level neutrino kinematic distributions for the $\rho\rho$ channel. The three columns correspond to transverse momentum ($p_T$), pseudorapidity ($\eta$), and azimuthal angle ($\phi$). Each plot includes the reconstructed (solid) and truth-level (dashed) distributions for neutrinos originating from $\tau^+$ (blue) and $\tau^-$ (red). The middle panels show the ratio of reconstructed to truth values, while the bottom panels contain two-dimensional correlation contour plots illustrating the linear dependency between reconstructed and truth values. The contour lines represent density levels, with numerical labels indicating the percentage of total data enclosed within each contour. 
    }
    \label{fig:neutrino_kinematics_rho_rho}
\end{figure}

\begin{figure}[h]
    \centering
    \begin{subfigure}[b]{0.32\textwidth}
        \centering
        \includegraphics[width=\textwidth]
        {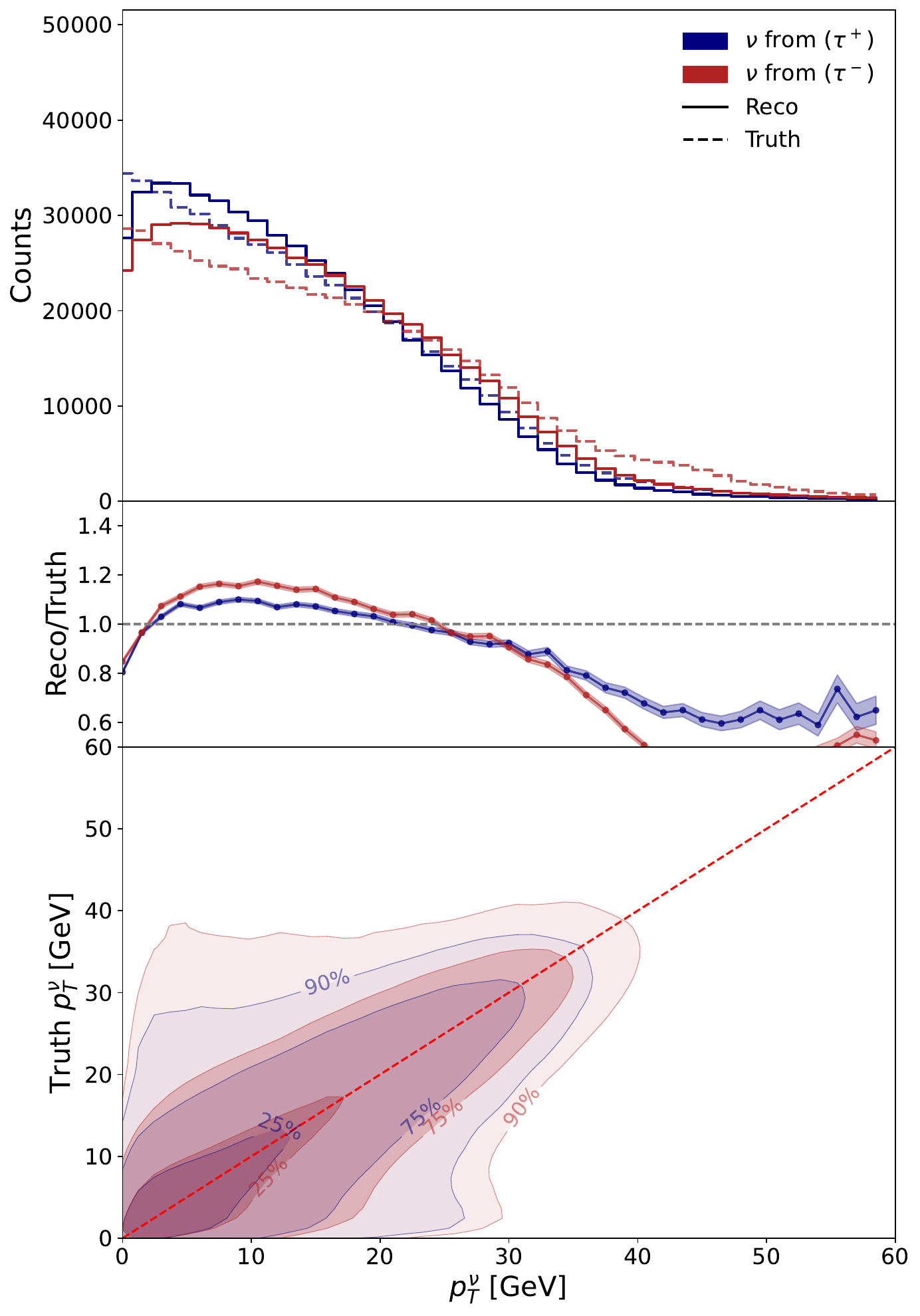}
    \end{subfigure}
    \hfill
    \begin{subfigure}[b]{0.32\textwidth}
        \centering
        \includegraphics[width=\textwidth]
        {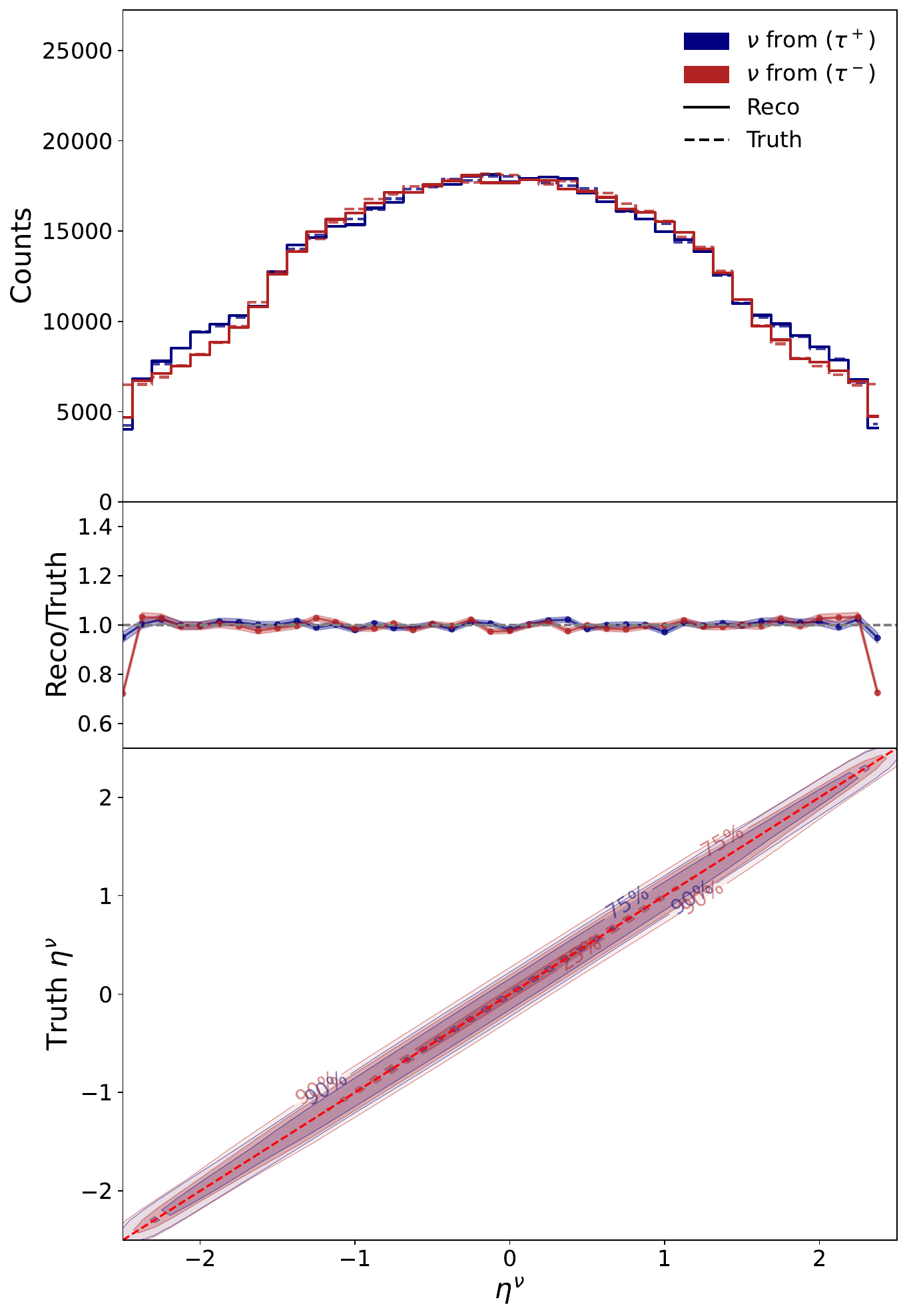}
    \end{subfigure}
    \hfill
    \begin{subfigure}[b]{0.32\textwidth}
        \centering
        \includegraphics[width=\textwidth]
        {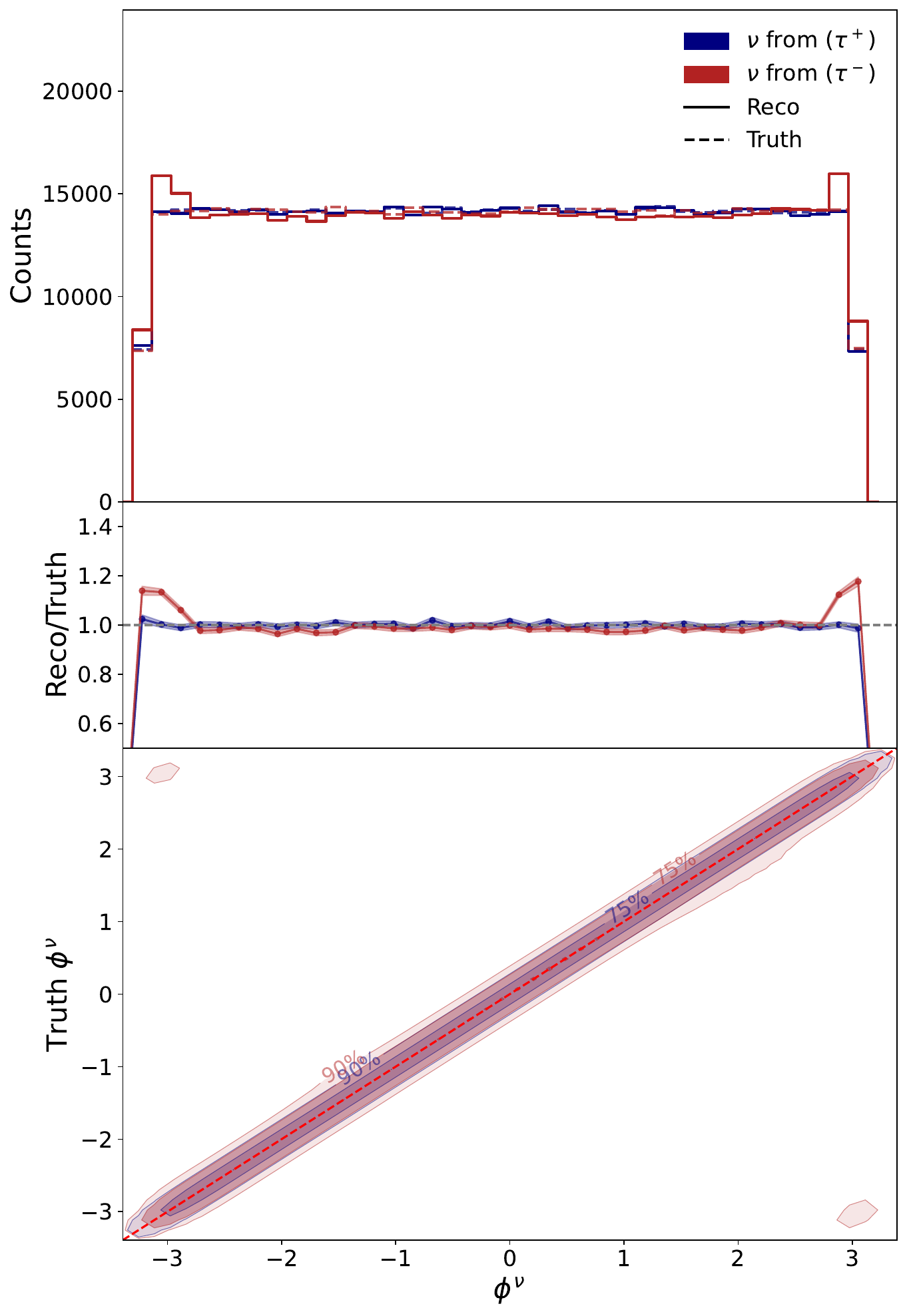}
    \end{subfigure}
    \caption{
     Comparison of reconstructed and truth-level neutrino kinematic distributions for the $\pi\rho$ channel. The three columns correspond to transverse momentum ($p_T$), pseudorapidity ($\eta$), and azimuthal angle ($\phi$). Each plot includes the reconstructed (solid) and truth-level (dashed) distributions for neutrinos originating from $\tau^+$ (blue) and $\tau^-$ (red). The middle panels show the ratio of reconstructed to truth values, while the bottom panels contain two-dimensional correlation contour plots illustrating the linear dependency between reconstructed and truth values. The contour lines represent density levels, with numerical labels indicating the percentage of total data enclosed within each contour. 
    }
    \label{fig:neutrino_kinematics_pi_rho}
\end{figure}

\begin{figure}[h]
    \centering
    \begin{subfigure}[b]{0.32\textwidth}
        \centering
        \includegraphics[width=\textwidth]
        {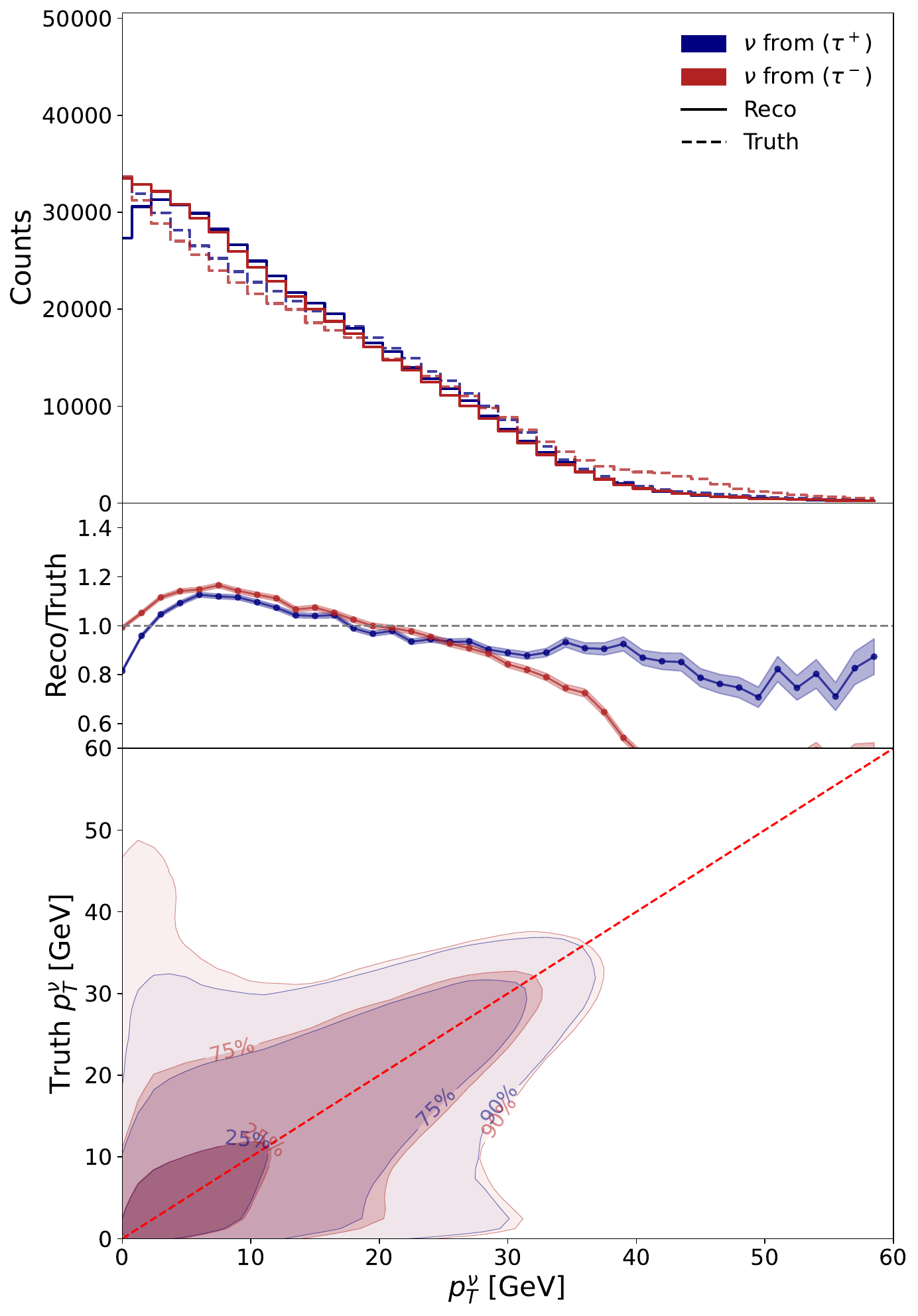}
    \end{subfigure}
    \hfill
    \begin{subfigure}[b]{0.32\textwidth}
        \centering
        \includegraphics[width=\textwidth]
        {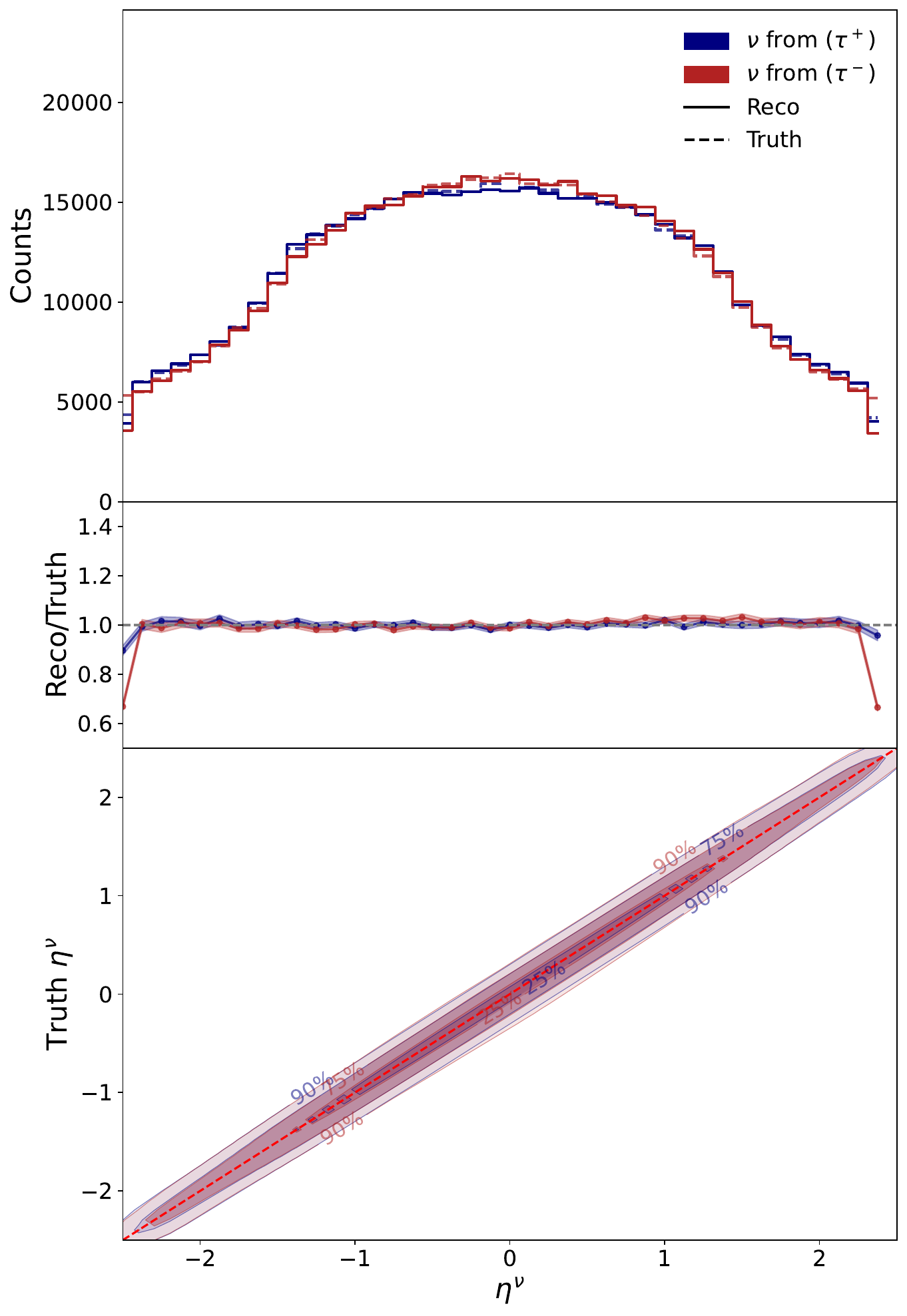}
    \end{subfigure}
    \hfill
    \begin{subfigure}[b]{0.32\textwidth}
        \centering
        \includegraphics[width=\textwidth]
        {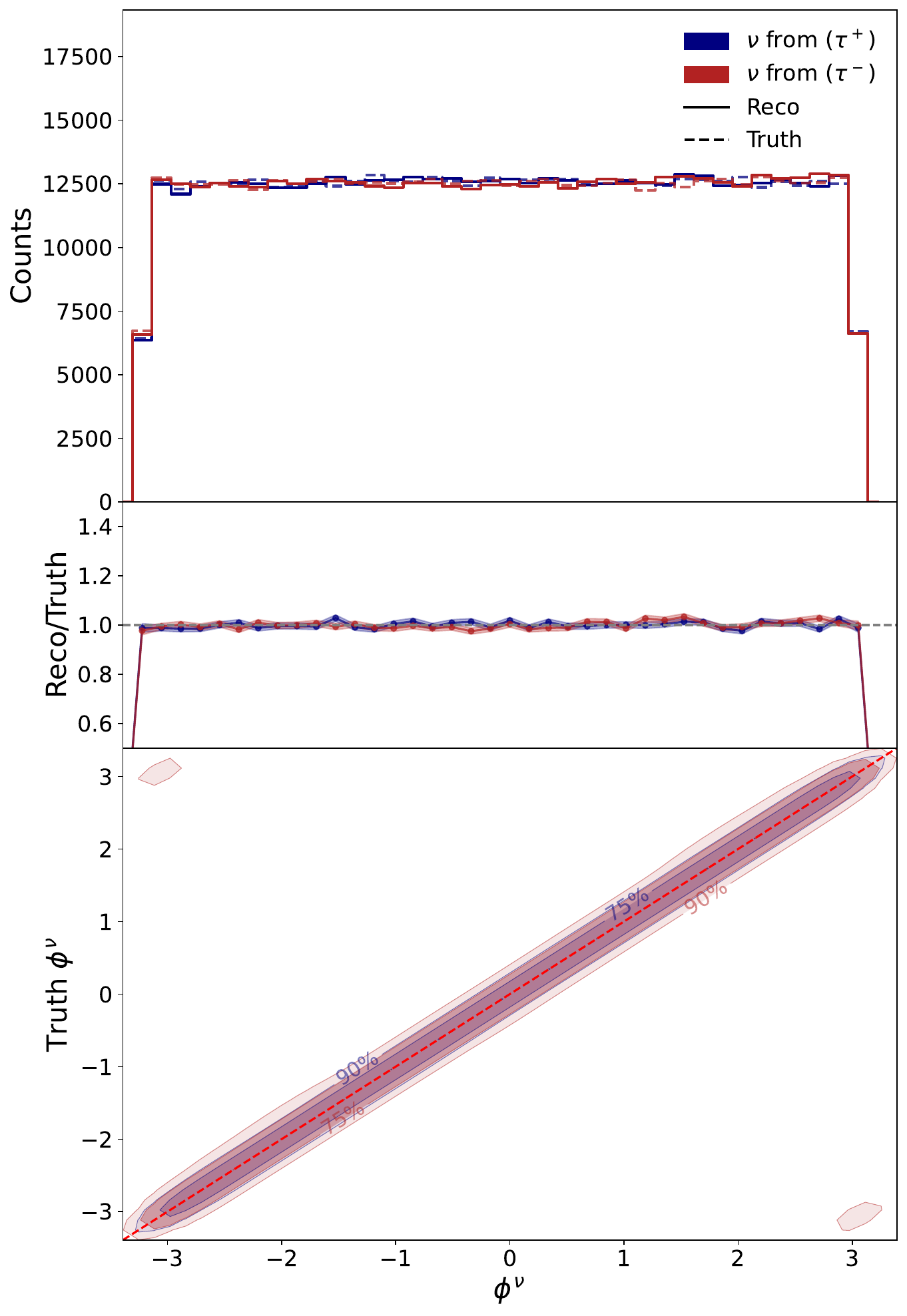}
    \end{subfigure}
    \caption{
     Comparison of reconstructed and truth-level neutrino kinematic distributions for the $\pi\pi$ channel. The three columns correspond to transverse momentum ($p_T$), pseudorapidity ($\eta$), and azimuthal angle ($\phi$). Each plot includes the reconstructed (solid) and truth-level (dashed) distributions for neutrinos originating from $\tau^+$ (blue) and $\tau^-$ (red). The middle panels show the ratio of reconstructed to truth values, while the bottom panels contain two-dimensional correlation contour plots illustrating the linear dependency between reconstructed and truth values. The contour lines represent density levels, with numerical labels indicating the percentage of total data enclosed within each contour. 
    }
    \label{fig:neutrino_kinematics_pi_pi}
\end{figure}

\begin{figure}[h]
    \centering
    \begin{subfigure}[b]{0.32\textwidth}
        \centering
        \includegraphics[width=\textwidth]
        {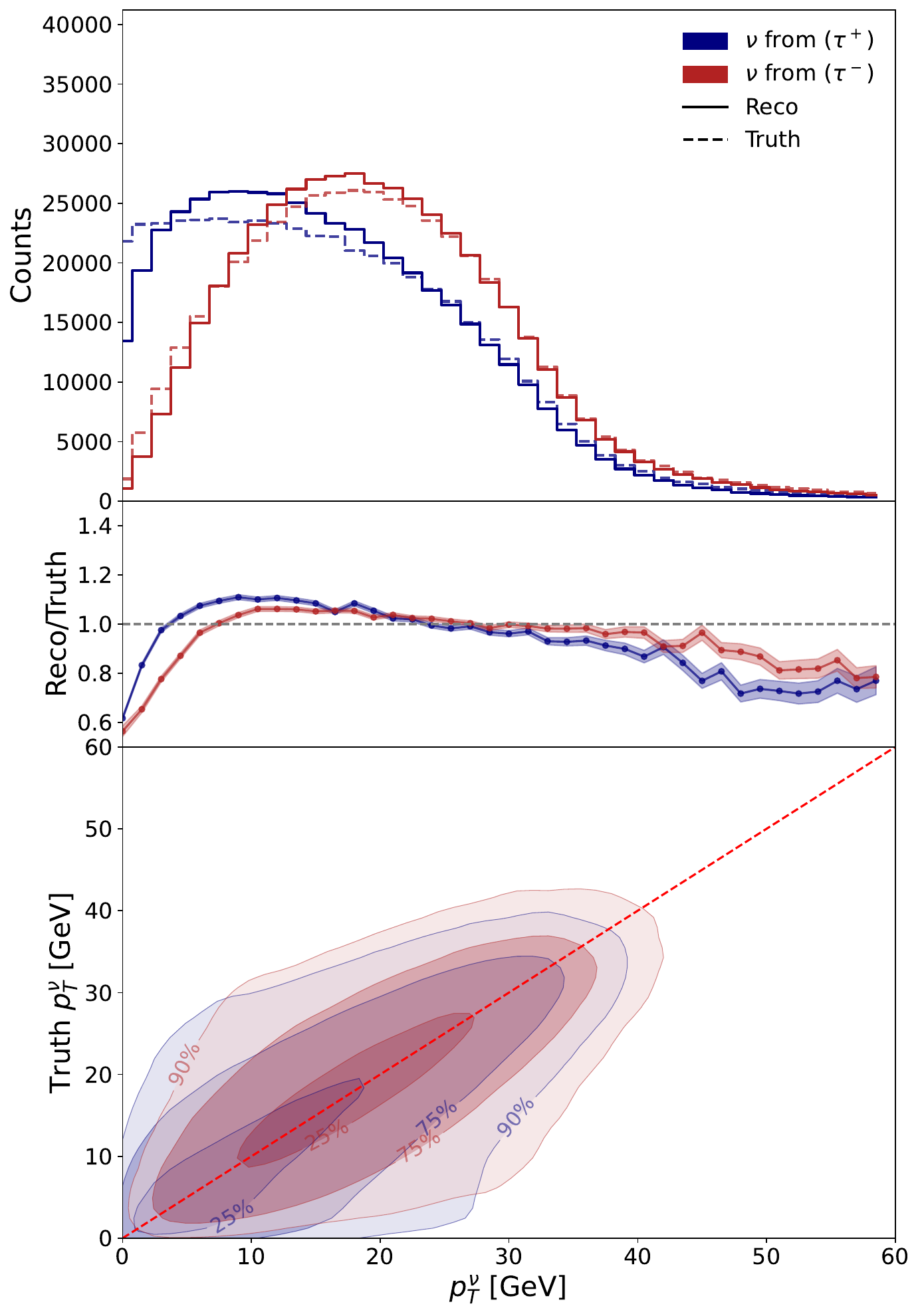}
    \end{subfigure}
    \hfill
    \begin{subfigure}[b]{0.32\textwidth}
        \centering
        \includegraphics[width=\textwidth]
        {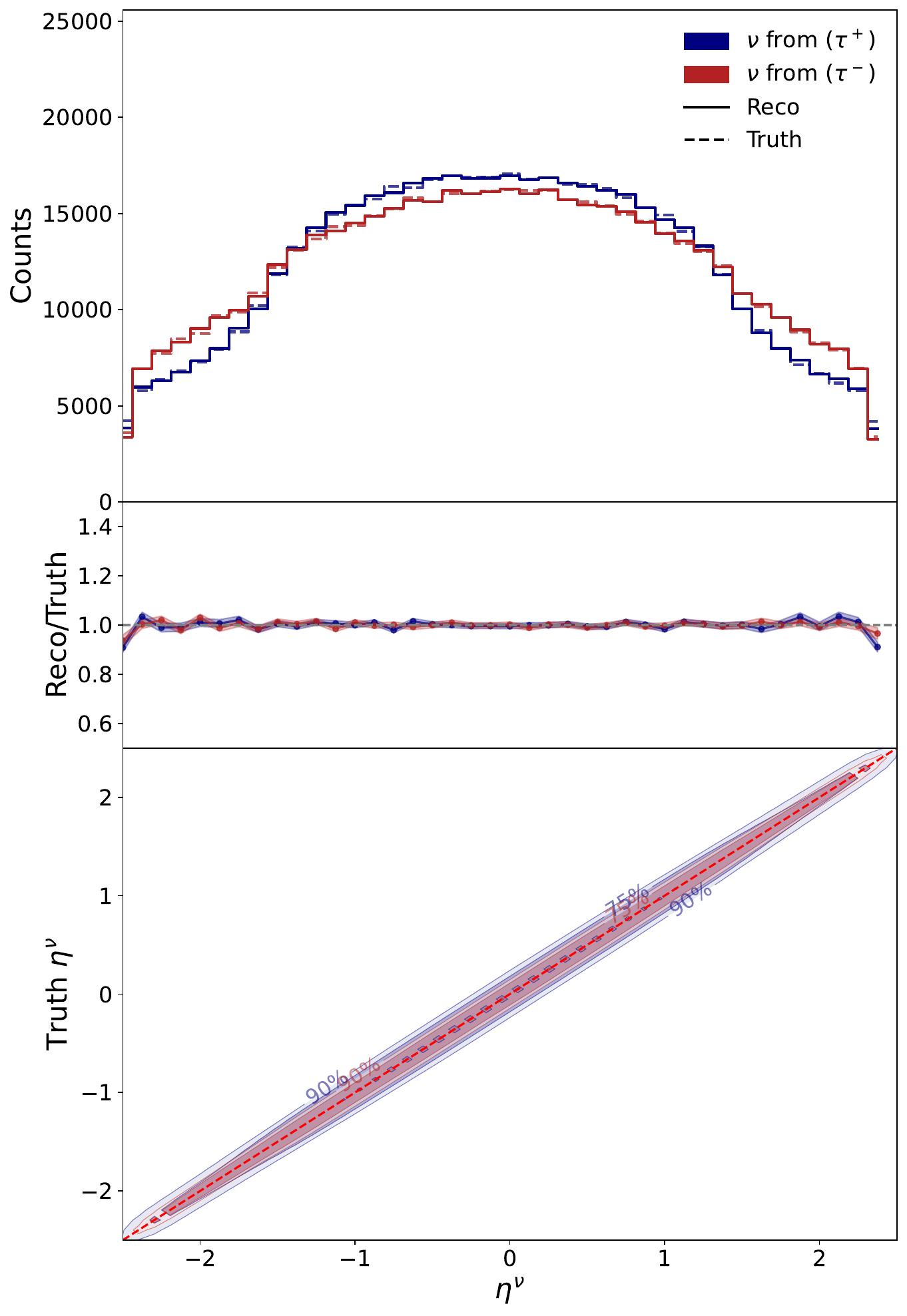}
    \end{subfigure}
    \hfill
    \begin{subfigure}[b]{0.32\textwidth}
        \centering
        \includegraphics[width=\textwidth]
        {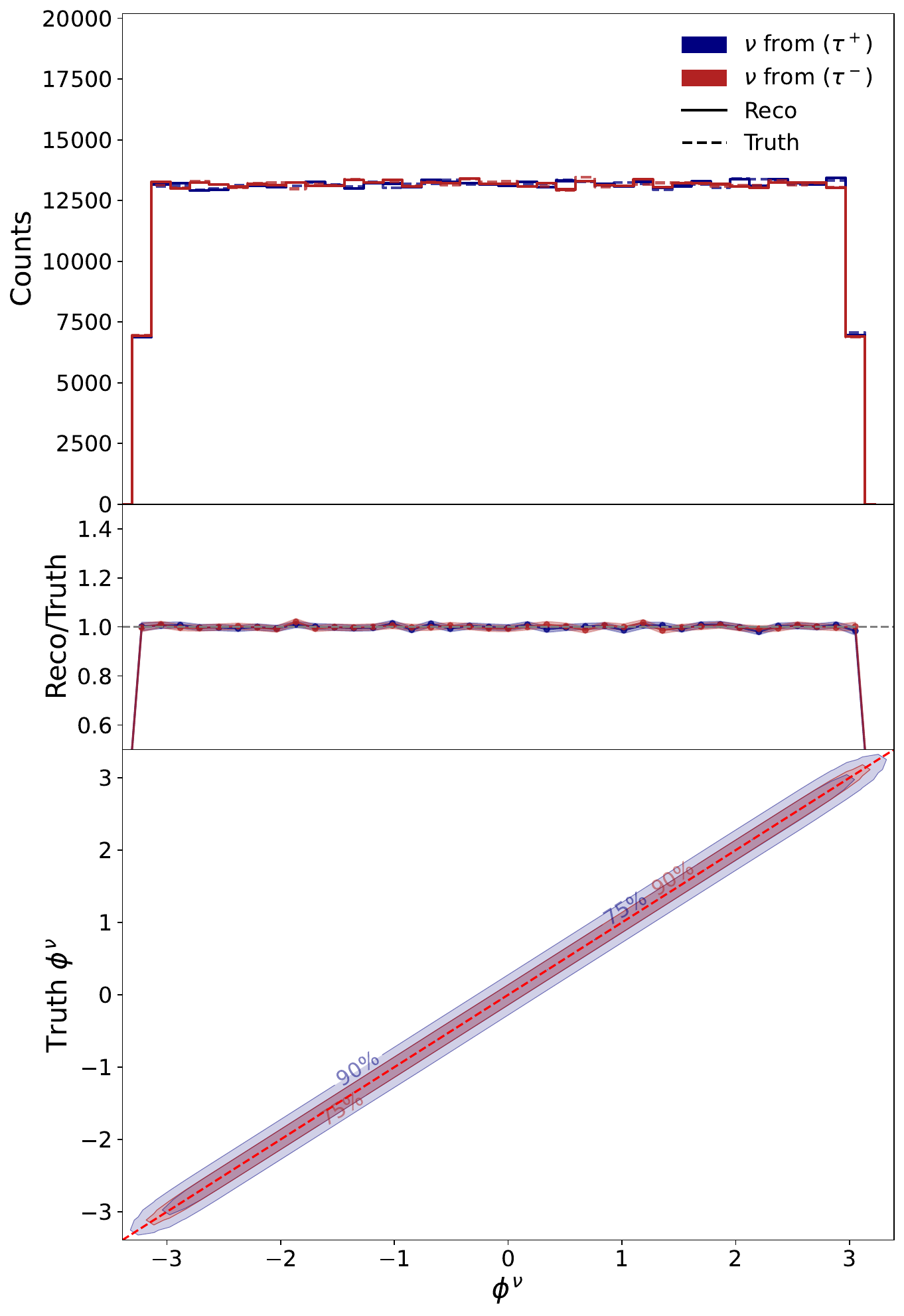}
    \end{subfigure}
    \caption{
     Comparison of reconstructed and truth-level neutrino kinematic distributions for the $\ell\pi$ channel. The three columns correspond to transverse momentum ($p_T$), pseudorapidity ($\eta$), and azimuthal angle ($\phi$). Each plot includes the reconstructed (solid) and truth-level (dashed) distributions for neutrinos originating from $\tau^+$ (blue) and $\tau^-$ (red). The middle panels show the ratio of reconstructed to truth values, while the bottom panels contain two-dimensional correlation contour plots illustrating the linear dependency between reconstructed and truth values. The contour lines represent density levels, with numerical labels indicating the percentage of total data enclosed within each contour. 
    }
    \label{fig:neutrino_kinematics_lep_pi}
\end{figure}

Figures~\ref{fig:neutrino_kinematics_rho_rho}, \ref{fig:neutrino_kinematics_pi_rho}, \ref{fig:neutrino_kinematics_pi_pi}, and~\ref{fig:neutrino_kinematics_lep_pi} show the comparison between reconstructed and truth-level neutrino kinematic distributions for the $\rho\rho$, $\pi\rho$, $\pi\pi$, and $\ell\pi$ channels, respectively. Each figure includes three columns representing the transverse momentum ($p_T$), pseudorapidity ($\eta$), and azimuthal angle ($\phi$) of the neutrinos. The top panels display the reconstructed (solid) and truth-level (dashed) distributions for neutrinos originating from $\tau^+$ (blue) and $\tau^-$ (red). The middle panels show the ratio of reconstructed to truth values, while the bottom panels present two-dimensional correlation contour plots between reconstructed and truth-level quantities. The contour lines reflect density levels, with annotations indicating the percentage of total events enclosed. These plots provide a detailed assessment of the reconstruction performance across different hadronic and semileptonic $\tau^+\tau^-$ final states.

\subsection{Density Matrix}

Table~\ref{tab:full_spin_density_result_decay_method} presents the extracted values of the polarization terms ($B^\pm_i$) and spin correlation terms ($C_{ij}$) obtained from the decay method in the signal region. The table includes the central values along with their respective uncertainties.

In the kinematic method, spin correlations are extracted without relying on a template fit. Instead, the method directly utilizes kinematic observables, specifically the invariant mass of the tau pair ($m_{\tau\tau}$) and the tau decay angle in the center-of-mass frame ($\theta_\tau$), to compute the relevant correlation coefficients. All event samples, including signal and background contributions, are merged into a single dataset before evaluating the spin correlation matrices ($C_{ij}$) and ($B^{\pm}_i$).
Systematic uncertainties are estimated by varying the extraction function $f(m_{\tau\tau}, \theta_\tau)$ within its uncertainty range and taking half of the absolute difference between the upper and lower variations. This variation is propagated through to the final spin density matrix. Statistical uncertainties are determined based on Poisson statistics.

The final results of the kinematic method, including both systematic and statistical uncertainties, are presented in Table~\ref{tab:full_spin_density_result_kinematic_method}.

\begin{landscape}
\begin{table}
    \centering
    \caption{
    Summary of the fitted values for the polarization terms ($B^\pm_i$) and spin correlation terms ($C_{ij}$) in the signal region using decay method. Each entry represents the central fitted value with its corresponding upper and lower uncertainties. Uncertainties are multiplied by 100 for better readability due to their small magnitudes.
    }
    \label{tab:full_spin_density_result_decay_method}
    \renewcommand{\arraystretch}{1.05} 
    \resizebox{0.95\linewidth}{!}{
        \begin{tabular}{c | c c c c c c c c | c c c c c c c c }
            \toprule

& \multicolumn{7}{c}{Decay Method (SR Only)} & \multicolumn{7}{c}{Decay Method (SR \& Trigger)} \\ \midrule
 & $\pi \pi$ & $\pi \rho$ & $\rho \rho$ & $e \pi$ & $\mu \pi$ & $e \rho$ & $\mu \rho$ & $\pi \pi$ & $\pi \rho$ & $\rho \rho$ & $e \pi$ & $\mu \pi$ & $e \rho$ & $\mu \rho$ \\ \midrule
$B^{-}_{k}$ & $\mathbf{-0.21}^{+\mathbf{ 1.75}}_{-\mathbf{ 1.71}}$ & $\mathbf{-0.21}^{+\mathbf{ 1.05}}_{-\mathbf{ 1.05}}$ & $\mathbf{-0.21}^{+\mathbf{ 1.73}}_{-\mathbf{ 1.75}}$ & $\mathbf{-0.21}^{+\mathbf{ 1.61}}_{-\mathbf{ 1.61}}$ & $\mathbf{-0.21}^{+\mathbf{ 1.57}}_{-\mathbf{ 1.58}}$ & $\mathbf{-0.20}^{+\mathbf{ 0.96}}_{-\mathbf{ 0.98}}$ & $\mathbf{-0.21}^{+\mathbf{ 0.94}}_{-\mathbf{ 0.96}}$ & $\mathbf{-0.21}^{+\mathbf{ 8.88}}_{-\mathbf{ 8.45}}$ & $\mathbf{-0.21}^{+\mathbf{ 3.65}}_{-\mathbf{ 3.58}}$ & $\mathbf{-0.21}^{+\mathbf{ 4.27}}_{-\mathbf{ 4.19}}$ & $\mathbf{-0.21}^{+\mathbf{ 1.83}}_{-\mathbf{ 1.83}}$ & $\mathbf{-0.21}^{+\mathbf{ 1.42}}_{-\mathbf{ 1.42}}$ & $\mathbf{-0.20}^{+\mathbf{ 1.33}}_{-\mathbf{ 1.36}}$ & $\mathbf{-0.21}^{+\mathbf{ 0.89}}_{-\mathbf{ 0.91}}$ \\
$B^{-}_{n}$ & $\phantom{-} 0.00^{+ 0.00}_{- 0.00}$ & $-0.00^{+ 0.02}_{- 0.02}$ & $-0.01^{+ 0.06}_{- 0.06}$ & $-0.00^{+ 0.14}_{- 0.13}$ & $\phantom{-} 0.00^{+ 0.31}_{- 0.31}$ & $-0.00^{+ 0.06}_{- 0.06}$ & $\phantom{-} 0.01^{+ 0.02}_{- 0.02}$ & $\phantom{-} 0.00^{+ 0.01}_{- 0.01}$ & $-0.00^{+ 0.04}_{- 0.04}$ & $-0.01^{+ 0.06}_{- 0.06}$ & $-0.00^{+ 0.11}_{- 0.11}$ & $\phantom{-} 0.00^{+ 0.30}_{- 0.30}$ & $-0.00^{+ 0.06}_{- 0.06}$ & $\phantom{-} 0.01^{+ 0.02}_{- 0.02}$ \\
$B^{-}_{r}$ & $\phantom{-} 0.00^{+ 0.01}_{- 0.01}$ & $\phantom{-} 0.00^{+ 0.01}_{- 0.01}$ & $\phantom{-} 0.00^{+ 0.02}_{- 0.02}$ & $-0.00^{+ 0.05}_{- 0.05}$ & $-0.01^{+ 0.07}_{- 0.07}$ & $\phantom{-} 0.00^{+ 0.09}_{- 0.09}$ & $-0.01^{+ 0.02}_{- 0.02}$ & $\phantom{-} 0.00^{+ 0.04}_{- 0.03}$ & $\phantom{-} 0.00^{+ 0.02}_{- 0.02}$ & $\phantom{-} 0.00^{+ 0.03}_{- 0.03}$ & $-0.00^{+ 0.04}_{- 0.04}$ & $-0.01^{+ 0.07}_{- 0.07}$ & $\phantom{-} 0.00^{+ 0.10}_{- 0.10}$ & $-0.01^{+ 0.02}_{- 0.02}$ \\
$B^{+}_{k}$ & $\mathbf{-0.21}^{+\mathbf{ 1.78}}_{-\mathbf{ 1.74}}$ & $\mathbf{-0.21}^{+\mathbf{ 3.27}}_{-\mathbf{ 3.22}}$ & $\mathbf{-0.21}^{+\mathbf{ 1.94}}_{-\mathbf{ 1.96}}$ & $\mathbf{-0.21}^{+\mathbf{ 1.34}}_{-\mathbf{ 1.34}}$ & $\mathbf{-0.21}^{+\mathbf{ 1.10}}_{-\mathbf{ 1.10}}$ & $\mathbf{-0.21}^{+\mathbf{ 2.64}}_{-\mathbf{ 2.67}}$ & $\mathbf{-0.21}^{+\mathbf{ 2.24}}_{-\mathbf{ 2.26}}$ & $\mathbf{-0.21}^{+\mathbf{ 15.40}}_{-\mathbf{ 10.44}}$ & $\mathbf{-0.21}^{+\mathbf{ 9.57}}_{-\mathbf{ 8.36}}$ & $\mathbf{-0.21}^{+\mathbf{ 5.79}}_{-\mathbf{ 5.63}}$ & $\mathbf{-0.21}^{+\mathbf{ 1.82}}_{-\mathbf{ 1.83}}$ & $\mathbf{-0.21}^{+\mathbf{ 1.67}}_{-\mathbf{ 1.67}}$ & $\mathbf{-0.21}^{+\mathbf{ 2.44}}_{-\mathbf{ 2.50}}$ & $\mathbf{-0.21}^{+\mathbf{ 1.75}}_{-\mathbf{ 1.78}}$ \\
$B^{+}_{n}$ & $\phantom{-} 0.00^{+ 0.00}_{- 0.00}$ & $-0.00^{+ 0.24}_{- 0.23}$ & $\phantom{-} 0.00^{+ 0.01}_{- 0.01}$ & $\phantom{-} 0.00^{+ 0.02}_{- 0.02}$ & $\phantom{-} 0.00^{+ 0.02}_{- 0.02}$ & $\phantom{-} 0.00^{+ 0.02}_{- 0.02}$ & $\phantom{-} 0.00^{+ 0.02}_{- 0.02}$ & $\phantom{-} 0.00^{+ 0.02}_{- 0.01}$ & $-0.00^{+ 0.70}_{- 0.44}$ & $\phantom{-} 0.00^{+ 0.02}_{- 0.02}$ & $\phantom{-} 0.00^{+ 0.03}_{- 0.03}$ & $\phantom{-} 0.00^{+ 0.02}_{- 0.02}$ & $\phantom{-} 0.00^{+ 0.03}_{- 0.03}$ & $\phantom{-} 0.00^{+ 0.02}_{- 0.02}$ \\
$B^{+}_{r}$ & $\phantom{-} 0.00^{+ 0.01}_{- 0.01}$ & $-0.00^{+ 0.02}_{- 0.02}$ & $\phantom{-} 0.00^{+ 0.00}_{- 0.00}$ & $\phantom{-} 0.00^{+ 0.01}_{- 0.01}$ & $\phantom{-} 0.00^{+ 0.01}_{- 0.01}$ & $\phantom{-} 0.01^{+ 0.00}_{- 0.00}$ & $\phantom{-} 0.01^{+ 0.00}_{- 0.00}$ & $\phantom{-} 0.00^{+ 0.09}_{- 0.03}$ & $-0.00^{+ 0.04}_{- 0.03}$ & $\phantom{-} 0.00^{+ 0.00}_{- 0.00}$ & $\phantom{-} 0.00^{+ 0.01}_{- 0.01}$ & $\phantom{-} 0.00^{+ 0.01}_{- 0.01}$ & $\phantom{-} 0.01^{+ 0.00}_{- 0.00}$ & $\phantom{-} 0.01^{+ 0.00}_{- 0.00}$ \\\midrule
$C_{kk}$ & $\mathbf{\phantom{-} 1.01}^{+\mathbf{ 0.02}}_{-\mathbf{ 0.02}}$ & $\mathbf{\phantom{-} 1.03}^{+\mathbf{ 0.07}}_{-\mathbf{ 0.07}}$ & $\mathbf{\phantom{-} 1.02}^{+\mathbf{ 0.04}}_{-\mathbf{ 0.04}}$ & $\mathbf{\phantom{-} 1.02}^{+\mathbf{ 0.14}}_{-\mathbf{ 0.14}}$ & $\mathbf{\phantom{-} 1.01}^{+\mathbf{ 0.01}}_{-\mathbf{ 0.01}}$ & $\mathbf{\phantom{-} 0.99}^{+\mathbf{ 0.15}}_{-\mathbf{ 0.15}}$ & $\mathbf{\phantom{-} 1.01}^{+\mathbf{ 0.24}}_{-\mathbf{ 0.25}}$ & $\mathbf{\phantom{-} 1.01}^{+\mathbf{ 0.13}}_{-\mathbf{ 0.08}}$ & $\mathbf{\phantom{-} 1.03}^{+\mathbf{ 0.22}}_{-\mathbf{ 0.20}}$ & $\mathbf{\phantom{-} 1.02}^{+\mathbf{ 0.04}}_{-\mathbf{ 0.04}}$ & $\mathbf{\phantom{-} 1.02}^{+\mathbf{ 0.30}}_{-\mathbf{ 0.30}}$ & $\mathbf{\phantom{-} 1.01}^{+\mathbf{ 0.04}}_{-\mathbf{ 0.04}}$ & $\mathbf{\phantom{-} 0.99}^{+\mathbf{ 0.11}}_{-\mathbf{ 0.11}}$ & $\mathbf{\phantom{-} 1.01}^{+\mathbf{ 0.24}}_{-\mathbf{ 0.24}}$ \\
$C_{nn}$ & $\mathbf{\phantom{-} 0.77}^{+\mathbf{ 4.58}}_{-\mathbf{ 4.34}}$ & $\mathbf{\phantom{-} 0.79}^{+\mathbf{ 4.48}}_{-\mathbf{ 4.39}}$ & $\mathbf{\phantom{-} 0.80}^{+\mathbf{ 3.43}}_{-\mathbf{ 3.45}}$ & $\mathbf{\phantom{-} 0.78}^{+\mathbf{ 5.57}}_{-\mathbf{ 5.51}}$ & $\mathbf{\phantom{-} 0.78}^{+\mathbf{ 3.46}}_{-\mathbf{ 3.43}}$ & $\mathbf{\phantom{-} 0.80}^{+\mathbf{ 2.95}}_{-\mathbf{ 2.98}}$ & $\mathbf{\phantom{-} 0.77}^{+\mathbf{ 4.59}}_{-\mathbf{ 4.67}}$ & $\mathbf{\phantom{-} 0.77}^{+\mathbf{ 19.14}}_{-\mathbf{ 11.32}}$ & $\mathbf{\phantom{-} 0.79}^{+\mathbf{ 9.16}}_{-\mathbf{ 8.06}}$ & $\mathbf{\phantom{-} 0.80}^{+\mathbf{ 6.94}}_{-\mathbf{ 6.73}}$ & $\mathbf{\phantom{-} 0.78}^{+\mathbf{ 9.73}}_{-\mathbf{ 9.61}}$ & $\mathbf{\phantom{-} 0.78}^{+\mathbf{ 6.98}}_{-\mathbf{ 6.82}}$ & $\mathbf{\phantom{-} 0.80}^{+\mathbf{ 4.29}}_{-\mathbf{ 4.39}}$ & $\mathbf{\phantom{-} 0.77}^{+\mathbf{ 5.94}}_{-\mathbf{ 6.09}}$ \\
$C_{rr}$ & $\mathbf{-0.77}^{+\mathbf{ 3.69}}_{-\mathbf{ 3.53}}$ & $\mathbf{-0.80}^{+\mathbf{ 3.30}}_{-\mathbf{ 3.26}}$ & $\mathbf{-0.80}^{+\mathbf{ 4.58}}_{-\mathbf{ 4.62}}$ & $\mathbf{-0.77}^{+\mathbf{ 4.71}}_{-\mathbf{ 4.67}}$ & $\mathbf{-0.78}^{+\mathbf{ 4.56}}_{-\mathbf{ 4.50}}$ & $\mathbf{-0.80}^{+\mathbf{ 5.55}}_{-\mathbf{ 5.66}}$ & $\mathbf{-0.80}^{+\mathbf{ 3.23}}_{-\mathbf{ 3.25}}$ & $\mathbf{-0.77}^{+\mathbf{ 30.04}}_{-\mathbf{ 14.20}}$ & $\mathbf{-0.80}^{+\mathbf{ 8.02}}_{-\mathbf{ 7.16}}$ & $\mathbf{-0.80}^{+\mathbf{ 8.25}}_{-\mathbf{ 7.95}}$ & $\mathbf{-0.77}^{+\mathbf{ 7.58}}_{-\mathbf{ 7.49}}$ & $\mathbf{-0.78}^{+\mathbf{ 8.63}}_{-\mathbf{ 8.36}}$ & $\mathbf{-0.80}^{+\mathbf{ 11.62}}_{-\mathbf{ 12.64}}$ & $\mathbf{-0.80}^{+\mathbf{ 4.01}}_{-\mathbf{ 4.06}}$ \\
$C_{kn}$ & $\phantom{-} 0.00^{+ 0.01}_{- 0.01}$ & $\phantom{-} 0.00^{+ 0.06}_{- 0.06}$ & $-0.01^{+ 0.10}_{- 0.10}$ & $-0.00^{+ 0.02}_{- 0.02}$ & $\phantom{-} 0.01^{+ 0.09}_{- 0.09}$ & $\phantom{-} 0.02^{+ 0.30}_{- 0.31}$ & $\phantom{-} 0.03^{+ 0.34}_{- 0.35}$ & $\phantom{-} 0.00^{+ 0.16}_{- 0.13}$ & $\phantom{-} 0.00^{+ 0.29}_{- 0.25}$ & $-0.01^{+ 0.16}_{- 0.16}$ & $-0.00^{+ 0.02}_{- 0.02}$ & $\phantom{-} 0.01^{+ 0.06}_{- 0.06}$ & $\phantom{-} 0.02^{+ 0.20}_{- 0.21}$ & $\phantom{-} 0.03^{+ 0.29}_{- 0.29}$ \\
$C_{kr}$ & $-0.00^{+ 0.00}_{- 0.00}$ & $-0.00^{+ 0.06}_{- 0.06}$ & $\phantom{-} 0.01^{+ 0.12}_{- 0.12}$ & $-0.00^{+ 0.14}_{- 0.14}$ & $-0.01^{+ 0.01}_{- 0.01}$ & $\phantom{-} 0.01^{+ 0.19}_{- 0.19}$ & $-0.00^{+ 0.26}_{- 0.26}$ & $-0.00^{+ 0.06}_{- 0.04}$ & $-0.00^{+ 0.35}_{- 0.29}$ & $\phantom{-} 0.01^{+ 0.17}_{- 0.17}$ & $-0.00^{+ 0.10}_{- 0.10}$ & $-0.01^{+ 0.01}_{- 0.01}$ & $\phantom{-} 0.01^{+ 0.15}_{- 0.15}$ & $-0.00^{+ 0.19}_{- 0.19}$ \\
$C_{nr}$ & $\phantom{-} 0.02^{+ 0.08}_{- 0.07}$ & $\phantom{-} 0.02^{+ 0.10}_{- 0.10}$ & $\phantom{-} 0.03^{+ 0.03}_{- 0.03}$ & $\phantom{-} 0.01^{+ 0.03}_{- 0.03}$ & $\phantom{-} 0.03^{+ 0.16}_{- 0.16}$ & $\phantom{-} 0.03^{+ 0.15}_{- 0.15}$ & $\phantom{-} 0.03^{+ 0.17}_{- 0.18}$ & $\phantom{-} 0.02^{+ 0.66}_{- 0.28}$ & $\phantom{-} 0.02^{+ 0.21}_{- 0.18}$ & $\phantom{-} 0.03^{+ 0.03}_{- 0.03}$ & $\phantom{-} 0.01^{+ 0.03}_{- 0.04}$ & $\phantom{-} 0.03^{+ 0.23}_{- 0.23}$ & $\phantom{-} 0.03^{+ 0.27}_{- 0.28}$ & $\phantom{-} 0.03^{+ 0.23}_{- 0.24}$ \\
$C_{rk}$ & $-0.00^{+ 0.01}_{- 0.01}$ & $\phantom{-} 0.00^{+ 0.09}_{- 0.09}$ & $\phantom{-} 0.01^{+ 0.09}_{- 0.09}$ & $-0.00^{+ 0.06}_{- 0.06}$ & $-0.00^{+ 0.05}_{- 0.05}$ & $-0.03^{+ 0.24}_{- 0.24}$ & $\phantom{-} 0.00^{+ 0.01}_{- 0.01}$ & $-0.00^{+ 0.17}_{- 0.06}$ & $\phantom{-} 0.00^{+ 0.19}_{- 0.16}$ & $\phantom{-} 0.01^{+ 0.21}_{- 0.20}$ & $-0.00^{+ 0.08}_{- 0.08}$ & $-0.00^{+ 0.05}_{- 0.05}$ & $-0.03^{+ 0.43}_{- 0.44}$ & $\phantom{-} 0.00^{+ 0.01}_{- 0.01}$ \\
$C_{rn}$ & $\phantom{-} 0.02^{+ 0.10}_{- 0.10}$ & $\phantom{-} 0.03^{+ 0.01}_{- 0.01}$ & $\phantom{-} 0.02^{+ 0.26}_{- 0.26}$ & $\phantom{-} 0.02^{+ 0.06}_{- 0.06}$ & $\phantom{-} 0.02^{+ 0.05}_{- 0.05}$ & $\phantom{-} 0.01^{+ 0.28}_{- 0.28}$ & $\phantom{-} 0.00^{+ 0.13}_{- 0.13}$ & $\phantom{-} 0.02^{+ 0.80}_{- 0.39}$ & $\phantom{-} 0.03^{+ 0.03}_{- 0.03}$ & $\phantom{-} 0.02^{+ 0.30}_{- 0.30}$ & $\phantom{-} 0.02^{+ 0.08}_{- 0.08}$ & $\phantom{-} 0.02^{+ 0.07}_{- 0.07}$ & $\phantom{-} 0.01^{+ 0.31}_{- 0.32}$ & $\phantom{-} 0.00^{+ 0.18}_{- 0.19}$ \\
$C_{nk}$ & $-0.00^{+ 0.03}_{- 0.03}$ & $\phantom{-} 0.00^{+ 0.09}_{- 0.09}$ & $\phantom{-} 0.00^{+ 0.24}_{- 0.24}$ & $\phantom{-} 0.00^{+ 0.05}_{- 0.05}$ & $\phantom{-} 0.00^{+ 0.11}_{- 0.11}$ & $-0.01^{+ 0.43}_{- 0.44}$ & $-0.00^{+ 0.01}_{- 0.01}$ & $-0.00^{+ 0.52}_{- 0.16}$ & $\phantom{-} 0.00^{+ 0.34}_{- 0.20}$ & $\phantom{-} 0.00^{+ 0.68}_{- 0.65}$ & $\phantom{-} 0.00^{+ 0.06}_{- 0.06}$ & $\phantom{-} 0.00^{+ 0.13}_{- 0.12}$ & $-0.01^{+ 0.56}_{- 0.58}$ & $-0.00^{+ 0.01}_{- 0.01}$ \\

            \bottomrule
        \end{tabular}%
    }

\end{table}
\end{landscape}

\begin{landscape}
\begin{table}
    \centering
    \caption{
    Summary of the fitted values for the polarization terms ($B^\pm_i$) and spin correlation terms ($C_{ij}$) in the signal region using kinematic method. Each entry represents the central fitted value with its corresponding upper and lower uncertainties. Uncertainties are multiplied by 100 for better readability due to their small magnitudes.
    }
    \label{tab:full_spin_density_result_kinematic_method}
    \renewcommand{\arraystretch}{1.05} 
    \resizebox{0.95\linewidth}{!}{%
        \begin{tabular}{c | c c c c c c c c | c c c c c c c c }
            \toprule

 & \multicolumn{7}{c}{Kinematic Method (SR Only)} & \multicolumn{7}{c}{Kinematic Method (SR \& Trigger)} \\ \midrule
 & $\pi \pi$ & $\pi \rho$ & $\rho \rho$ & $e \pi$ & $\mu \pi$ & $e \rho$ & $\mu \rho$ & $\pi \pi$ & $\pi \rho$ & $\rho \rho$ & $e \pi$ & $\mu \pi$ & $e \rho$ & $\mu \rho$ \\ \midrule
$B^{-}_{k}$ & $\mathbf{-0.17}^{+\mathbf{ 0.08}}_{-\mathbf{ 0.08}}$ & $\mathbf{-0.18}^{+\mathbf{ 0.05}}_{-\mathbf{ 0.05}}$ & $\mathbf{-0.20}^{+\mathbf{ 0.06}}_{-\mathbf{ 0.06}}$ & $\mathbf{-0.19}^{+\mathbf{ 0.06}}_{-\mathbf{ 0.06}}$ & $\mathbf{-0.19}^{+\mathbf{ 0.09}}_{-\mathbf{ 0.09}}$ & $\mathbf{-0.20}^{+\mathbf{ 0.08}}_{-\mathbf{ 0.08}}$ & $\mathbf{-0.20}^{+\mathbf{ 0.06}}_{-\mathbf{ 0.06}}$ & $\mathbf{-0.20}^{+\mathbf{ 0.17}}_{-\mathbf{ 0.17}}$ & $\mathbf{-0.20}^{+\mathbf{ 0.15}}_{-\mathbf{ 0.15}}$ & $\mathbf{-0.21}^{+\mathbf{ 0.09}}_{-\mathbf{ 0.09}}$ & $\mathbf{-0.20}^{+\mathbf{ 0.08}}_{-\mathbf{ 0.08}}$ & $\mathbf{-0.20}^{+\mathbf{ 0.11}}_{-\mathbf{ 0.11}}$ & $\mathbf{-0.21}^{+\mathbf{ 0.08}}_{-\mathbf{ 0.08}}$ & $\mathbf{-0.21}^{+\mathbf{ 0.06}}_{-\mathbf{ 0.06}}$ \\
$B^{-}_{n}$ & $\phantom{-} 0.00^{+ 0.00}_{- 0.00}$ & $\phantom{-} 0.00^{+ 0.00}_{- 0.00}$ & $\phantom{-} 0.00^{+ 0.00}_{- 0.00}$ & $\phantom{-} 0.00^{+ 0.00}_{- 0.00}$ & $\phantom{-} 0.00^{+ 0.00}_{- 0.00}$ & $\phantom{-} 0.00^{+ 0.00}_{- 0.00}$ & $\phantom{-} 0.00^{+ 0.00}_{- 0.00}$ & $\phantom{-} 0.00^{+ 0.00}_{- 0.00}$ & $\phantom{-} 0.00^{+ 0.00}_{- 0.00}$ & $\phantom{-} 0.00^{+ 0.00}_{- 0.00}$ & $\phantom{-} 0.00^{+ 0.00}_{- 0.00}$ & $\phantom{-} 0.00^{+ 0.00}_{- 0.00}$ & $\phantom{-} 0.00^{+ 0.00}_{- 0.00}$ & $\phantom{-} 0.00^{+ 0.00}_{- 0.00}$ \\
$B^{-}_{r}$ & $\phantom{-} 0.00^{+ 0.00}_{- 0.00}$ & $-0.00^{+ 0.00}_{- 0.00}$ & $-0.00^{+ 0.00}_{- 0.00}$ & $-0.00^{+ 0.00}_{- 0.00}$ & $-0.00^{+ 0.00}_{- 0.00}$ & $-0.00^{+ 0.00}_{- 0.00}$ & $-0.00^{+ 0.00}_{- 0.00}$ & $-0.00^{+ 0.00}_{- 0.00}$ & $-0.00^{+ 0.00}_{- 0.00}$ & $-0.00^{+ 0.00}_{- 0.00}$ & $-0.00^{+ 0.00}_{- 0.00}$ & $-0.00^{+ 0.00}_{- 0.00}$ & $-0.00^{+ 0.00}_{- 0.00}$ & $-0.00^{+ 0.00}_{- 0.00}$ \\
$B^{+}_{k}$ & $\mathbf{-0.17}^{+\mathbf{ 0.08}}_{-\mathbf{ 0.08}}$ & $\mathbf{-0.18}^{+\mathbf{ 0.05}}_{-\mathbf{ 0.05}}$ & $\mathbf{-0.20}^{+\mathbf{ 0.06}}_{-\mathbf{ 0.06}}$ & $\mathbf{-0.19}^{+\mathbf{ 0.06}}_{-\mathbf{ 0.06}}$ & $\mathbf{-0.19}^{+\mathbf{ 0.09}}_{-\mathbf{ 0.09}}$ & $\mathbf{-0.20}^{+\mathbf{ 0.08}}_{-\mathbf{ 0.08}}$ & $\mathbf{-0.20}^{+\mathbf{ 0.06}}_{-\mathbf{ 0.06}}$ & $\mathbf{-0.20}^{+\mathbf{ 0.17}}_{-\mathbf{ 0.17}}$ & $\mathbf{-0.20}^{+\mathbf{ 0.15}}_{-\mathbf{ 0.15}}$ & $\mathbf{-0.21}^{+\mathbf{ 0.09}}_{-\mathbf{ 0.09}}$ & $\mathbf{-0.20}^{+\mathbf{ 0.08}}_{-\mathbf{ 0.08}}$ & $\mathbf{-0.20}^{+\mathbf{ 0.11}}_{-\mathbf{ 0.11}}$ & $\mathbf{-0.21}^{+\mathbf{ 0.08}}_{-\mathbf{ 0.08}}$ & $\mathbf{-0.21}^{+\mathbf{ 0.06}}_{-\mathbf{ 0.06}}$ \\
$B^{+}_{n}$ & $\phantom{-} 0.00^{+ 0.00}_{- 0.00}$ & $\phantom{-} 0.00^{+ 0.00}_{- 0.00}$ & $\phantom{-} 0.00^{+ 0.00}_{- 0.00}$ & $\phantom{-} 0.00^{+ 0.00}_{- 0.00}$ & $\phantom{-} 0.00^{+ 0.00}_{- 0.00}$ & $\phantom{-} 0.00^{+ 0.00}_{- 0.00}$ & $\phantom{-} 0.00^{+ 0.00}_{- 0.00}$ & $\phantom{-} 0.00^{+ 0.00}_{- 0.00}$ & $\phantom{-} 0.00^{+ 0.00}_{- 0.00}$ & $\phantom{-} 0.00^{+ 0.00}_{- 0.00}$ & $\phantom{-} 0.00^{+ 0.00}_{- 0.00}$ & $\phantom{-} 0.00^{+ 0.00}_{- 0.00}$ & $\phantom{-} 0.00^{+ 0.00}_{- 0.00}$ & $\phantom{-} 0.00^{+ 0.00}_{- 0.00}$ \\
$B^{+}_{r}$ & $\phantom{-} 0.00^{+ 0.00}_{- 0.00}$ & $-0.00^{+ 0.00}_{- 0.00}$ & $-0.00^{+ 0.00}_{- 0.00}$ & $-0.00^{+ 0.00}_{- 0.00}$ & $-0.00^{+ 0.00}_{- 0.00}$ & $-0.00^{+ 0.00}_{- 0.00}$ & $-0.00^{+ 0.00}_{- 0.00}$ & $-0.00^{+ 0.00}_{- 0.00}$ & $-0.00^{+ 0.00}_{- 0.00}$ & $-0.00^{+ 0.00}_{- 0.00}$ & $-0.00^{+ 0.00}_{- 0.00}$ & $-0.00^{+ 0.00}_{- 0.00}$ & $-0.00^{+ 0.00}_{- 0.00}$ & $-0.00^{+ 0.00}_{- 0.00}$ \\\midrule
$C_{kk}$ & $\mathbf{\phantom{-} 1.00}^{+\mathbf{ 0.00}}_{-\mathbf{ 0.00}}$ & $\mathbf{\phantom{-} 1.00}^{+\mathbf{ 0.00}}_{-\mathbf{ 0.00}}$ & $\mathbf{\phantom{-} 1.00}^{+\mathbf{ 0.00}}_{-\mathbf{ 0.00}}$ & $\mathbf{\phantom{-} 1.00}^{+\mathbf{ 0.00}}_{-\mathbf{ 0.00}}$ & $\mathbf{\phantom{-} 1.00}^{+\mathbf{ 0.00}}_{-\mathbf{ 0.00}}$ & $\mathbf{\phantom{-} 1.00}^{+\mathbf{ 0.00}}_{-\mathbf{ 0.00}}$ & $\mathbf{\phantom{-} 1.00}^{+\mathbf{ 0.00}}_{-\mathbf{ 0.00}}$ & $\mathbf{\phantom{-} 1.00}^{+\mathbf{ 0.00}}_{-\mathbf{ 0.00}}$ & $\mathbf{\phantom{-} 1.00}^{+\mathbf{ 0.00}}_{-\mathbf{ 0.00}}$ & $\mathbf{\phantom{-} 1.00}^{+\mathbf{ 0.00}}_{-\mathbf{ 0.00}}$ & $\mathbf{\phantom{-} 1.00}^{+\mathbf{ 0.00}}_{-\mathbf{ 0.00}}$ & $\mathbf{\phantom{-} 1.00}^{+\mathbf{ 0.00}}_{-\mathbf{ 0.00}}$ & $\mathbf{\phantom{-} 1.00}^{+\mathbf{ 0.00}}_{-\mathbf{ 0.00}}$ & $\mathbf{\phantom{-} 1.00}^{+\mathbf{ 0.00}}_{-\mathbf{ 0.00}}$ \\
$C_{nn}$ & $\mathbf{\phantom{-} 0.76}^{+\mathbf{ 0.06}}_{-\mathbf{ 0.06}}$ & $\mathbf{\phantom{-} 0.76}^{+\mathbf{ 0.04}}_{-\mathbf{ 0.04}}$ & $\mathbf{\phantom{-} 0.77}^{+\mathbf{ 0.03}}_{-\mathbf{ 0.03}}$ & $\mathbf{\phantom{-} 0.77}^{+\mathbf{ 0.05}}_{-\mathbf{ 0.05}}$ & $\mathbf{\phantom{-} 0.77}^{+\mathbf{ 0.04}}_{-\mathbf{ 0.04}}$ & $\mathbf{\phantom{-} 0.77}^{+\mathbf{ 0.04}}_{-\mathbf{ 0.04}}$ & $\mathbf{\phantom{-} 0.78}^{+\mathbf{ 0.03}}_{-\mathbf{ 0.03}}$ & $\mathbf{\phantom{-} 0.80}^{+\mathbf{ 0.23}}_{-\mathbf{ 0.23}}$ & $\mathbf{\phantom{-} 0.80}^{+\mathbf{ 0.22}}_{-\mathbf{ 0.22}}$ & $\mathbf{\phantom{-} 0.81}^{+\mathbf{ 0.23}}_{-\mathbf{ 0.23}}$ & $\mathbf{\phantom{-} 0.79}^{+\mathbf{ 0.08}}_{-\mathbf{ 0.08}}$ & $\mathbf{\phantom{-} 0.79}^{+\mathbf{ 0.07}}_{-\mathbf{ 0.07}}$ & $\mathbf{\phantom{-} 0.79}^{+\mathbf{ 0.06}}_{-\mathbf{ 0.06}}$ & $\mathbf{\phantom{-} 0.79}^{+\mathbf{ 0.03}}_{-\mathbf{ 0.03}}$ \\
$C_{rr}$ & $\mathbf{-0.76}^{+\mathbf{ 0.06}}_{-\mathbf{ 0.06}}$ & $\mathbf{-0.76}^{+\mathbf{ 0.04}}_{-\mathbf{ 0.04}}$ & $\mathbf{-0.77}^{+\mathbf{ 0.03}}_{-\mathbf{ 0.03}}$ & $\mathbf{-0.77}^{+\mathbf{ 0.05}}_{-\mathbf{ 0.05}}$ & $\mathbf{-0.77}^{+\mathbf{ 0.04}}_{-\mathbf{ 0.04}}$ & $\mathbf{-0.77}^{+\mathbf{ 0.04}}_{-\mathbf{ 0.04}}$ & $\mathbf{-0.78}^{+\mathbf{ 0.03}}_{-\mathbf{ 0.03}}$ & $\mathbf{-0.80}^{+\mathbf{ 0.23}}_{-\mathbf{ 0.23}}$ & $\mathbf{-0.80}^{+\mathbf{ 0.22}}_{-\mathbf{ 0.22}}$ & $\mathbf{-0.81}^{+\mathbf{ 0.23}}_{-\mathbf{ 0.23}}$ & $\mathbf{-0.79}^{+\mathbf{ 0.08}}_{-\mathbf{ 0.08}}$ & $\mathbf{-0.79}^{+\mathbf{ 0.07}}_{-\mathbf{ 0.07}}$ & $\mathbf{-0.79}^{+\mathbf{ 0.06}}_{-\mathbf{ 0.06}}$ & $\mathbf{-0.79}^{+\mathbf{ 0.03}}_{-\mathbf{ 0.03}}$ \\
$C_{kn}$ & $\phantom{-} 0.00^{+ 0.00}_{- 0.00}$ & $\phantom{-} 0.00^{+ 0.00}_{- 0.00}$ & $\phantom{-} 0.00^{+ 0.00}_{- 0.00}$ & $\phantom{-} 0.00^{+ 0.00}_{- 0.00}$ & $\phantom{-} 0.00^{+ 0.00}_{- 0.00}$ & $\phantom{-} 0.00^{+ 0.00}_{- 0.00}$ & $\phantom{-} 0.00^{+ 0.00}_{- 0.00}$ & $\phantom{-} 0.00^{+ 0.00}_{- 0.00}$ & $\phantom{-} 0.00^{+ 0.00}_{- 0.00}$ & $\phantom{-} 0.00^{+ 0.00}_{- 0.00}$ & $\phantom{-} 0.00^{+ 0.00}_{- 0.00}$ & $\phantom{-} 0.00^{+ 0.00}_{- 0.00}$ & $\phantom{-} 0.00^{+ 0.00}_{- 0.00}$ & $\phantom{-} 0.00^{+ 0.00}_{- 0.00}$ \\
$C_{kr}$ & $\phantom{-} 0.00^{+ 0.01}_{- 0.01}$ & $\phantom{-} 0.00^{+ 0.00}_{- 0.00}$ & $\phantom{-} 0.00^{+ 0.00}_{- 0.00}$ & $\phantom{-} 0.00^{+ 0.00}_{- 0.00}$ & $\phantom{-} 0.00^{+ 0.01}_{- 0.01}$ & $\phantom{-} 0.00^{+ 0.01}_{- 0.01}$ & $\phantom{-} 0.00^{+ 0.00}_{- 0.00}$ & $\phantom{-} 0.00^{+ 0.01}_{- 0.01}$ & $\phantom{-} 0.00^{+ 0.01}_{- 0.01}$ & $\phantom{-} 0.01^{+ 0.00}_{- 0.00}$ & $\phantom{-} 0.00^{+ 0.01}_{- 0.01}$ & $\phantom{-} 0.00^{+ 0.01}_{- 0.01}$ & $\phantom{-} 0.01^{+ 0.00}_{- 0.00}$ & $\phantom{-} 0.01^{+ 0.00}_{- 0.00}$ \\
$C_{nr}$ & $\phantom{-} 0.00^{+ 0.00}_{- 0.00}$ & $\phantom{-} 0.00^{+ 0.00}_{- 0.00}$ & $\phantom{-} 0.00^{+ 0.00}_{- 0.00}$ & $\phantom{-} 0.00^{+ 0.00}_{- 0.00}$ & $\phantom{-} 0.00^{+ 0.00}_{- 0.00}$ & $\phantom{-} 0.00^{+ 0.00}_{- 0.00}$ & $\phantom{-} 0.00^{+ 0.00}_{- 0.00}$ & $\phantom{-} 0.00^{+ 0.00}_{- 0.00}$ & $\phantom{-} 0.00^{+ 0.00}_{- 0.00}$ & $\phantom{-} 0.00^{+ 0.00}_{- 0.00}$ & $\phantom{-} 0.00^{+ 0.00}_{- 0.00}$ & $\phantom{-} 0.00^{+ 0.00}_{- 0.00}$ & $\phantom{-} 0.00^{+ 0.00}_{- 0.00}$ & $\phantom{-} 0.00^{+ 0.00}_{- 0.00}$ \\
$C_{rk}$ & $\phantom{-} 0.00^{+ 0.01}_{- 0.01}$ & $\phantom{-} 0.00^{+ 0.00}_{- 0.00}$ & $\phantom{-} 0.00^{+ 0.00}_{- 0.00}$ & $\phantom{-} 0.00^{+ 0.00}_{- 0.00}$ & $\phantom{-} 0.00^{+ 0.01}_{- 0.01}$ & $\phantom{-} 0.00^{+ 0.01}_{- 0.01}$ & $\phantom{-} 0.00^{+ 0.00}_{- 0.00}$ & $\phantom{-} 0.00^{+ 0.01}_{- 0.01}$ & $\phantom{-} 0.00^{+ 0.01}_{- 0.01}$ & $\phantom{-} 0.01^{+ 0.00}_{- 0.00}$ & $\phantom{-} 0.00^{+ 0.01}_{- 0.01}$ & $\phantom{-} 0.00^{+ 0.01}_{- 0.01}$ & $\phantom{-} 0.01^{+ 0.00}_{- 0.00}$ & $\phantom{-} 0.01^{+ 0.00}_{- 0.00}$ \\
$C_{rn}$ & $\phantom{-} 0.00^{+ 0.00}_{- 0.00}$ & $\phantom{-} 0.00^{+ 0.00}_{- 0.00}$ & $\phantom{-} 0.00^{+ 0.00}_{- 0.00}$ & $\phantom{-} 0.00^{+ 0.00}_{- 0.00}$ & $\phantom{-} 0.00^{+ 0.00}_{- 0.00}$ & $\phantom{-} 0.00^{+ 0.00}_{- 0.00}$ & $\phantom{-} 0.00^{+ 0.00}_{- 0.00}$ & $\phantom{-} 0.00^{+ 0.00}_{- 0.00}$ & $\phantom{-} 0.00^{+ 0.00}_{- 0.00}$ & $\phantom{-} 0.00^{+ 0.00}_{- 0.00}$ & $\phantom{-} 0.00^{+ 0.00}_{- 0.00}$ & $\phantom{-} 0.00^{+ 0.00}_{- 0.00}$ & $\phantom{-} 0.00^{+ 0.00}_{- 0.00}$ & $\phantom{-} 0.00^{+ 0.00}_{- 0.00}$ \\
$C_{nk}$ & $\phantom{-} 0.00^{+ 0.00}_{- 0.00}$ & $\phantom{-} 0.00^{+ 0.00}_{- 0.00}$ & $\phantom{-} 0.00^{+ 0.00}_{- 0.00}$ & $\phantom{-} 0.00^{+ 0.00}_{- 0.00}$ & $\phantom{-} 0.00^{+ 0.00}_{- 0.00}$ & $\phantom{-} 0.00^{+ 0.00}_{- 0.00}$ & $\phantom{-} 0.00^{+ 0.00}_{- 0.00}$ & $\phantom{-} 0.00^{+ 0.00}_{- 0.00}$ & $\phantom{-} 0.00^{+ 0.00}_{- 0.00}$ & $\phantom{-} 0.00^{+ 0.00}_{- 0.00}$ & $\phantom{-} 0.00^{+ 0.00}_{- 0.00}$ & $\phantom{-} 0.00^{+ 0.00}_{- 0.00}$ & $\phantom{-} 0.00^{+ 0.00}_{- 0.00}$ & $\phantom{-} 0.00^{+ 0.00}_{- 0.00}$ \\

            \bottomrule
        \end{tabular}%
    }

\end{table}
\end{landscape}

\subsection{Results and Systematics}

\begin{figure}[!htp]
    \centering
    \includegraphics[width=0.76\textwidth]{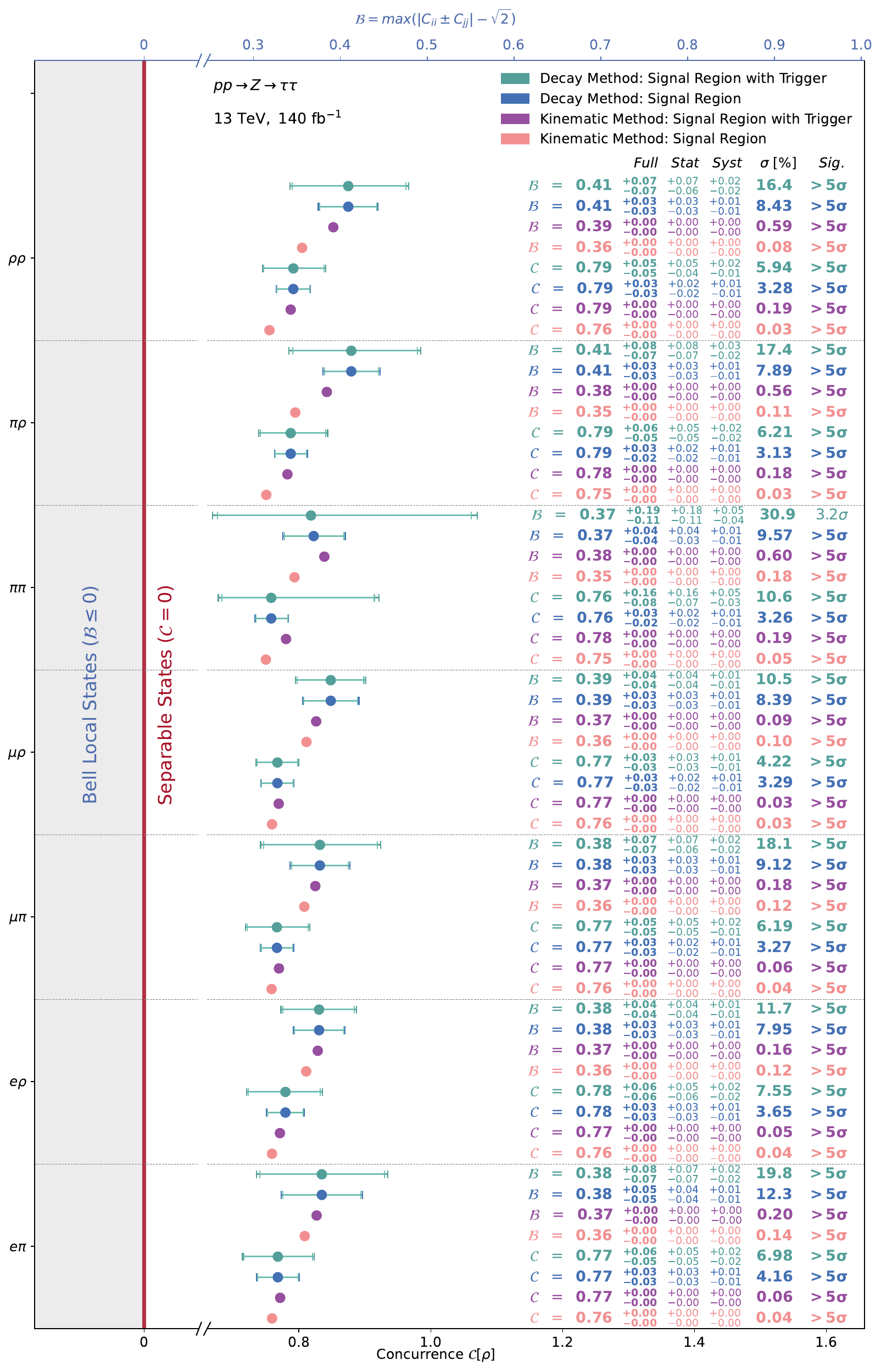}
    \caption{Results for concurrence (bottom axis) and Bell’s nonlocality (top axis).  Results are shown for an integrated luminosity of $140~\text{fb}^{-1}$.}
    \label{fig:poi_results_full}
\end{figure}

Figure~\ref{fig:poi_results_full} shows the full set of results for concurrence and the Bell variable separated by channel.

\begin{figure}[!htp]
    \centering
    \includegraphics[width=0.95\textwidth]{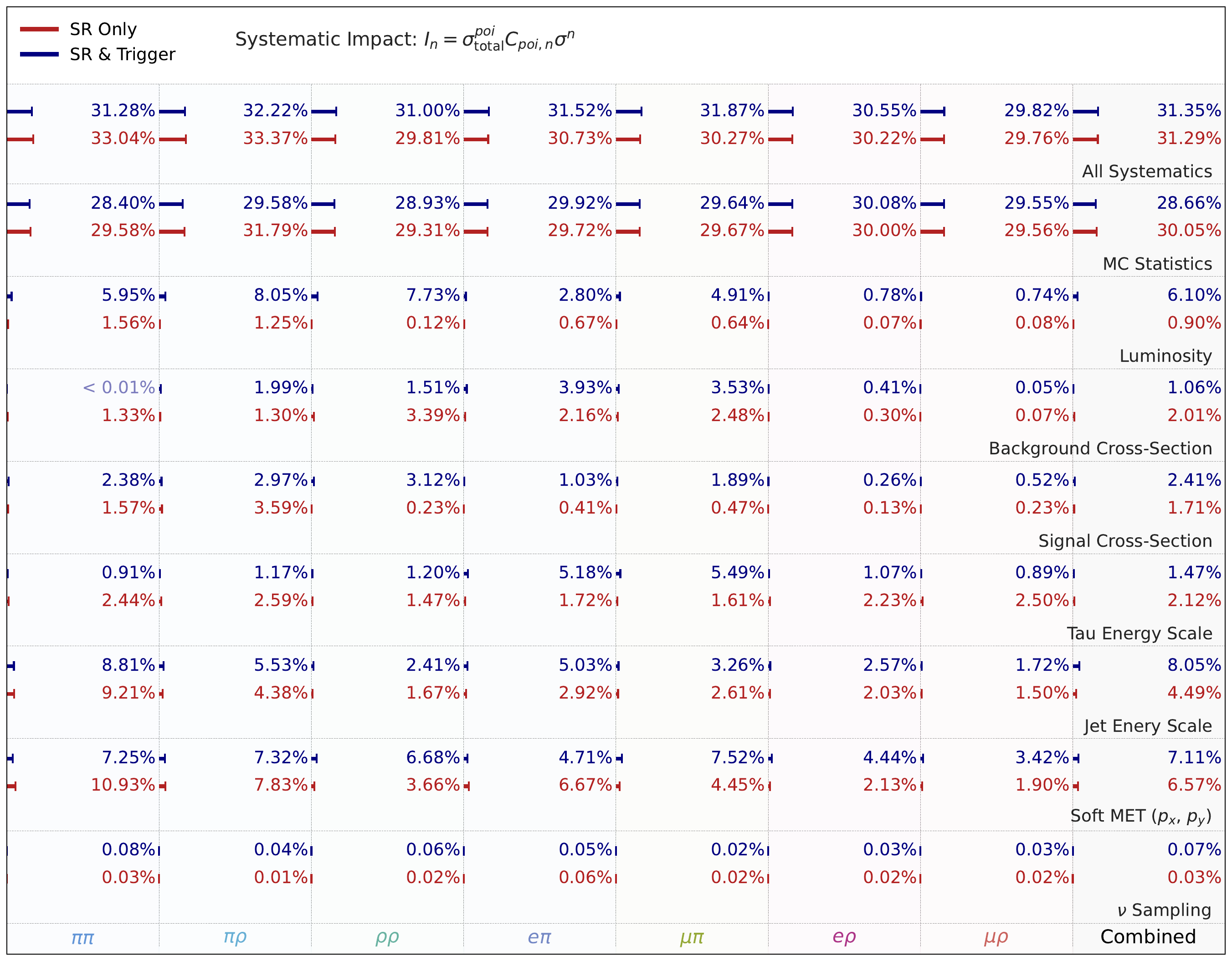}
    \caption{Post-fit impact of individual nuisance parameters on the concurrence and Bell parameter measurements, ranked by their relative effects with respect to the total uncertainty. The impact is computed using the covariance matrix method described in Sec.~\ref{sec:syst_impact}. Blue and red bars correspond to SR only and SR with trigger requirement selections, respectively.}
    \label{fig:merged_systematic_impact}
\end{figure}

A detailed visualization of the post-fit impact of each nuisance parameter on the measured concurrence and Bell parameter is provided in Figure~\ref{fig:merged_systematic_impact}. This summary complements the numerical results shown in Table~\ref{tab:systematics_impact}, illustrating the relative contributions from different sources of systematic uncertainty in the decay-based template fit. The ranking is based on the absolute post-fit impact, as defined by the covariance-matrix-based method discussed in Sec.~\ref{sec:syst_impact}.

\bibliographystyle{jhep}

\providecommand{\href}[2]{#2}\begingroup\raggedright\endgroup

\end{document}